\documentclass[a4paper,fleqn,usenatbib]{aa}
\usepackage{txfonts}
\usepackage{pdflscape}
\usepackage{ae,aecompl}
\usepackage{lscape}
\usepackage{graphicx}
\usepackage{txfonts}
\usepackage{amsmath}	
\usepackage{amssymb}	
\usepackage{orcidlink}
\usepackage{hyperref}
\usepackage{longtable}

\newcommand{\eps}{erg s$^{-1}$~}

\begin{document} 
\title{Multi-wavelength properties of changing-state active galactic nuclei: I. the evolution of soft excess and X-ray continuum}
\author{Arghajit Jana\inst{1, 2}$^{\orcidlink{0000-0001-7500-5752}}$\thanks{E-mail: arghajit.jana@mail.udp.cl},
Claudio Ricci\inst{3, 1, 4}$^{\orcidlink{0000-0001-5231-2645}}$,
Alessia Tortosa\inst{5}$^{\orcidlink{0000-0003-3450-6483}}$,
George Dimopoulos\inst{1}$^{\orcidlink{0009-0002-4945-5121}}$, \\
Benny Trakhtenbrot\inst{6, 7, 8}$^{\orcidlink{0000-0002-3683-7297}}$,
Franz E. Bauer\inst{9}$^{\orcidlink{0000-0002-8686-8737}}$,
Matthew J. Temple\inst{10}$^{\orcidlink{0000-0001-8433-550X}}$,
Michael Koss\inst{11, 12}$^{\orcidlink{0000-0002-7998-9581}}$, \\
Kriti Kamal Gupta\inst{13, 14}$^{\orcidlink{0009-0007-9018-1077}}$,
Hsian-Kuang Chang\inst{15}$^{\orcidlink{0000-0002-5617-3117}}$,
Yaherlyn Diaz\inst{1}$^{\orcidlink{0000-0002-8604-1158}}$,
Dragana Illic\inst{16, 17}$^{\orcidlink{0000-0002-1134-4015}}$,
Krist\'{i}na Kallov\'{a}\inst{1}$^{\orcidlink{0009-0008-8860-0372}}$,
Elena Shablovinskaya\inst{18}$^{\orcidlink{0000-0003-2914-2507}}$ \\ 
}
\institute{$^{1}$ Instituto de Estudios Astrof\'isicos, Facultad de Ingenier\'ia y Ciencias, Universidad Diego Portales, Av. Ej\'ercito Libertador 441, Santiago, Chile \\
$^{2}$ Department of Physics, SRM University-AP, Amaravati 522240, Andhra Pradesh, India \\
$^{3}$ Department of Astronomy, University of Geneva, ch. d’Ecogia 16, 1290, Versoix, Switzerland \\
$^{4}$ Kavli Institute for Astronomy and Astrophysics, Peking University, Beijing 100871, People's Republic of China \\
$^{5}$ INAF-Osservatorio Astronomico di Roma, via di Frascati 33, I-00078 Monte Porzio Catone, Italy\\
$^{6}$ School of Physics and Astronomy, Tel Aviv University, Tel Aviv 69978, Israel \\
$^{7}$ Max-Planck-Institut f{\"u}r extraterrestrische Physik, Gie\ss{}enbachstra\ss{}e 1, 85748 Garching, Germany \\
$^{8}$ Excellence Cluster ORIGINS, Boltzmannsstra\ss{}e 2, 85748, Garching, Germany \\
$^{9}$ Instituto de Alta Investigaci{\'{o}}n, Universidad de Tarapac{\'{a}}, Casilla 7D, Arica, Chile \\
$^{10}$ Centre for Extragalactic Astronomy, Department of Physics, Durham University, South Road, Durham DH1 3LE, UK \\
$^{11}$ Eureka Scientific, 2452 Delmer Street Suite 100, Oakland, CA 94602-3017, USA \\ 
$^{12}$ Space Science Institute, 4750 Walnut Street, Suite 205, Boulder, Colorado 80301, USA \\
$^{13}$ STAR Institute, Li\`ege Universit\'e, Quartier Agora - All\'ee du six Ao\^ut, 19c B-4000 Li\`ege, Belgium \\
$^{14}$ Sterrenkundig Observatorium, Universiteit Gent, Krijgslaan 281 S9, B-9000 Gent, Belgium \\
$^{15}$ Institute of Astronomy, National Tsing Hua University, Hsinchu, 300044, Taiwan\\
$^{16}$ Department of Astronomy, Faculty of Mathematics, University of Belgrade, Studentski trg 16, 11000 Belgrade, Serbia\\
$^{17}$ Hamburger Sternwarte, Universitat Hamburg, Gojenbergsweg 112, D-21029 Hamburg, Germany \\
$^{18}$ Humboldt Research Fellow, Max-Planck-Institut f\"{u}r Radioastronomie, Auf dem Hügel 69, Bonn D-53121, Germany 
}



\date{Received --; accepted --}
\abstract
{Changing-state active galactic nuclei (CSAGNs) exhibit rapid variability, with mass accretion rates that can change by several orders of magnitude in a few years. This provides us with a unique opportunity to study the evolution of the inner accretion flow almost in real time. Here, we used over 1,000 observations to study the broadband X‐ray spectra of a sample of five CSAGNs, spanning three orders of magnitude in Eddington ratio ($\lambda_{\rm Edd}$), using phenomenological models to trace the evolution of key spectral components. We derive several fundamental parameters, such as the photon index, soft excess strength, reflection strength, and luminosities of the soft excess and primary continuum. We find that the soft excess and primary continuum emissions show a very strong positive correlation ($p \ll 10^{-10}$), suggesting a common physical origin. The soft excess strength does not show any dependency on the reflection parameter, suggesting that in these objects the soft excess is not dominated by a blurred ionized reflection process. On the other hand, the strength of the soft excess is found to be strongly positively correlated with the Eddington ratio ($p \ll 10^{-10}$), and we find that the soft excess vanishes below $\log \lambda_{\rm Edd} \sim -2.5$. Moreover, we find a clear `V'-shaped relation for $\Gamma-\lambda_{\rm Edd}$, with a break at $\log \lambda_{\rm Edd}=-2.47\pm0.09$.  Our findings indicate a change in the geometry of the inner accretion flow at low Eddington ratios, and that the soft excess is primarily produced via warm Comptonization.
}
\keywords{accretion, accretion disks – galaxies: active – galaxies: nuclei – quasars: supermassive black holes – X-rays: galaxies}

\titlerunning{SE CSAGNs}
\authorrunning{Jana et al.}
\maketitle
%

\section{Introduction}
\label{sec:intro}

Active galactic nuclei (AGNs) are powered by accreting supermassive black holes (SMBHs) located at the center of galaxies \citep{Rees1988}. 
Matter from the surrounding medium accretes onto the black hole through a geometrically thin, optically thick accretion disk. As the material spirals inward, its gravitational potential energy is efficiently converted into radiation, which is emitted across the entire electromagnetic spectrum. The inner accretion flow consists of a geometrically thin, optically thick disk and a compact X-ray corona. Such a disk emits mainly in UV/optical, producing the characteristic `big blue bump' in the spectral energy distribution \citep[SED; ][]{SS73,Malkan1982,Koratkar1995}, while inverse Compton scattering of the UV photons by hot electrons in the corona produces a power-law X-ray continuum \citep{ST80,HM1991}. An additional excess emission above the power-law continuum is commonly observed below $\sim 1-2$ keV, known as the soft excess \citep[SE,][]{Singh1985,Arnaud1985}. The origin of this SE remains the subject of debate. Two primary scenarios have been proposed to explain the SE: (i) ionized reflection from the inner accretion disk \citep{Crummy2006,Walton2013,Dauser2014} and (ii) thermal Comptonization in a warm corona \citep{Done2012,Petrucci2013}. In the ionized reflection model, soft excess arises from blurred ionized reflection, where fluorescent lines from reprocessed X-ray emission are smeared due to the strong gravitational field near the SMBH \citep[e.g.,][]{Ross2005,Garcia2010,Dauser2016,Ding2024}. In contrast, the warm corona model attributes the SE to thermal Comptonization of seed photons from the AD in an optically thick, warm plasma \citep[e.g.,][]{Magdziarz1998,Petrucci2018}, which is distinct from the hot corona ($kT_{\rm h}\sim 50-100$\,keV; $\tau_{\rm h} \lesssim 1 $) responsible for the primary X-ray power-law emission. The warm corona is typically characterized by an electron temperature of $kT_{\rm w} \sim 0.1-0.2$ keV and an optical depth of $\tau \sim 10-20$ \citep{Petrucci2018}. A hybrid origin involving both components is also proposed \citep{Garcia2019,Laha2022,Xiang2022,Chen2025}.

Due to the high mass of the SMBHs ($M_{\rm BH} \sim 10^6-10^9$\,$M_{\odot}$), the timescales expected for significant optical/UV variations in AGNs are expected to be $\sim 10^4-10^7$\,yr \citep[e.g.,][]{Frank2002,Netzer2013}. Thus, major changes in accretion disk (AD) emission are in principle not expected to be observable on timescales of a few years. This limitation has been overcome by studying large samples of AGN, which includes objects with very different accretion rates. This can offer some insights into the inner accretion flow dynamics across a large range of accretion rates \citep[e.g.,][]{Krolik1999,Netzer2013}.
A complementary approach involves studying black hole X-ray binaries (BHXBs), which can transition between different accretion states on timescales of months, allowing a detailed investigation of accretion physics \citep[e.g.,][]{RM06,Done2007}. Despite the large difference in mass, AGNs might share similar accretion mechanisms with BHXBs \citep[e.g.,][]{Merloni2003,McHardy2006}, as suggested by their comparable spectral and timing properties \citep[e.g.,][]{Sobolewska2011}. Yet, some observational differences remain. For example, AGNs exhibit a soft X-ray excess (SE) below $\sim 1-2$\,keV not seen in BHXBs \citep{Gierlinski2004,Done2012,Ricci2017apjs}, for which the soft X-ray emission is often dominated by the accretion disk \citep[e.g.,][]{RM06}. Additionally, AGNs display a dichotomy in radio loudness, with some sources launching powerful relativistic jets, while others remain radio quiet, even at similar accretion rates (e.g., \citealp{Sikora2007}, see also \citealp{Svoboda2017}) while BHXBs show tighter correlations between jet activity and accretion state \citep{Merloni2003,Fender2004}.

Changing-state AGNs (CSAGNs) provide a unique opportunity to probe AGN accretion physics, as their accretion rates can evolve by $\sim 1-2$ orders of magnitude within weeks to years \citep[e.g.,][]{MacLeod2016,Ruan2019,Trakhtenbrot2019,Temple2023,Ricci2021,Ricci2023Nat}. In UV/optical wavebands, CSAGNs switch between type\,1 (with broad emission lines; BELs) and type\,2 states (without broad emission lines), on timescales ranging from a few months to a few years \citep{Stern2018,Noda2018}. These transitions are primarily attributed to significant changes in the accretion rate \citep{Sheng2017,Noda2018,Ricci2023Nat,AJ2025cl}.

Accretion rate changes in CSAGNs are attributed to the disk instabilities \citep{Noda2018} or external perturbation, such as tidal disruptions \citep{Merloni2015,Ricci2020}. \cite{Noda2018} suggested that the SE ionizes the BLR, leading to BELs in type\,1 states; however, in type\,2 state, as SE intensity diminishes, BELs disappear. In Mrk\,1018, the CS transition was accompanied by a decrease in the Eddington ratio ($\lambda_{\rm Edd} = L_{\rm bol}/L_{\rm Edd}$)
from $\sim 0.08$ to $\sim 0.006$, while the primary X-ray continuum (PC) and SE flux decreased by factors of $\sim 7$ and $\sim 60$, respectively. SE evolution has been observed in other CSAGNs \citep[e.g.,][]{Laha2022,Tripathi2022,Layek2024}, resembling soft-to-hard state transitions in BHXBs around $\lambda_{\rm Edd} \sim 0.01-0.02$ \citep[e.g.,][]{Maccarone2003,Done2007,Yang2015}. A systematic study of optically-identified CLAGNs using \emph{Swift}/BAT light curves by \cite{Temple2023} showed significant $14-195$\,keV flux changes in most CLAGNs during optical transitions, suggesting these events are largely driven by accretion state changes. Using long-term optical and X-ray quasi-simultaneous observation, \cite{AJ2025cl} confirmed this picture, finding that transitions typically occur around $\lambda_{\rm Edd}^{\rm tr} \sim 0.01$, consistent with other studies \citep[e.g.,][]{Ruan2019,Ai2020},

The inner accretion geometry is expected to change at a few percent of the Eddington ratio \citep[e.g.,][]{Esin1997,Yuan2014}.
This change is thought to be imprinted on the observed `V'-shaped photon index ($\Gamma$)--$\lambda_{\rm Edd}$ relation \citep[e.g.,][and references therein]{Shemmer2006,Emmanoulopoulos2012,Trakhtenbrot2017,She2018}. At $\lambda_{\rm Edd} > 0.01$, the accretion disk likely extends close the innermost stable circular orbit (ISCO), accompanied by a compact X-ray corona. At $\lambda_{\rm Edd}<0.01$, the disk is thought to be truncated at larger radii, possibly due to evaporation of the inner region. This creates a hot, radiatively inefficient advective flow (RIAF) that replaces the inner disk and may increase the size of the X-ray emitting region \citep{CT95,Reis2013,Yuan2014,Yang2015}. Although, the exact geometry of the X-ray emitting region is still debated.
The positive $\Gamma-\lambda_{\rm Edd}$ correlation at high accretion rates ($\lambda_{\rm Edd}>0.01$) can be explained by thermal Comptonization, where disk photons are up-scattered in the corona, producing a power-law X-ray spectrum. As $\lambda_{\rm Edd}$ rises, the increased photon flux cools the corona, resulting in a softer spectrum (higher $\Gamma$). Additionally, a higher Eddington ratio increases the compactness of the coronae, and enhances the pair production \citep{Ricci2018}. As the source is expected to remain below the pair line, the increase in compactness would result in a decrease in $kT_{\rm e}$, leading to softer spectra.
At lower accretion rates ($\lambda_{\rm Edd}<0.01$), synchrotron emission from the RIAF or jet base may dominate as the primary seed photons. With decreasing $\lambda_{\rm Edd}$, the density and optical depth of the flow decrease, which would weaken synchrotron self-absorption. This would lead to the production of more seed photons with respect to the power dissipated in the flow, leading to a softening of the spectrum \citep{Zdziarski2014,Yang2015}.

CSAGNs provide therefore a unique opportunity to probe the inner accretion flow evolution on observable timescales. Here, we study the broadband X-ray properties of five CSAGNs to understand how the inner accretion flow with the accretion rates. Additionally, we also explore the evolution of the SE emission to understand its nature and origin. In forthcoming publication, we will investigate the UV-to-X-ray spectral energy distributions (SEDs) of CSAGNs to characterize the disk-corona connection across Eddington ratios.

The current paper is organized as follows. Section~\ref{sec:obs} describes sample selection, observations, and data reduction. Section~\ref{sec:analysis} presents the analysis methods. Section~\ref{sec:res} discusses our findings. Finally, we summarize the results in Section~\ref{sec:summary}. Throughout, we adopt a $\Lambda$CDM cosmology with $H_0 = 70$\,km\,s$^{-1}$\,Mpc$^{-1}$, $\Omega_{\rm M} = 0.3$, and $\Omega_\Lambda = 0.7$.


\section{Sample, Observations, and Data}
\label{sec:obs}

\subsection{Sample selection}
\label{subsec:sample}
The primary aim of this work is to investigate the soft X-ray properties of CSAGNs, especially the soft excess. We draw our sample of CSAGNs from the work of \cite{AJ2025cl}, which studied a sample of optically-identified CLAGNs\citep{Temple2023} selected from the BAT AGN Spectroscopic Survey (BASS) project\footnote{\url{https://www.bass-survey.com/}}. We selected our sample based on two criteria. i) The CSAGNs showed both type\,1 and type\,2 spectral states in the optical regime in post-2000. ii) The CSAGNs have simultaneous multi-epochs and multi-wavelength observations, from UV/optical to X-rays. Based on this, we found nine CSAGNs that meet both criteria. However, we exclude four sources, namely, NGC\,1365, NGC\,3516, NGC\,4151, and NGC\,5548 from our sample due to the presence of warm absorbers and disk-winds \citep{Risaliti2005,Risaliti2009,Mehdipour2022,Beuchert2017,Edelson2017,Mehdipour2022n55}. The presence of warm absorbers and/or disk winds would make it difficult to infer the intrinsic SE properties. The final sample consists of five CSAGNs, namely NGC\,1566, NGC\,2617, Mrk\,590, Mrk\,1018, and IRAS\,23226--3843. NGC\,1566 showed signature of possible outflows in two observations, which we did not consider for our study.
These sources are unobscured, hence, absorption is unlikely to affect the SE. Some of the main properties of these five sources are presented in Table~\ref{tab:info}.

For our sample, we primarily adopt black hole masses from the BASS data release 2 (DR2) catalog, which estimate the $M_{\rm BH}$ in a consistent way using uniform scaling relations and fitting procedures for the broad and narrow components \citep[see][for details]{Koss2022}.
The only exception is NGC\,2617, for which we use the reverberation mapping estimation of $M_{\rm BH}$ \citep{Feng2021}. In BASS DR2, the $M_{\rm BH}$ for NGC\,1566, Mrk\,590, and Mrk\,1018 are estimated using single-epoch measurements based on broad emission lines, while the mass for IRAS\,23226–3843 is obtained via the $M_{\rm BH}$–$\sigma_*$ relation \citep{koss2022b}. Although different mass estimations exist in the literature for these sources, we rely on the BASS DR2 values wherever available to maintain consistency across the sample. We stress that these masses are not homogeneous in the sense of being measured at comparable luminosity or $\lambda_{\rm Edd}$, and systematic uncertainties related to accretion state may affect single epoch virial estimates \citep[see, e.g.,][]{Panda2024}.

\subsection{Data reduction process}
In the present work, we utilized the data obtained from \emph{Swift} \citep{Burrows2005}, \emph{XMM-Newton} \citep{Jansen2001}, \emph{Suzaku} \citep{Koyama2007,Takahashi2007} and \emph{NuSTAR} \citep{Harrison2013}. The observation log is presented in Table~\ref{tab:log}.

\subsubsection{Swift}
\label{subsec:swift}
\emph{Swift} observed the five CSAGNs in our sample 1021 times between 2005 and 2024. We used all available \emph{Swift}/XRT observations, taken in both photon-counting and window-timing modes. To improve the signal-to-noise ratio for spectral analysis and enable detection of soft excess emission, we combined observations under two conditions: (i) consecutive epochs with fluxes consistent within 10\% and total exposure $\geq 3000$\,s or S/N $\geq20$; or (ii) when the next observation was more than 6 months apart. The $0.5-10$\,keV spectra were generated using the online tools from the UK Swift Science Data Centre\footnote{\url{https://www.swift.ac.uk/user_objects/}} \citep{Evans2009}. Source spectra were extracted from circular regions with radii between $28^{\prime\prime}.3$ and $59^{\prime\prime}.0$, depending on source brightness, and background spectra from a $260^{\prime\prime}.0$ region \citep[see][]{Vasudevan2007}. We re-binned the spectra with a minimum of one count per bin using the \textsc{grppha} task.

\emph{Swift}/UVOT provides data in three optical (V, B, U) and three UV (UVW1, UVM2, UVW2) filters. We reduced the level 2 image files and performed photometry with \textsc{uvotsource}, adopting a 5$^{\prime\prime}$ circular aperture centered on the source and a 20$^{\prime\prime}$ background region free of contaminating sources. This yielded source and background counts, fluxes, and magnitudes. UVOT observations obtained simultaneously with XRT exposures were merged for consistency. We estimated the host-galaxy corrected UV fluxes by subtracting the host galaxy fluxes from \cite{Gupta2024}.

\subsubsection{XMM--Newton}
\label{subsec:xmm}
We analyzed a total 42 \emph{XMM–Newton}-EPIC/PN observations of our five CSAGNs in the $0.5-10$ keV energy range. Data reduction was performed using the Standard Analysis Software (SAS) version 20.0.0. The raw PN event files were processed with the \textsc{epchain} task, and particle background flares in the $10-12$ keV energy range were inspected. Good Time Intervals (GTIs) were generated using the \textsc{tabgtigen} task.

Using the \textsc{especget} task, source and background spectra were extracted from a circular region with a radius of $30^{\prime\prime}$, centered on the position of the optical counterpart, and away from the X-ray source on the same CCD, respectively. With the \textsc{epatplot} task, we checked for pileup. We removed the pile-up by adjusting the inner and outer radii of the annular region for the source extraction. The response files were generated using the \textsc{SAS} tasks \textsc{rmfgen} and \textsc{arfgen}. We rebinned the spectra with a minimum of 20 counts per bin using the \textsc{grppha} task.

\subsubsection{NuSTAR}
\label{subsec:nustar}
We utilized a total of 22 \emph{NuSTAR} \citep{Harrison2013} observations for our sample of CSAGNs in the $3-78$\,keV energy range. Using NuSTAR Data Analysis Software \texttt{NuSTARDAS} (version 1.4.1), we reprocessed the data with the latest calibration files available in the NuSTAR calibration database. Clean event files were generated using the \texttt{nupipeline} task, applying the standard filtering criteria. For source and background extraction, we used circular regions with radii of $60^{\prime\prime}$ and $90^{\prime\prime}$, respectively. The source region was centered at the position of the optical counterpart, whereas the background region was chosen away from the source. Spectra were extracted using the \texttt{nuproducts} task and rebinned to ensure a minimum of 20 counts per bin using the \textsc{grppha} tool.

\subsubsection{Suzaku}
\label{subsec:suzaku}
We used four \emph{Suzaku} observations for our analysis, utilizing data from both the X-ray Imaging Spectrometer (XIS) and the Hard X-ray Detector (HXD). The XIS consisted of four CCDs; XIS-0, XIS-2, and XIS-3 (front-illuminated), and XIS-1 (back-illuminated). Since XIS-2 was non-operational, only XIS-0, XIS-1, and XIS-3 were used. Standard data reduction was performed using \textsc{ftools} v6.25 with the latest calibration files. Source and background spectra were extracted from a 250$^{\prime\prime}$ circular region centered on the source, away from the source, respectively. Response files were generated using \textsc{xisrmfgen} and \textsc{xisarfgen}. The XIS-0 and XIS-3 spectra (2--10\,keV) were combined using \textsc{addascaspec}, while the XIS-1 spectrum ($0.5-10$\,keV) was analyzed separately. The $1.6-2$\,keV range was excluded due to the known Si edge. All spectra were binned to a minimum of 20 counts per bin using \textsc{grppha}.

For the HXD/PIN data, cleaned event files were processed with \textsc{aepipeline}, and deadtime-corrected spectra were generated using \textsc{hxdpinxbpi}, incorporating both the non-X-ray background \citep{Fukazawa2009} and simulated cosmic X-ray background \citep{Gruber1999}. HXD/PIN spectra in the $15-40$\,keV range were used for analysis.

\begin{table*}[]
\centering
\caption{General properties of the sample of CSAGNs}
\begin{tabular}{ccccccccc}
\hline
No. & Name & BAT ID & RA  & Dec & z & $d_{\rm L}$ & $\log (M_{\rm BH}/M_{\odot})$ & Ref\\
    &      &        &     &     &   & (Mpc)       &   &             \\
(1) &   (2)& (3)    & (4) & (5) & (6)& (7)        & (8) & (9) \\
\hline
\hline
1 & NGC\,1566  & 216  & 65.002	 & --54.938 & 0.0047 & 17.9 & $6.83$ & A \\
2 & NGC\,2617  & 1327 & 128.912  & --4.088  & 0.0142 & 61.5 & $7.32\pm0.08$ & B \\
3 & Mrk\,590   & 116  & 33.639   & --0.767  & 0.026  & 115.7 & $7.57$ & A \\
4 & Mrk\,1018  & 106  & 31.567   & --0.291  & 0.042  & 188.5 & $7.81$ & A \\
5 & IRAS\,23226--3843& 1194 & 351.359& --38.471 & 0.035 & 158.7 & $7.83$ & A\\
\hline\hline
\end{tabular}
\leftline{Columns: (2) common name of the sources, (3) BAT ID, (4) \& (5) source position in J2000 epoch, (6) redshift }
\leftline{of the source, (7) luminosity distance of the sources, (8) black hole mass, (9) references for mass.} 
\leftline{Mass reference: (A) \cite{koss2022b}, (B) \cite{Feng2021}.}
\label{tab:info}
\end{table*}

\section{Data analysis}
\label{sec:analysis}

\subsection{Swift/XRT}
\label{subsec:xrt}

We performed the spectral analysis using \textsc{xspec} version 12.13.1 \citep{Arnaud1996}. The $0.5-10$ keV \emph{Swift}/XRT spectra were modeled with three components which are modified by absorption. We used \textsc{blackbody (zBB)}, \textsc{cutoffpl (zCUT)}, and \textsc{PEXRAV} \citep{Magdziarz1998} models for the soft X-ray emission, primary continuum emission with high energy cutoff, and reprocessed emission, respectively. We employed two absorption components to account for both Galactic and intrinsic obscuration. Both absorption components were modeled with \textsc{PHABS} models. We also added a Gaussian line (GA) at $\sim 6.4$~keV for the Fe K-line if present. In \textsc{xspec} terminology, the full model is : \textsc{PHABS*zPHABS*(zBB + zCUT + PEXRAV + zGA)}. 

During the analysis, we fixed the high-energy cutoff $E_{\rm cut}$ at 200\,keV, as it is expected to be higher than the \emph{Swift}/XRT coverage. We tied the $\Gamma$, $E_{\rm cut}$ and the normalization of \textsc{PEXRAV} model to that of the \textsc{zCUT} model. We fixed the inclination angle ($i$) at $30^{\circ{}}$, iron ($A_{\rm Fe}$) and metal abundances ($A_{\rm M}$) at Solar value (i.e. 1). The only free parameter of the \textsc{PEXRAV} model is therefore the reflection fraction ($R_{\rm f}$).

For some observations, we could not constrain the \textsc{zBB} model parameters and fixed the blackbody temperature ($kT_{\rm BB}$) at 120\,eV, which is the median for our sample. We obtained an upper limit of the BB flux from these observations. The Gaussian line width (LW) was fixed based on initial fits. LW was set to 0.1, 0.05, or 0.01\,keV for initial values in the ranges $\sim$0.07--0.12, 0.03--0.07, and 0.005--0.03\,keV, respectively. We used Cash statistics for spectral fitting over the $0.5-8$\,keV XRT band and estimated 90\% confidence intervals ($1.6\sigma$) using the \textsc{error} command in \textsc{xspec}. All spectra yielded good fits ($C/{\rm dof} \sim 1$). The detailed result of \emph{Swift}/XRT analysis is available online through CDS.

\begin{figure*}
\centering
\includegraphics[width=0.45\linewidth]{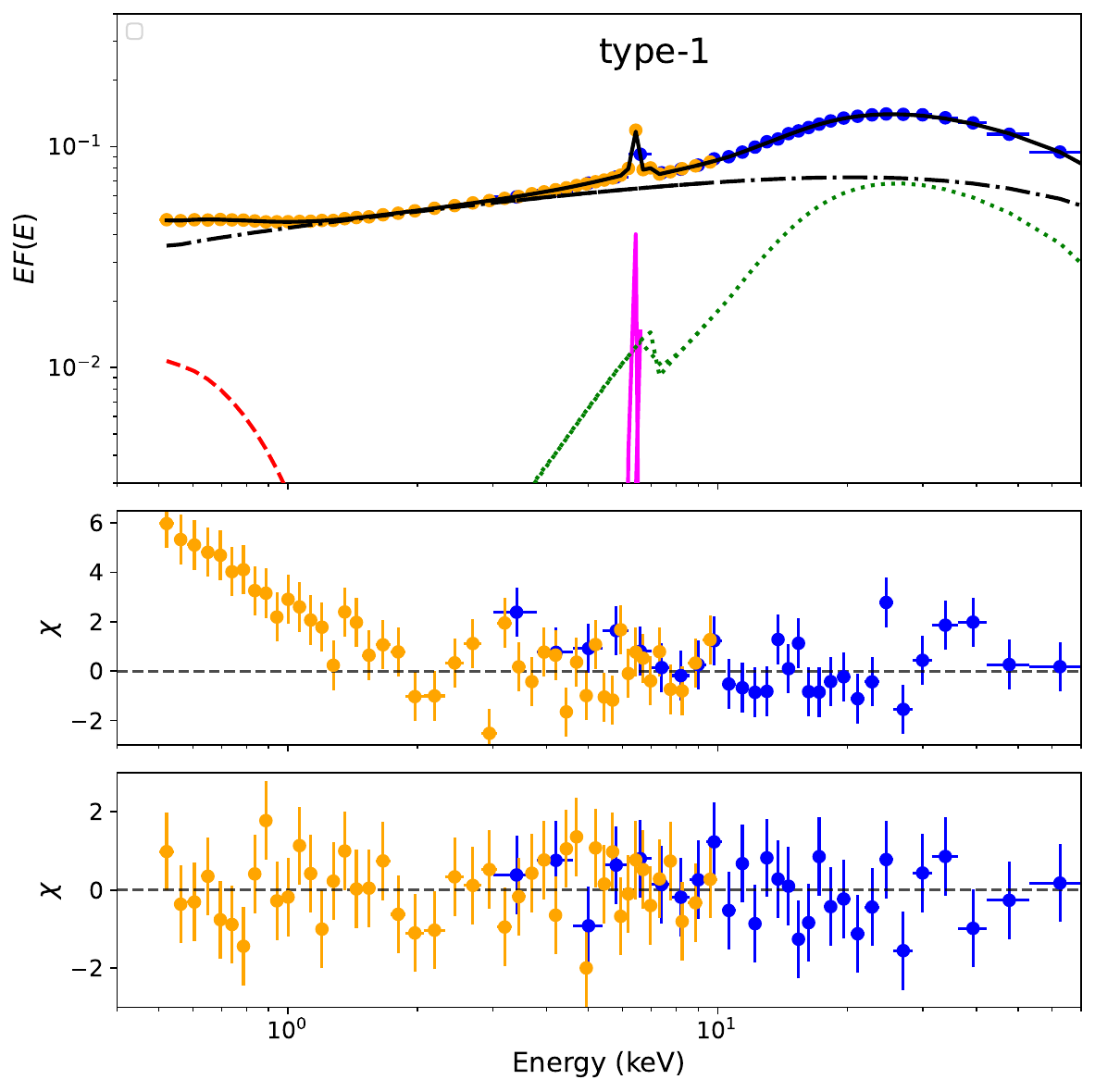}
\includegraphics[width=0.45\linewidth]{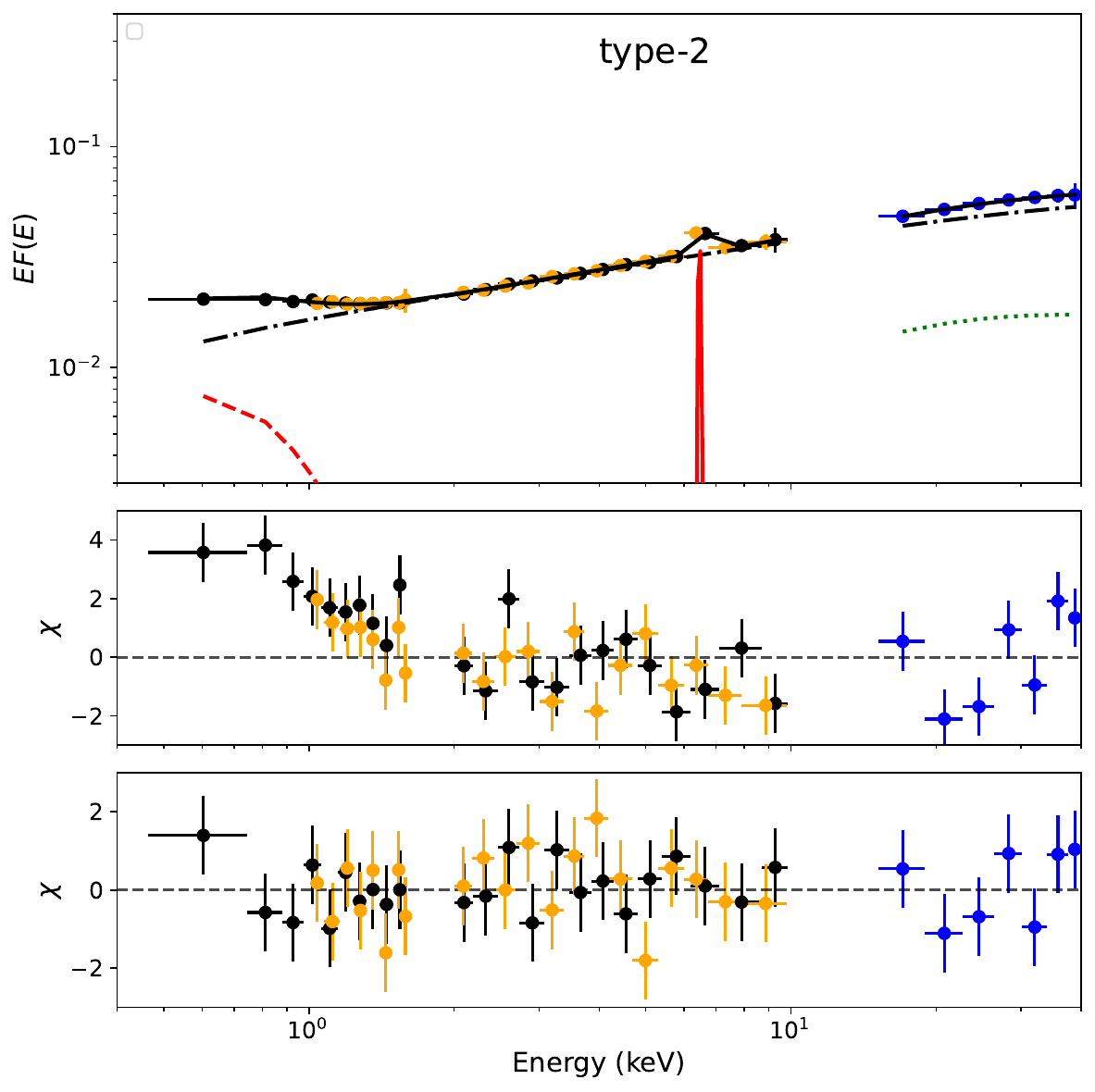}
\caption{Representative unfolded broadband X-ray spectra of NGC\,1566 in type\,1 and type\,2 state, in the left and right panel, respectively. In the left panel, the orange and blue points represent the data from \emph{XMM-Newton}/EPIC-PN and \emph{NuSTAR}/FPMA observations, respectively. 
In the right panels, the black, orange, and blue points represent the \emph{Suzaku}/XIS1, \emph{Suzaku}/XIS0+XIS3, and \emph{Suzaku}/HXD-PIN observations, respectively. The solid black, dashed-dot black, red dashed, solid magenta, and dotted green lines represent the total, continuum, soft excess, iron K-line and reprocessed emission, respectively. The middle figures of each panel show the residual while data are fitted without the `soft-excess' components. The bottom panels show the residuals for the full models.
}
\label{fig:spec}
\end{figure*}

\hspace{-10cm}

\subsection{Broadband X-ray analysis}
\label{subsec:broad_xray}

For the broadband X-ray analysis, we used \emph{NuSTAR} and \emph{Suzaku}/HXD-PIN for the hard X-ray band, and \emph{Swift}/XRT, \emph{Suzaku}/XIS, or \emph{XMM-Newton}/EPIC-PN for the soft band. We applied the same spectral model as in the \emph{Swift}/XRT analysis. A cross-normalization factor ($C$) was included to account for instrument differences. 
Unlike in the XRT analysis, we allowed $E_{\rm cut}$ in the \textsc{zCUT} model to vary. The $\Gamma$, $E_{\rm cut}$, and normalization of \textsc{pexrav} were tied to \textsc{zCUT}. The inclination ($i$), iron ($A_{\rm Fe}$), and metal abundances ($A_{\rm M}$) were fixed at $30^\circ$, 1, and 1, respectively. When the \textsc{blackbody} parameters could not be constrained, we fixed $kT_{\rm BB}$ at 120\,eV. 

We used $\chi^2$ statistics for spectral fitting and estimated 90\% confidence intervals ($1.6\sigma$) using the \textsc{error} command in \textsc{xspec}. Results of the broadband fits are detailed in Section~\ref{subsec:broad_xray}. Figure~\ref{fig:spec} shows representative spectra of NGC\,1566 in both type\,1 and type\,2 states in the left and right panels, respectively. The middle-bottom panels show residuals without SE components while the bottom panels show residuals from the full model.

\subsection{Eddington ratio and bolometric luminosity}
\label{subsec:edd}

Using \emph{Swift}/UVOT and \emph{Swift}/XRT observations, we study the UV-to-X-ray SEDs using three components models: diskbb, blackbody and cut-off powerlaw model for disk, soft-excess and continuum emission, respectively. From the spectral modeling, we estimated the disk luminosity ($L_{\rm disk}$) in $10^{-7}-0.5$\,keV, SE luminosity ($L_{\rm SE}$) in 0.001--10\,keV and continuum luminosity ($L_{\rm PL}$) in 0.1--500\,keV. Then, we calculated the bolometric luminosity as $L_{\rm bol}=L_{\rm disk} + L_{\rm SE} + L_{\rm PL}$. Once, we calculated $L_{\rm bol}$, the Eddington ratio is estimated as $\lambda_{\rm Edd}=L_{\rm bol}/L_{\rm Edd}$, where $L_{\rm Edd} = 1.5\times 10^{38}(M_{\rm BH}/M_{\odot})$ erg s$^{-1}$ is Eddington luminosity. Our results are in good agreement with \citet{Gupta2024}, who derived $\lambda_{\rm Edd}$-dependent bolometric correction factors ($\kappa_{\rm 2-10}$) from SED analysis of unobscured AGNs in the BASS sample. The detailed SED fitting will be presented in a forthcoming publication.

\begin{figure*}
\centering
\includegraphics[width=0.95\linewidth]{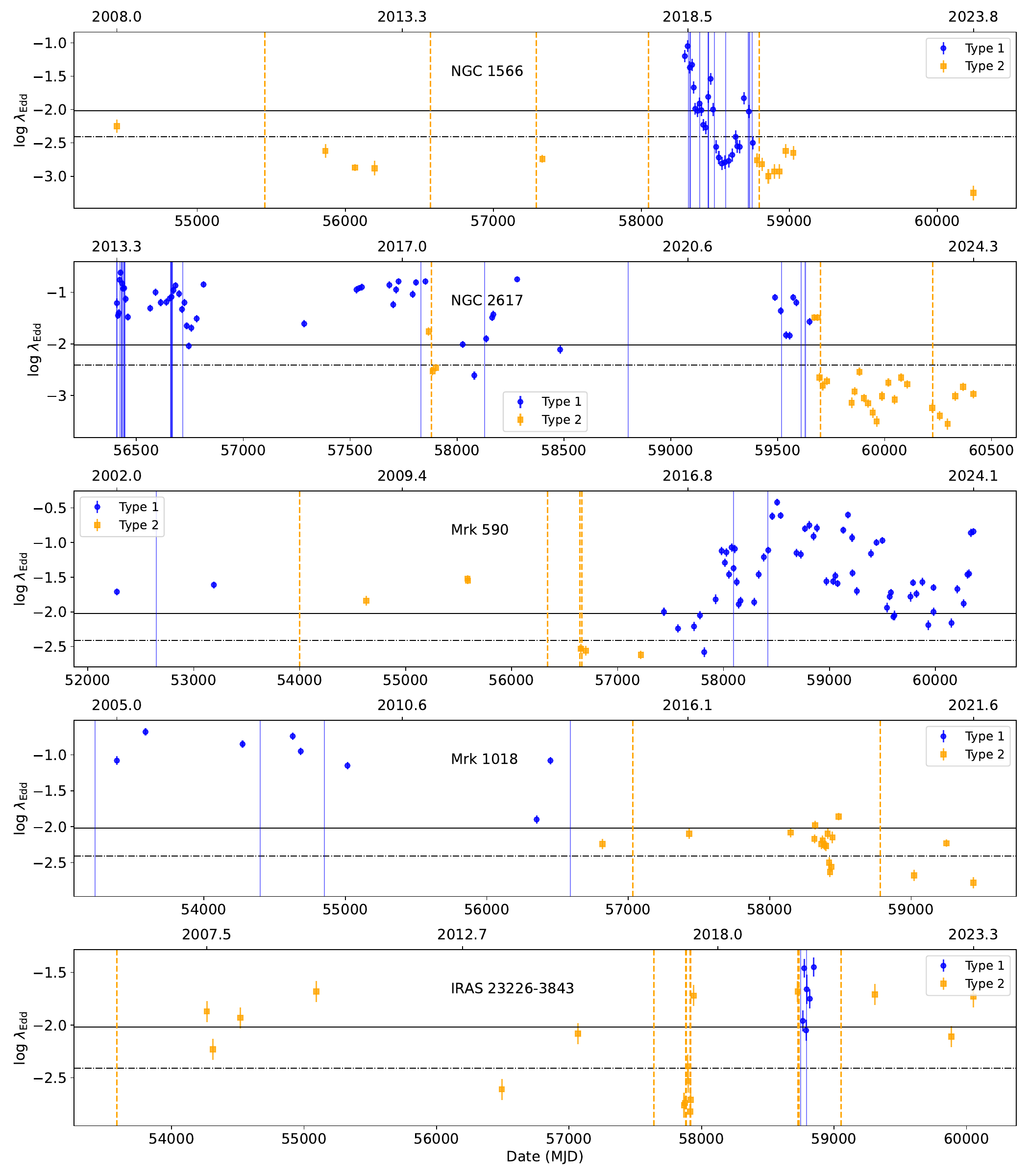}
\caption{Light-curves of all five CSAGNs in different panels. Blue circles and orange diamonds mark the type\,1 and type\,2 states, respectively. The vertical lines in each panel represent the optical observations, where solid blue and orange dot-dashed lines represent the type\,1 and type\,2 spectral states, respectively. The solid horizontal line in each panel represent the median of transition Eddington ratio $\log \lambda_{\rm Edd}^{\rm tr}$ at $-2.01 \pm 0.23$ from \citet{AJ2025cl}. The dot-dashed horizontal line represent the break-Eddington ratio ($\log \lambda_{\rm Edd}^{\rm break}=-2.47\pm0.098$.)}
\label{fig:lc}
\end{figure*}

\subsection{Parameter estimation}
\label{subsec:param}

We obtained several parameters from the spectral analysis, namely, $kT_{\rm BB}$, $\Gamma$, $E_{\rm cut}$, and $R_{\rm f}$. We also calculated the soft excess flux ($F_{\rm SE}^{0.5-2}$), continuum flux ($F_{\rm PC}^{2-10}$), and reflection flux ($F_{\rm ref}^{10-40}$), using the \textsc{blackbody} component in the 0.5--2 keV range, \textsc{cutoffpl} in 2--10 keV, and \textsc{pexrav} in 10--40 keV, respectively. We chose the 0.5--2 keV energy range to represent $F_{\rm SE}^{0.5-2}$ as \textsc{blackbody} with $kT_{\rm BB} \sim 0.1-0.2$ keV is unlikely to contribute to the flux above 2 keV significantly. We represent the continuum emission in the 2--10 keV energy range, as this energy range has historically been used to present the continuum emission. As the reflection hump is generally observed to be most dominant in the $\sim$ 10--40 keV energy range, we used this energy range to represent $F_{\rm ref}^{10-40}$.

We defined the soft excess strength ($Q$) as the ratio between the SE and the continuum luminosity,
\begin{equation}\label{Eq:SE}
Q=L_{\rm SE}^{\rm 0.5-2}/L_{\rm PC}^{\rm 2-10}.
\end{equation}
We also calculated the reflection strength as $R_{\rm S}=F_{\rm ref}^{10-40}/F_{\rm PC}^{2-10}$. 

Following \cite{AJ2025cl}, we marked spectral type\,1--1.5 as type\,1 and type\,1.8--2 as type\,2 in this work. Since simultaneous optical spectra are scarce, we assigned the optical state of each observation based on the nearest available optical spectral classification. Figure~\ref{fig:lc} displays the lightcurve of all five CSAGNs in different panels. Blue circles and orange diamonds mark the type\,1 and type\,2 states, respectively. The vertical lines in each panel represent the optical observations. The solid horizontal lines in each panel represents the median of transition Eddington ratio, $\log \lambda_{\rm Edd}^{\rm tr}$ at $-2.01 \pm 0.23$ (from \citealp{AJ2025cl}). The dot-dashed horizontal lines represent the break-Eddington ratio in the `V'-shaped $\lambda_{\rm Edd}-\Gamma$ relation ($\log \lambda_{\rm Edd}^{\rm break}=-2.47\pm0.09$, see Section~\ref{subsec:gam_edd} for details). 

We caution that some observations show type\,2 state at high-$\lambda_{\rm Edd}$, while some exhibit type\,1 state at low $\lambda_{\rm Edd}$. This discrepancy may reflect the response time of the BLR to changes in the accretion rate. Moreover, the lack of simultaneous optical data for all X-ray observations may lead to misclassifications, especially during the state transitions.

\subsection{Distribution of parameters}
\label{subsec:param}

In our sample, $L_{\rm PC}^{\rm 2-10}$ was found to be in the range $\sim 10^{41}-10^{44}$ \eps. The median of $\log L_{\rm PC}^{\rm 2-10}$ for type\,1 and type\,2 states was found to be $42.89\pm0.03$ and $42.35\pm0.16$, respectively. The SE spanned four orders of magnitude, ranging $\sim10^{39-43}$ \eps. 

In several observations, only upper limits on SE flux were available. For these, we performed 1000 Monte Carlo simulations by sampling the \textsc{blackbody} normalization ($N_{\rm BB}$) uniformly between 0 and its upper limit, and computing the SE flux in each trial. The median was estimated using bootstrapping \citep[see ][for details]{Ricci2018,Gupta2021}. The resulting median $\log L_{\rm SE}^{\rm 0.5-2}$ values are $41.80\pm0.07$ and $40.72\pm0.20$ in type\,1 and type\,2 states, respectively.

\begin{figure}
\centering
\includegraphics[width=0.95\linewidth]{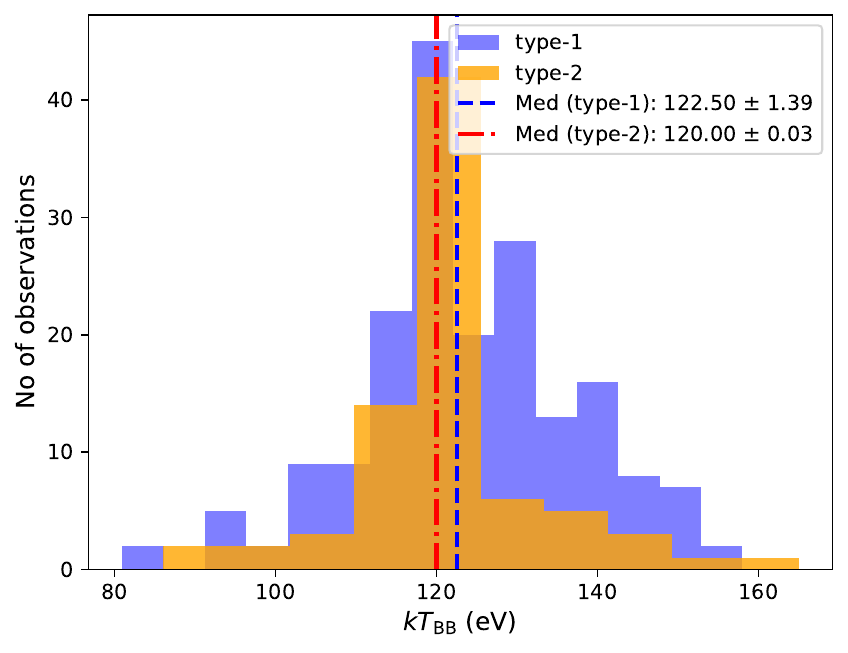}
\caption{Distribution of blackbody temperature ($kT_{\rm BB}$). The blue dashed and red dash-dot lines represent the median value for type\,1 and type\,2 states, respectively.}
\label{fig:hist-bb-gam}
\end{figure}

Figure~\ref{fig:hist-bb-gam} shows the distribution of $kT_{\rm BB}$, which lies in the $\sim 80-160$\,eV range and is similar across states, with medians at $122.5\pm1.4$\,eV and $120.0\pm0.1$\,eV in in type\,1 and type\,2, respectively. The right panel shows the photon index $\Gamma$, ranging in $\sim 1.5-2.1$, with medians of $1.72\pm0.01$ and $1.69\pm0.02$ in type\,1 and type\,2 states, respectively.

$\log \lambda_{\rm Edd}$ spans $\sim -3.6$ to $-0.5$ in our sample. The median of $\log \lambda_{\rm Edd}$ for the two AGN types is obtained at $-1.63\pm0.06$ and $-2.67\pm0.05$, respectively. In type\,1, $\log Q$ peaks around $\sim -0.75$ to $-1.25$, while in type\,2, it is more evenly spread from $\sim -2.4$ to $-1.25$. Median values of $\log Q$ are $-1.10\pm0.03$ and $-1.73\pm0.07$ in type\,1 and type\,2 states, respectively. Median values of all parameters are listed in Table~\ref{tab:median}.

To investigate the correlation between different parameters, we used Spearman rank correlation. The correlation index and p-values are listed in Table~\ref{tab:corr}.
To investigate correlations between various parameters, we performed linear fitting in logarithmic space. 
To properly account for censored data, we employed a bootstrap approach. For each upper limit, a random value was drawn from a uniform distribution between zero and the upper limit, and for each measured value, a random sample was drawn from a Gaussian distribution using the measurement uncertainty as the standard deviation. This procedure produced a simulated dataset, which was fitted using \textsc{linmix} \citep{kelly2009} to obtain the best fit. The procedure was repeated 1000 times, and the average parameters were adopted as the final best-fit values.

For visualization, we estimated binned medians using survival analysis with the \textsc{scikit-survival} package \citep{sksurv2020}, which implements the Kaplan–Meier product-limit (KMPL) estimator \citep{Feigelson1985,Shimizu2017}. This non-parametric method allows for robust estimation of median values in the presence of censored data. For each correlation, we computed KMPL-based medians in bins of the independent variable, ensuring at least 20 data points per bin.

\begin{table}[]
\centering
\caption{Median of the X-ray spectral fit parameters}
\begin{tabular}{ccc}
\hline
Parameters & Type\,1 & Type\,2  \\
\hline
$\log L_{\rm PC}^{\rm 2-10}$ (erg s$^{-1}$) & $42.89\pm0.03$ & $42.35\pm0.16$ \\
$\log L_{\rm SE}^{\rm 0.5-2}$ (erg s$^{-1}$) & $41.80\pm0.07$ & $40.72\pm 0.20$ \\
$kT_{\rm BB}$ (eV) & $122.5\pm1.4$ & $120.0\pm0.1$ \\
$\Gamma$ & $1.72\pm 0.01$ & $1.69\pm0.02$ \\
$\log \lambda_{\rm Edd}$ & $-1.63\pm0.06$ & $-2.67\pm0.05$ \\
$\log Q$ & $-1.10\pm0.03$ & $-1.73\pm0.07$ \\
\hline
\end{tabular}
\label{tab:median}
\end{table}

\begin{table}
\centering
\caption{Spearman correlation analysis result}
\begin{tabular}{cccc}
\hline
Parameter 1 & Parameter 1 & $\rho$ & p-value \\
\hline
$F_{\rm SE}^{\rm 0.5-2}$ & $F_{\rm PC,~2-10}$ & 0.93 & $\ll 10^{-10}$ \\
$L_{\rm SE}^{\rm 0.5-2}$ & $L_{\rm PC}^{\rm 2-10}$ & 0.93 & $\ll 10^{-10}$ \\
$Q$ & $\lambda_{\rm Edd }$ & 0.86 & $\ll 10^{-10}$ \\
$Q$ & $R_{\rm S}$ & --0.04 & 0.83 \\
\hline    
\end{tabular}
\label{tab:corr}
\end{table}

\section{Results and discussion}
\label{sec:res}

We studied five CSAGNs to investigate the origin of the SE and its relationship with primary continuum emission. As CSAGNs can show a wide range of $\lambda_{\rm Edd}$ on short timescales, we took advantage of CSAGNs to explore SE emission and its connection to the inner parts of the accretion flows in AGN. 

\subsection{Soft-excess temperature}
\label{subsec:kt}
We modeled the SE emission using a phenomenological blackbody component. From the spectral fits, we obtained the blackbody temperature ($kT_{\rm BB}$) for each observations. To investigate whether $kT_{\rm BB}$ is related to the global properties of AGNs, we examined its dependence on several key parameters, namely $L_{\rm bol}$, $L_{\rm PC}^{\rm 2-10}$, $L_{\rm SE}^{\rm 0.5-2}$, Q, $\lambda_{\rm Edd}$, and $M_{\rm BH}$. We found no statistically significant correlations between $kT_{\rm BB}$ and any of these AGN parameters. The soft-excess temperature remains remarkably uniform across a wide range of AGN properties, consistent with previous studies \citep[e.g.,][]{Gierlinski2004,Petrucci2018}.

\begin{figure*}
\centering
\includegraphics[width=0.8\linewidth]{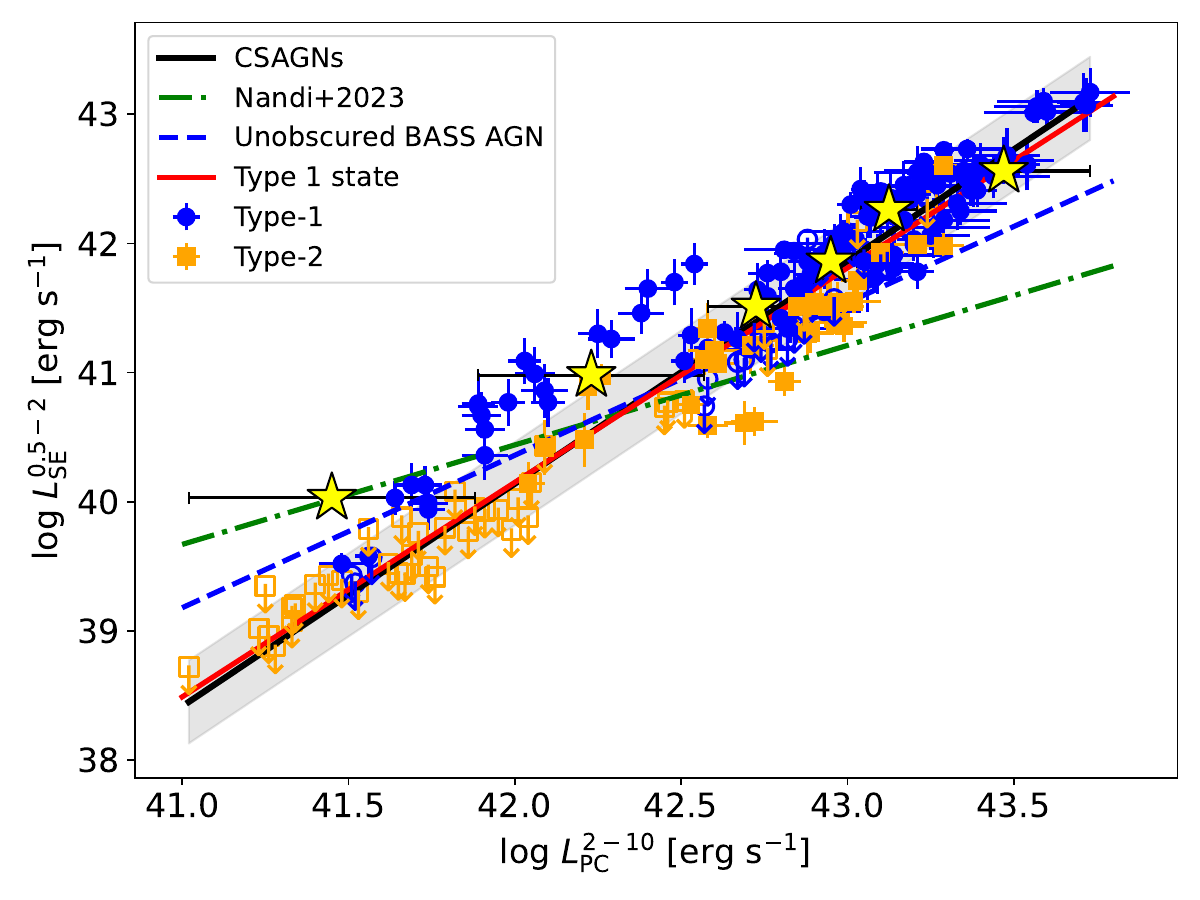}
\caption{Variation of the soft-excess luminosity ($L_{\rm SE}^{\rm 0.5-2}$) in 0.5--2~keV energy range as a function of continuum luminosity ($L_{\rm PC}^{\rm 2-10}$) in 2--10~keV flux. The filled blue circles and orange squares represent the data from type\,1 and type\,2 states, respectively. The hollow blue circles and orange squares represent the upper limit from type\,1 and type\,2 state, respectively. The yellow stars represent the binned data points. The solid black line represents the linear best-fit. The gray regions mark the $1\sigma$ scatter. The red solid line represents the linear best-fit, considering only type\,1 state. The green dashed-dot and blue dashed lines represent the linear best-fit for bare AGNs from \cite{Nandi2023} and unobscured BASS AGNs from Jana et al. (in prep).}
\label{fig:lx-se}
\end{figure*}

\subsection{Primary continuum and soft-excess relation}
\label{subsec:cont_se}

We find a strong positive correlation between SE and PC emission ($p \ll 10^{-10}$). Figure~\ref{fig:lx-se} displays the variation of $L_{\rm SE}^{\rm 0.5-2}$ as a function of $L_{\rm PC}^{\rm 2-10}$. Using linear regression analysis in logarithmic space, we obtained 
\begin{equation}
\log L_{\rm SE,~42}^{\rm 0.5-2}=(-1.77\pm 0.08) + (1.71\pm 0.17) \log L_{\rm PC,~42}^{\rm 2-10},
\end{equation}
with an intrinsic scatter of 0.22\,dex. Here, $L_{\rm SE,~42}^{\rm 0.5-2}=(L_{\rm SE}^{\rm 0.5-2}~{\rm erg~s^{-1}})/10^{42}$ and $L_{\rm PC,~42}^{\rm 2-10}=(L_{\rm PC}^{\rm 2-10}~{\rm erg~s^{-1}})/10^{42}$. We also obtained similar results for $F_{\rm SE}^{\rm 0.5-2}-F_{\rm PC}^{\rm 2-10}$.

Previous studies have reported a strong positive correlation between SE and continuum flux in AGNs \citep[e.g.,][]{Waddell2020,Nandi2023}. Our fit yields a steep slope of $\sim 1.71$, suggesting that SE emission increases more rapidly with PC emission than found in earlier works. For instance, \cite{Nandi2023} analyzed 20 unobscured AGNs and reported a shallower slope of $1.10 \pm 0.04$, using a power-law model over 0.5--10\,keV. However, it is critical to stress that their study used a power-law model to fit the SE component and considered the 0.5--10\,keV energy range for both SE and PC emission. We converted those to 0.5--2\,keV for the SE and 2--10\,keV for the PC emission, for direct comparison with our work. By doing this, we obtained a revised slope of $0.77 \pm 0.04$, which is significantly flatter than the value we derived from our own sample of CSAGNs. 

A similar trend is found for unobscured AGNs in the BASS sample (Jana et al., in prep), which also uses a \textsc{blackbody} model for the SE. The slope for BASS AGNs is $0.73\pm 0.04$, still flatter than our sample of CSAGN. These consistently flatter slopes across different AGN samples, are possibly due to differences in their accretion properties, black hole masses, or spectral modeling approaches. Furthermore, both \cite{Nandi2023} and Jana et al. (in prep) mainly focused on type\,1 AGNs with $\lambda_{\rm Edd}>0.01$, whereas our sample spans a broader range of $\lambda_{\rm Edd}$ ($\sim 0.0003-0.3$) and includes both type\,1 and type\,2 states.

To test the impact of AGN classification, we repeated our analysis using only CSAGNs in the type\,1 state. The slope remains steep at $1.66 \pm 0.21$, which is consistent with the whole sample, and remains significantly steeper than the results of previous studies of type\,1 AGNs. The strong SE-PC correlation across a large range of $\lambda_{\rm Edd}$ supports a common origin, possibly low-temperature Comptonization in a warm corona \citep[e.g.,][]{Petrucci2018,Kubota2018}. 

\begin{figure*}
\centering
\includegraphics[width=0.8\linewidth]{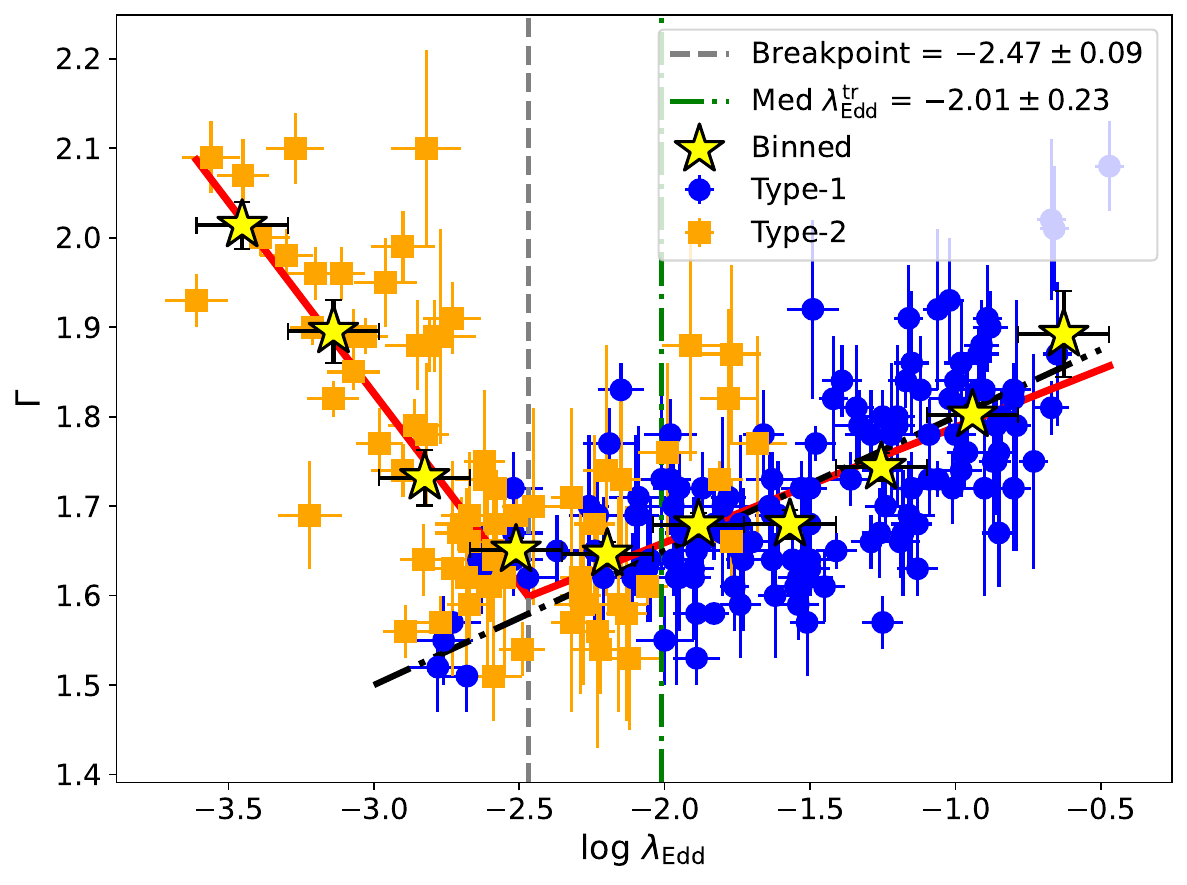}
\caption{Relation between $\Gamma$ and $\lambda_{\rm Edd}$. 
The blue circles and orange squares represent the data from type\,1 and type\,2 states, respectively. The yellow stars represent the binned data points. The red lines represent the linear best-fit of the dataset with a break at $\log \lambda_{\rm Edd}=-2.47\pm 0.09$. The break-point is marked by the vertical gray dashed line. The vertical green dash-dot line represent the median of the transition Eddington ratio ($\lambda_{\rm Edd}^{\rm tr}$) for CSAGNs, which is $\log \lambda_{\rm Edd}^{\rm tr} = -2.01 \pm 0.23$. The $\lambda_{\rm Edd}^{\rm tr}$ is taken from \citet{AJ2025cl}. The black dashed-dot line represents the $\Gamma-\lambda_{\rm Edd}$ relation of BASS AGNs, adopted from \cite{Trakhtenbrot2017}.}
\label{fig:gam_ed}
\end{figure*}

\subsection{$\Gamma-\lambda_{\rm Edd}$ relation and CS transitions}
\label{subsec:gam_edd}

The $\Gamma-\lambda_{\rm Edd}$ relation provides important insights into AGN accretion and disk-corona coupling (\citealp{Shemmer2006,Brightman2013,Yang2015,Trakhtenbrot2017} and references therein). Figure~\ref{fig:gam_ed} shows this relation for our sample. A single-component linear fit of the form $\Gamma = A + B \log \lambda_{\rm Edd}$ fails to capture the observed trend. A two-component linear model with a break at $\log \lambda_{\rm Edd}^{\rm break} = -2.47\pm0.09$ significantly improves the fit, indicating a fundamental change in the accretion properties of AGNs at this $\lambda_{\rm Edd}$.

For $\lambda_{\rm Edd} > \lambda_{\rm Edd}^{\rm break}$, we found a positive correlation between $\Gamma$ and $\lambda_{\rm Edd}$, with a slope of $0.13 \pm 0.07$, indicating that AGNs in this regime exhibit a softer X-ray spectrum at higher accretion rates. This trend is consistent with findings obtained by previous studies focused on nearby BASS AGN (e.g., \citealp{Trakhtenbrot2017}), which also reported a similar positive $\Gamma$–$\lambda_{\rm Edd}$ relation with a slope of $\sim 0.15$ for $\log \lambda_{\rm Edd}>-2.5$. In contrast, for AGNs with $\lambda_{\rm Edd} < \lambda_{\rm Edd}^{\rm break}$, we observed a negative correlation, with a slope of $-0.44 \pm 0.11$. We analyzed the variation of individual CSAGNs in our sample, which revealed a range of $\lambda_{\rm Edd}^{\rm break}$ values, from $0.003$ to $0.006$ (see Appendix~\ref{sec:sources}). This variation in $\lambda_{\rm Edd}^{\rm break}$ suggests that the transition in the $\Gamma$–$\lambda_{\rm Edd}$ relation is a general feature in CSAGN and occurs around $\log \lambda_{\rm Edd}\sim $--2 to --3. 

Previous studies reported a possible `V'-shaped $\Gamma-\lambda_{\rm Edd}$ relation with breaks around $\log \lambda_{\rm Edd} \sim -2$ to $-2.5$ \citep{Gu2009,She2018,Diaz2023}. Recently, \cite{Diaz2023} studied a sample of low-luminosity AGNs and found a similar break at $\log \lambda_{\rm Edd} \sim -2.39$, which closely matches our findings. \cite{She2018} also reported a break at $\log \lambda_{\rm Edd} \sim -2.5$ while studying a sample of nearby AGN with {\it Chandra} observations. 
The break observed in the $\Gamma-\lambda_{\rm Edd}$ relation is consistent with a transition between two distinct accretion states, similar to the hard-to-soft spectral state transitions observed in BHXBs \citep[e.g.,][]{RM06,Gu2009,AJ2022a}. This is in agreement with theoretical expectations and recent observations, which suggest that CSAGNs are undergoing state transitions, analogous to the hard-to-soft transitions in BHXBs \citep[e.g..,][]{Noda2018,Ross2018,Ruan2019,Ai2020,Yan2020,Hagen2024,Kang2025}. In agreement with this idea, using a sample of AGN that includes the sources analyzed here, \cite{AJ2025cl} showed that the median of transition Eddington ratio ($\lambda_{\rm Edd}^{\rm tr}$; at which CSAGNs switch between type\,1 and type\,2 states) is $\log \lambda_{\rm Edd}^{\rm tr} = -2.01 \pm 0.23$, which is similar as $\log \lambda_{\rm Edd}^{\rm break} = -2.47\pm0.09$.

\subsection{The relation between Soft-excess and Eddington ratio}
\label{subsec:se_ed}

To understand how the SE emission changes in the variable objects studied here, we study the dependence of $Q$ (Eq.\,\ref{Eq:SE}) on different AGN parameters. Several studies suggest that soft excess emission varies with AGN properties, hinting to a close connection between soft excess and the accretion process \citep[e.g.,][]{Boissay2016,Noda2018,Ghosh2022,Mehdipour2023,Nandi2023,Chen2025b}. In some cases, particularly at low flux levels, SE emission has been observed to disappear completely, suggesting a strong connection between SE strength and the accretion process \citep[e.g.,][]{Noda2018,Ghosh2022}.

\begin{figure*}
\centering
\includegraphics[width=0.8\linewidth]{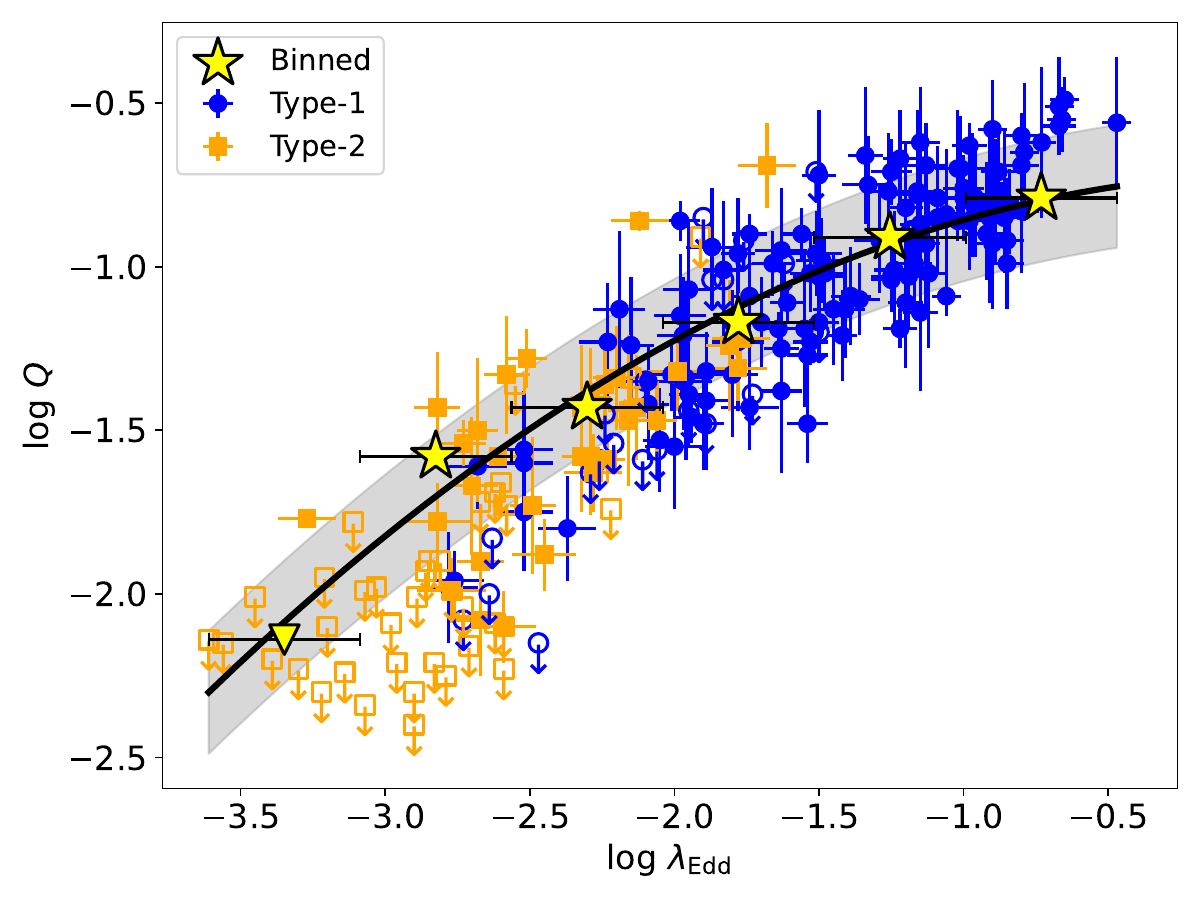}
\caption{The soft-excess strength ($Q$) as a function of Eddington ratio ($\lambda_{\rm Edd}$). The filled blue circles and orange squares represent the data from type\,1 and type\,2 states, respectively, while the hollow blue circles and orange squares represent the upper limit from type\,1 and type\,2 state, respectively. The yellow stars represent the binned data points while the yellow down triangle represent the upper limit. The black dashed line represent the best-fit. The gray region marks the $1\sigma$ scatter.}
\label{fig:q_ed}
\end{figure*}

We studied the dependence of $Q$ on $\lambda_{\rm Edd}$, a fundamental parameter thought to govern the physics of accretion in AGNs \citep[e.g.,][]{Done2007,Ricci2018,Gupta2025}. 
We find that as $\lambda_{\rm Edd}$ increases, the SE emission becomes significantly stronger.
Figure~\ref{fig:q_ed} shows a strong positive correlation between $Q$ and $\lambda_{\rm Edd}$ ($p \ll 10^{-10}$). 
The $Q-\lambda_{\rm Edd}$ relation is best described the second-order polynomial fit,
\begin{multline}
\log Q = (-0.078 \pm 0.019) [\log \lambda_{\rm Edd}]^2  \\
+ (0.382 \pm 0.072) \log \lambda_{\rm Edd} - (0.339 \pm 0.088),
\end{multline}
with an intrinsic scatter of 0.19\,dex.

A positive correlation between $Q$ and $\lambda_{\rm Edd}$ was originally reported by \citet{Boissay2016} for a sample of 102 hard X-ray selected Seyfert\,1 galaxies, although with a very large scatter and flatter slope. Similarly, \citet{Waddell2020} suggested the presence of this trend for broad-line Seyfert\,1 (BLS1) galaxies, but found that the correlation weakens for narrow-line Seyfert\,1 (NLS1) galaxies, which are typically high-$\lambda_{\rm Edd}$ sources. Unlike these studies, our sample includes lower-luminosity AGNs, enabling us to probe SE behavior at $\lambda_{\rm Edd} < 0.01$. In this regime, $Q$ declines sharply, and the soft excess is in many cases not detected, indicating that SE emission weakens or disappears at very low accretion rates. A similar behavior has also been observed in some AGNs where SE emission weakens at the low-accretion state \citep[e.g.,][]{Noda2018,Hagen2024,Chen2025b}

The $Q-\lambda_{\rm Edd}$ trend we find supports a scenario in which SE emission is closely linked to the inner accretion flow and corona. In both warm Comptonization and blurred reflection scenarios one could expect an enhanced SE at higher $\lambda_{\rm Edd}$ due to increased disk ionization/density or more efficient disk-corona coupling. In the reflection scenario, at low $\lambda_{\rm Edd}$, the disk may recede and the corona may expand, reducing the reprocessing efficiency and thus the SE emission \citep[e.g.,][]{Reis2013,Done2007,Wilkins2016}. At high accretion rates, the disk would be more ionized, with an inner radius closer to the ISCO, resulting in stronger reflection. On the other hand, in the warm Comptonization scenario at low $\lambda_{\rm Edd}$, the warm corona may become inefficient or the inner accretion disk may transition to a radiatively inefficient flow, leading to the suppression or disappearance of the soft excess emission \citep[e.g.,][]{Mehdipour2023,Layek2024}. The disappearance or weakening of SE emission at very low $\lambda_{\rm Edd}$ suggests that at $\lambda_{\rm Edd} < 0.01$ the conditions required to produce SE are no longer sustained.

\begin{figure}
\centering
\includegraphics[width=1.0\linewidth]{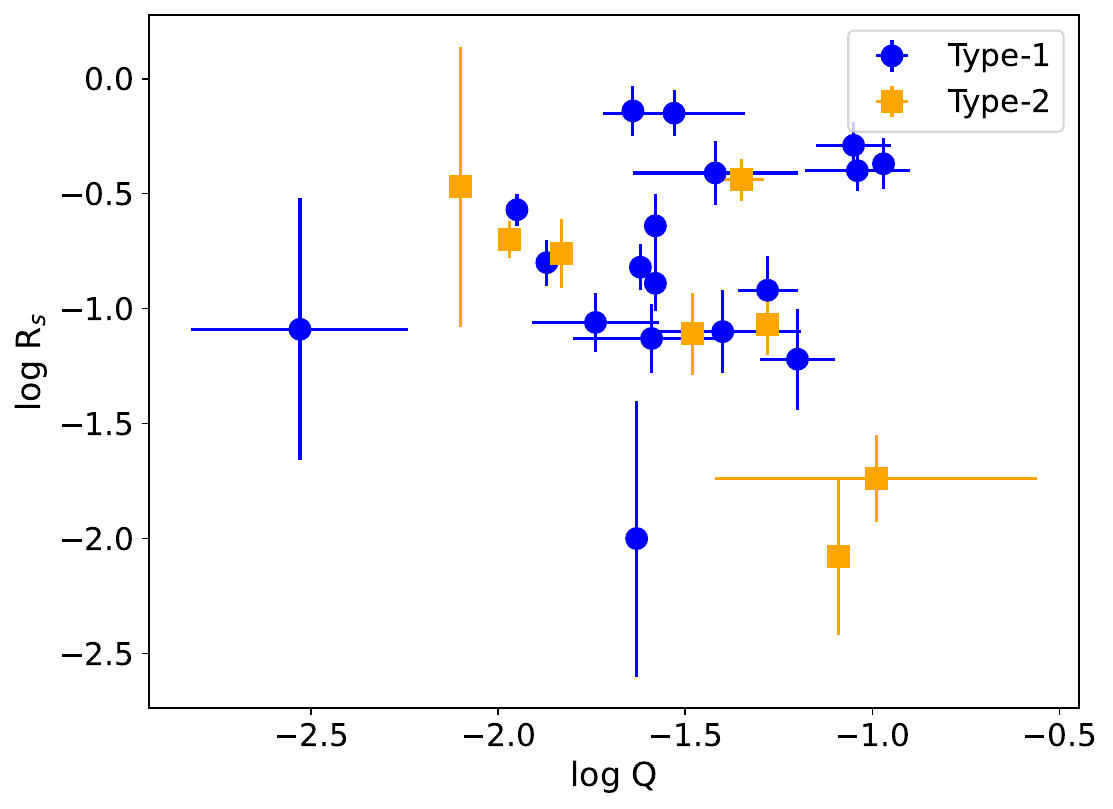}
\caption{The soft-excess strength ($Q$) as a function of reflection strength ($R_{\rm S}$). 
The blue circles and orange squares represent the data from type\,1 and type\,2 states, respectively. }
\label{fig:r_q}
\end{figure}

\subsection{The relation between Soft excess and Compton hump}
\label{subsec:se_r}

To quantify the influence of the Compton hump, we parameterized it using the reflection strength ($R_{\rm S}$). Figure~\ref{fig:r_q} illustrates the variation of $R_{\rm S}$ as a function of $Q$. We did not find any significant correlation between these two quantities, with a p-value of 0.83, suggesting that the reflection strength in CSAGNs is not strongly dependent on $Q$. If the SE originates primarily from blurred reflection, a positive correlation between $Q$ and $R_{\rm S}$ is expected \citep[e.g.,][]{Vasudevan2014,Boissay2016}. However, our results do not support this scenario, which implies that SE and reflection arise from different physical processes. A study by \citet{Waddell2020} found a positive correlation between $Q$ and the hard X-ray excess (similar to $R_{\rm S}$) in a sample of NLS1 galaxies but no correlation in BLS1 galaxies. The authors suggested that in NLS1s, SE originates from reflection, whereas in BLS1s, alternative mechanisms such as warm Comptonization or ionized absorption may be responsible. 

We note that the neutral reflection from the torus and BLR also could contribute to the Compton hump, along with the the reflection from the disk. The reflection properties of our sample have been extensively investigated in previous studies, which consistently report very weak or negligible reflection signatures \citep[e.g.,][]{Parker2019,AJ2021,Ghosh2022,Tripathi2022,Veronese2024}. In many cases, the spectra can be adequately modeled without requiring any reflection component. Even if the torus or the BLR contributes to the Compton hump, the contribution is minimal, and the overall Compton hump flux (torus + ionized disk) remains weak. The combination of a strong SE and weak or absent Compton hump disfavors ionized reflection as the dominant origin of the SE.

\begin{figure*}
\centering
\includegraphics[width=0.8\linewidth]{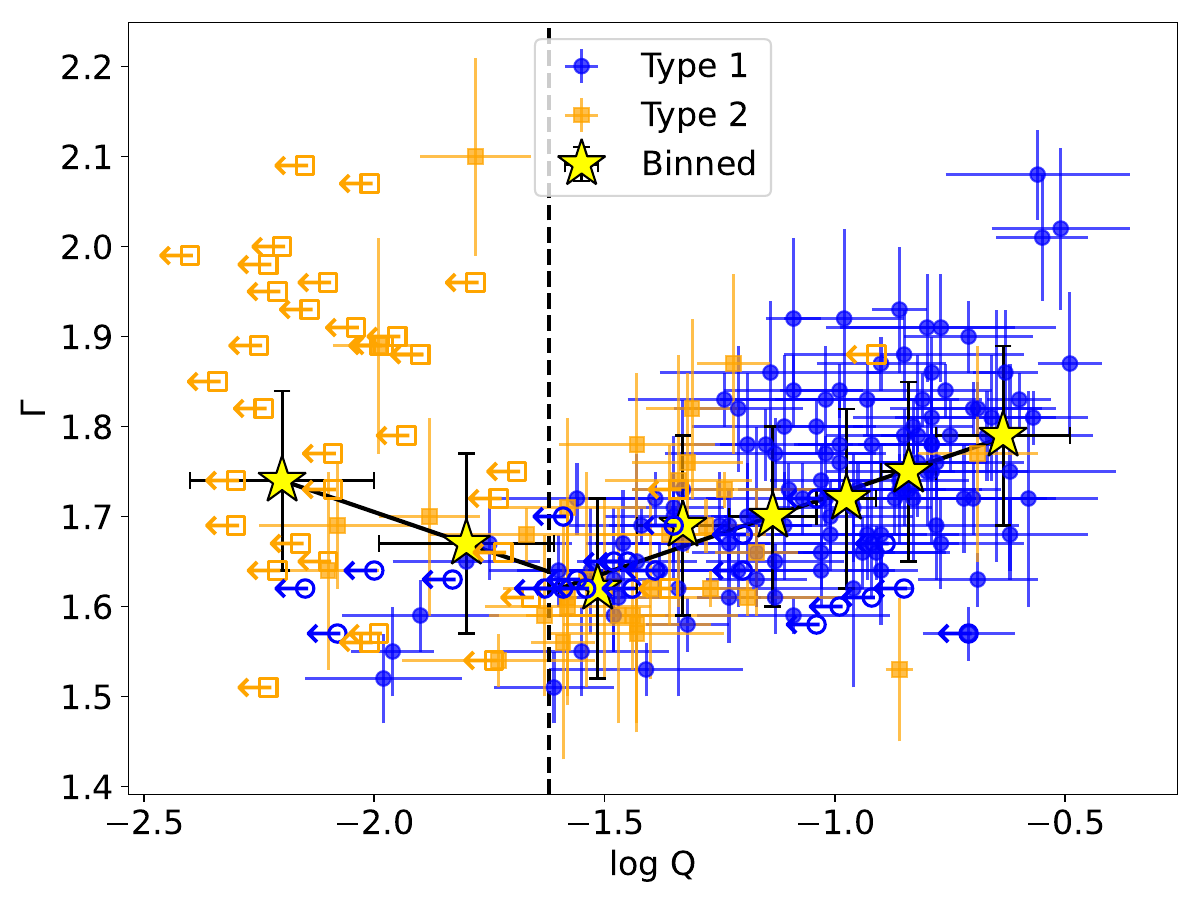}
\caption{The relation between photon index ($\Gamma$) and the soft-excess strength ($Q$) is shown. The blue circles and orange squares represent the data from type\,1 and type\,2 states, respectively. The hollow blue circles and orange squares represent the upper limit from type\,1 and type\,2 states, respectively. The yellow stars represent the binned data points. The black solid lines represent the linear best-fit of the dataset with a break at $\log Q=-1.62\pm 0.12$ which corresponds to $\log \lambda_{\rm Edd}=-2.60\pm0.14$. The break-point is marked by the vertical black dashed line. }
\label{fig:q_gam}
\end{figure*}

\subsection{Soft-excess and photon index relation}
\label{subsec:se_gam}

We explored the relationship between $\Gamma$ and $Q$ for our sample and found a `V'-shaped $\Gamma-Q$ relation (Figure~\ref{fig:q_gam}), similar to the relation between $\Gamma$ and $\lambda_{\rm Edd}$. A Spearman correlation analysis yielded a correlation coefficient of $-0.22$ with $p=0.33$ for $\log Q < -1.53$, and $0.71$ with $p \ll 10^{-5}$ for $\log Q > -1.53$. Given the strong correlation between $Q$ and $\lambda_{\rm Edd}$, the presence of this break is expected. 

\cite{Boissay2016} found a positive correlation between $Q$ and $\Gamma$. Their sample consisted of AGNs with $\log \lambda_{\rm Edd} > -2.2$, therefore our findings for $\log Q > -1.53$ (corresponding to $\log \lambda_{\rm Edd} \sim -2.6$) are consistent with their results. Additionally, \cite{Boissay2016} predicted a weak negative correlation between $\Gamma$ and $Q$ under an ionized reflection scenario using simulations carried out with the \texttt{relxillp\_ion} model. As the result of this study is consistent with the finding of \citet{Boissay2016}, 
our findings are inconsistent with the ionized reflection scenario for $\log \lambda_{\rm Edd} > -2.6$. For $\log Q < -1.53$ (i.e., $\log \lambda_{\rm Edd} < -2.2$), no correlation is observed between $\Gamma$ and $Q$, as we mostly observed an upper limit of $Q$. This suggests that, in this accretion regime, the warm corona may not be linked to the hot corona, or it becomes extremely weak.

\subsection{The origin of the soft excess}
\label{subsec:origin}

The soft excess has remained a long-standing puzzle for four decades after its discovery \citep{Arnaud1985,Singh1985}. Understanding the physical mechanisms responsible for the SE is crucial, as it directly impacts our understanding of accretion processes in AGNs. In the ionized reflection scenario, SE arises from reflection off an ionized AD close to the SMBH. Several studies, including those of \citet{Vasudevan2014} and \citet{Boissay2016}, have predicted specific correlations between spectral parameters within this framework. While a positive correlation between $R_{\rm S}$ and $Q$, as well as a negative correlation between $\Gamma$ and $Q$, are expected, we find no such trends (see Fig.~\ref{fig:q_gam} and Fig.~\ref{fig:r_q}). Moreover, our broadband spectra show weak or absent Compton humps, a key signature of reflection-dominated spectra \citep{Ross2005,Walton2013}. The absence of strong reprocessed emission across a wide range of $\lambda_{\rm Edd}$ further weakens the case for ionized reflection as the primary SE mechanism. This aligns with the findings of previous studies, which indicate that, while ionized reflection can contribute to the SE in certain AGNs, it is unlikely to be the main mechanism producing this component \citep{Gallo2019,Nandi2023}.

In contrast, our results show a tight correlation between SE and continuum flux (see Fig.~\ref{fig:lx-se}) across a wide range of $\lambda_{\rm Edd}$, favoring warm Comptonization. The tight $Q-\lambda_{\rm Edd}$ relation (see Fig.~\ref{fig:gam_ed}) implies that SE emission closely tracks accretion activity, consistent with models where UV disk photons are up-scattered in a warm, optically thick corona \citep{Crummy2006,Done2012,Kubota2018,Middei2020}.
Recently, \citet{Palit2024} suggested that the warm corona size increases with $\lambda_{\rm Edd}$, enhancing SE emission (see also \citealp{Chen2025b}), which is consistent with our findings.

Our study aligns with recent studies of CSAGNs, which suggest that warm Comptonization is the most likely origin for the SE \citep{AJ2021,Tripathi2022,Giustini2017,Veronese2024}. For Mrk\,590, \citet{Laha2022} found that both warm Comptonization and ionized reflection could fit the data. However, our analysis favors warm Comptonization as the leading explanation for SE in our sample. While ionized reflection can become significant in highly-accreting AGNs ($\lambda_{\rm Edd} \sim 1$), this regime is beyond the scope of our study and will be explored in future works. Our results add to the growing consensus that warm Comptonization is a key mechanism shaping the soft X-ray excess in AGNs, especially in systems with moderate to low accretion rates ($\lambda_{\rm Edd} = 0.001-0.3$).

\subsection{SE and CS transitions}
\label{subsec:se-cs}

The SE is believed to play a central role in regulating the ionization state of the BLR, with changes in its flux capable of substantially modifying the ionizing photon budget \citep{Noda2018}. Recent studies have shown that during CS transitions, the SE flux often varies more dramatically than the PC emission, suggesting a strong connection between SE variability and changes in AGN state\citep[e.g.,][]{Noda2018,Mehdipour2022}. For our sample, we find that the median SE and PC luminosities in the type\,1 state are $\sim 6.3 \times 10^{41}\,\rm erg\,s^{-1}$ and $\sim 7.8 \times 10^{42}\,\rm erg\,s^{-1}$, respectively. In contrast, during the type\,2 state, the corresponding median values drop to $\sim 5.2 \times 10^{40}\,\rm erg\,s^{-1}$ and $\sim 2.2 \times 10^{42}\,\rm erg\,s^{-1}$. This corresponds to a factor of $\sim 12$ change in the SE and a factor of $\sim 3.5$ in the PC between the two states, further indicating that the SE exhibits stronger variability across CS transitions. Except for IRAS\,23226--3843, the median SE luminosity changes more drastically than the PC luminosity between spectral states for all sources in our sample. 
This trend is consistent with the broader picture in which a decline in accretion rate reduces the EUV continuum, leading to fewer ionizing photons and, consequently, weaker or absent BELs \citep{Noda2018,Ruan2019}.Our finding is consistent with the previous studies of CSAGNs. For instance, in NGC\,1566, the SE and PC increased by factors of $\sim 200$ and $\sim 30$, respectively, as the source transitioned from a type\,2 to a type\,1 state \citep{Tripathi2022}. A similar trend has been observed in Mrk\,1018, where the SE exhibited stronger variability than the PC across spectral transitions \citep{Noda2018}.

This trend is not restricted to a few rare CSAGNs but observed in general AGN population. Few studies showed a sharp decline in the fraction of broad-line AGNs below $\lambda_{\rm Edd} \sim 0.01-0.02$ \citep{Trump2011,Mitchell2023,Hagen2024,Kang2025}, which coincides with the transition Eddington ratio for CSAGNs.
Additionally, we find that the median $Q$, is $\sim 0.08$ in the type\,1 state and $\sim 0.02$ in the type\,2 state. This result is consistent with the interpretation that the SE diminishes more rapidly than the PC as AGNs transition to type\,2 states. The strong dependence of both $Q$ and SE luminosity on spectral state suggests that the SE is intrinsically connected to CS transitions \citep{Noda2018}. Table~\ref{tab:factor} shows how the median values of different parameters change between the type\,1 and type\,2 states.

\begin{table}
\centering
\caption{Factor of change in the median value of fluxes and soft-excess strength between type\,1 \& type\,2 states.}
\begin{tabular}{cccccc}
\hline
Source &  $f_{\rm SE}$ & $f_{\rm PC}$ & $f_{\rm Q}$ \\
\hline
NGC\,1566     &  51 & 4 & 17 \\
NGC\,2617     &  295 & 34 & 21 \\
Mrk\,590      &  35 & 7 & 8 \\
Mrk\,1018     &  48 & 16 & 7 \\
IRAS\,23226--3843 & 2 & 2 & 1.4 \\
\hline
\end{tabular}
\label{tab:factor}
\leftline{$f_{\rm SE}=L_{\rm SE,~type~1}^{\rm 0.5-2}/L_{\rm SE,~type~2}^{\rm 0.5-2,}$;}
\leftline{$f_{\rm PC}=L_{\rm PC,~type~1}^{\rm 2-10}/L_{\rm PC,~type~2}^{\rm 2-10}$;}
\leftline{$f_{\rm Q}=Q_{\rm type~1}/Q_{\rm type~2}$.}
\end{table}

\section{Summary and conclusions}
\label{sec:summary}

We investigated the X-ray properties of a sample of five CSAGNs to understand the nature of their inner accretion flow, soft excess emission, and their connection to CS transitions. We used a total 42 \emph{XMM-Newton}, 4 \emph{Suzaku}, 22 \emph{NuSTAR}, and 1021 \emph{Swift}/XRT observations for our study. 
Our sample spans three orders of magnitude in Eddington ratios ($\lambda_{\rm Edd} \sim 0.0003-0.3$), covering both type\,1 and type\,2 states. Based on the broadband X-ray spectral study, we retrieve several accretion parameters, namely, photon index, blackbody temperature, soft excess strength, soft excess, and primary continuum luminosity. On the basis of multiple correlations, our findings suggest that SE emission originates from a warm corona across a broad range of accretion rates. In the following sections, we summarize our key findings.

\begin{enumerate}

\item We found that SE and PC emission are tightly correlated, with an intrinsic scatter of 0.22\,dex ($1\sigma$) over four orders of magnitude in luminosity (Fig.~\ref{fig:lx-se}). The soft excess luminosities cover a wider range than primary continuum luminosities, indicating that the SE is more variable than the continuum emission. 

\item The SE emission in CSAGNs varies more rapidly than in AGNs that do not undergo CS transitions. The steeper slope in the SE-PC relation for CSAGNs suggests that these objects experience more pronounced spectral variations, possibly driven by rapid changes in the accretion rate. We also found that the SE emission changes more compared to the PC emission during the CS transitions, suggesting that the SE is intrinsically connected to the optical state change.

\item We observed a very clear `V'-shaped $\Gamma-\lambda_{\rm Edd}$ relation with a break at $\log \lambda_{\rm Edd}=-2.47\pm0.09$ (Fig~\ref{fig:gam_ed}). This break is consistent with $\lambda_{\rm Edd}^{\rm tr}$ for changing-state transitions, which is generally observed at $\lambda_{\rm Edd}^{\rm tr} \sim 0.005-0.015$.

\item The soft-excess strength correlates positively with $\lambda_{\rm Edd}$ (Fig.~\ref{fig:q_ed}), indicating that SE emission increases significantly in high accretion states. This suggests a close connection between SE and the inner accretion disk.

\item We found no correlation between $Q$ and the reflection strength (Fig.~\ref{fig:r_q}), suggesting that SE emission is not dominated by blurred reflection.

\item At high accretion rates ($\lambda_{\rm Edd} > 0.001$), $Q$ and $\Gamma$ exhibit a positive correlation (Fig.~\ref{fig:q_gam}).

\end{enumerate}


Overall, from our sample, we found that the warm Comptonization scenario is the most likely origin of the SE, at least in the range of $\lambda_{\rm Edd}$ probed here ($\lambda_{\rm Edd}\sim 0.003-0.3$). CSAGNs offer crucial insight into the structure and evolution of the inner accretion flow in SMBHs. Our results indicate that SE emission is closely linked to the continuum and is likely dominated by warm Comptonization. This warm coronal emission appears tightly connected to the accretion disk, which radiates primarily in the UV/optical. In a forthcoming study, we will present the UV/optical to X-ray SED to further explore this connection.

Future work will extend the analysis to higher $\lambda_{\rm Edd}$ AGNs (Kallova et al., in prep.; Kumari et al., in prep.) to examine warm Comptonization under different accretion regimes. We find that SE varies more rapidly than the PC during state transitions. Follow-up studies will explore the connection between SE and BELs to understand the physical drivers of BELs using multi-wavelength data. To deepen our understanding of SE and its link to the primary X-ray continuum, future studies with larger AGN samples and high-resolution X-ray spectroscopy will be essential. Missions such as \emph{XPOSAT} \citep{Paul2022polix,Radhakrishna2025xspect} and \emph{NewAthena} \citep{Cruise2025newAthena} will help constrain the physical mechanisms behind SE. Additionally, proposed observatories like \emph{AXIS} \citep{Reynolds2023axis} will be critical for building a comprehensive model of AGN accretion physics and X-ray emission evolution across AGN populations.


\section*{Data availability}
All the data used in the paper are publicly available. The data are available at the CDS via anonymous ftp to cdsarc.u-strasbg.fr (130.79.128.5) or via http://cdsweb.u-strasbg.fr/cgi-bin/qcat?J/A+A/.

\begin{acknowledgements}
We acknowledge support from ANID grants FONDECYT Postdoctoral fellowship 3230303 (AJ) and 3230310 (YD), ANID-Chile BASAL CATA FB210003 (FEB), FONDECYT Regular 1241005 (FEB), and the Millennium Science Initiative, AIM23-0001 (FEB).
C.R.acknowledges support from SNSF Consolidator grant F01$-$13252, Fondecyt Regular grant 1230345, ANID BASAL project FB210003 and the China-Chile joint research fund. 
A.T. acknowledges financial support from the Bando Ricerca Fondamentale INAF 2022 Large Grant ``Toward an holistic view of the Titans: multi-band observations of $z > 6$ QSOs powered by greedy supermassive black holes."
B.T. acknowledges support from the European Research Council (ERC) under the European Union's Horizon 2020 research and innovation program (grant agreement number 950533). 
This research was supported by the Excellence Cluster ORIGINS which is funded by the Deutsche Forschungsgemeinschaft (DFG, German Research Foundation) under Germany's Excellence Strategy - EXC 2094 - 390783311.
B.T. also acknowledges the hospitality of the Instituto de Estudios Astrof\'isicos at Universidad Diego Portales, the Instituto de Astrof\'isica at Pontificia Universidad Cat\'olica de Chile, and the Institut d'Astrophysique de Paris, where parts of this study have been carried out.
MJT acknowledges support from UKRI ST/X001075/1.
M.K. acknowledges support from NASA through ADAP award 80NSSC22K1126.
KKG acknowledges financial support from the Belgian Federal Science Policy Office (BELSPO) in the framework of the PRODEX Programme of the European Space Agency.
HKC acknowledges supports the grant NSTC 113-2112-M-007-020.
D.I. acknowledges funding provided by the University of Belgrade, Faculty of Mathematics (the contract 451-03-136/2025-03/200104) through the grants by the Ministry of Science, Technological Development and Innovation of the Republic of Serbia. 
E.S. acknowledges the support of the Alexander von Humboldt Foundation.
This research has made use of data and/or software provided by the High Energy Astrophysics Science Archive Research Center (HEASARC), which is a service of the Astrophysics Science Division at NASA/GSFC and the High Energy Astrophysics Division of the Smithsonian Astrophysical Observatory. This work has made use of data obtained from the \emph{NuSTAR} mission, a project led by Caltech, funded by NASA and managed by NASA/JPL, and has utilized the NuSTARDAS software package, jointly developed by the ASDC, Italy and Caltech, USA. This research has made use of observations obtained with \emph{XMM–Newton}, an ESA science mission with instruments and contributions directly funded by ESA Member States and NASA. This work made use of XRT data supplied by the UK Swift Science Data Centre at the University of Leicester, UK. This work has made use of data obtained from \emph{Suzaku}, a collaborative mission between the space agencies of Japan (JAXA) and the USA (NASA).
\end{acknowledgements}

%
\bibliographystyle{aa} 
\bibliography{ref-clagn} 

@ARTICLE{Sheng2017,
       author = {{Sheng}, Zhenfeng and {Wang}, Tinggui and {Jiang}, Ning and {Yang}, Chenwei and {Yan}, Lin and {Dou}, Liming and {Peng}, Bo},
        title = "{Mid-infrared Variability of Changing-look AGNs}",
      journal = {\apjl},
         year = 2017,
        month = sep,
       volume = {846},
       number = {1},
          eid = {L7},
        pages = {L7},
          doi = {10.3847/2041-8213/aa85de},
archivePrefix = {arXiv},
       eprint = {1707.02686},
 primaryClass = {astro-ph.GA},
       adsurl = {https://ui.adsabs.harvard.edu/abs/2017ApJ...846L...7S}
}

@ARTICLE{Ricci2021,
       author = {{Ricci}, C. and {Loewenstein}, M. and {Kara}, E. and {Remillard}, R. and {Trakhtenbrot}, B. and {Arcavi}, I. and {Gendreau}, K.~C. and {Arzoumanian}, Z. and {Fabian}, A.~C. and {Li}, R. and {Ho}, L.~C. and {MacLeod}, C.~L. and {Cackett}, E. and {Altamirano}, D. and {Gandhi}, P. and {Kosec}, P. and {Pasham}, D. and {Steiner}, J. and {Chan}, C. -H.},
        title = "{The 450 Day X-Ray Monitoring of the Changing-look AGN 1ES 1927+654}",
      journal = {\apjs},
         year = 2021,
        month = jul,
       volume = {255},
       number = {1},
          eid = {7},
        pages = {7},
          doi = {10.3847/1538-4365/abe94b},
archivePrefix = {arXiv},
       eprint = {2102.05666},
 primaryClass = {astro-ph.HE},
       adsurl = {https://ui.adsabs.harvard.edu/abs/2021ApJS..255....7R}
}

@ARTICLE{Ross2018,
       author = {{Ross}, Nicholas P. and {Ford}, K.~E. Saavik and {Graham}, Matthew and {McKernan}, Barry and {Stern}, Daniel and {Meisner}, Aaron M. and {Assef}, Roberto J. and {Dey}, Arjun and {Drake}, Andrew J. and {Jun}, Hyunsung D. and {Lang}, Dustin},
        title = "{A new physical interpretation of optical and infrared variability in quasars}",
      journal = {\mnras},
         year = 2018,
        month = nov,
       volume = {480},
       number = {4},
        pages = {4468-4479},
          doi = {10.1093/mnras/sty2002},
archivePrefix = {arXiv},
       eprint = {1805.06921},
 primaryClass = {astro-ph.GA},
       adsurl = {https://ui.adsabs.harvard.edu/abs/2018MNRAS.480.4468R}
}

@ARTICLE{Ricci2023Nat,
       author = {{Ricci}, Claudio and {Trakhtenbrot}, Benny},
        title = "{Changing-look active galactic nuclei}",
      journal = {Nature Astronomy},
         year = 2023,
        month = nov,
       volume = {7},
        pages = {1282-1294},
          doi = {10.1038/s41550-023-02108-4},
archivePrefix = {arXiv},
       eprint = {2211.05132},
 primaryClass = {astro-ph.GA},
       adsurl = {https://ui.adsabs.harvard.edu/abs/2023NatAs...7.1282R}
}

@ARTICLE{kelly2009,
       author = {{Kelly}, Brandon C. and {Bechtold}, Jill and {Siemiginowska}, Aneta},
        title = "{Are the Variations in Quasar Optical Flux Driven by Thermal Fluctuations?}",
      journal = {\apj},
         year = 2009,
        month = jun,
       volume = {698},
       number = {1},
        pages = {895-910},
          doi = {10.1088/0004-637X/698/1/895},
archivePrefix = {arXiv},
       eprint = {0903.5315},
 primaryClass = {astro-ph.CO},
       adsurl = {https://ui.adsabs.harvard.edu/abs/2009ApJ...698..895K}
}

@ARTICLE{Waddell2020,
       author = {{Waddell}, S.~G.~H. and {Gallo}, L.~C.},
        title = "{A Suzaku sample of unabsorbed narrow-line and broad-line Seyfert 1 galaxies - I. X-ray spectral properties}",
      journal = {\mnras},
         year = 2020,
        month = nov,
       volume = {498},
       number = {4},
        pages = {5207-5226},
          doi = {10.1093/mnras/staa2783},
archivePrefix = {arXiv},
       eprint = {2009.04378},
 primaryClass = {astro-ph.HE},
       adsurl = {https://ui.adsabs.harvard.edu/abs/2020MNRAS.498.5207W}
}

@ARTICLE{Koratkar1995,
       author = {{Koratkar}, A. and {Deustua}, S.~E. and {Heckman}, T. and {Filippenko}, A.~V. and {Ho}, L.~C. and {Rao}, M.},
        title = "{Low-Luminosity Active Galaxies: Are They Similar to Seyfert Galaxies?}",
      journal = {\apj},
         year = 1995,
        month = feb,
       volume = {440},
        pages = {132},
          doi = {10.1086/175255},
       adsurl = {https://ui.adsabs.harvard.edu/abs/1995ApJ...440..132K}
}

@ARTICLE{Lyu2021,
       author = {{Lyu}, Bing and {Yan}, Zhen and {Yu}, Wenfei and {Wu}, Qingwen},
        title = "{Long-term and multiwavelength evolution of a changing-look AGN Mrk 1018}",
      journal = {\mnras},
         year = 2021,
        month = sep,
       volume = {506},
       number = {3},
        pages = {4188-4198},
          doi = {10.1093/mnras/stab1581},
archivePrefix = {arXiv},
       eprint = {2106.03059},
 primaryClass = {astro-ph.HE},
       adsurl = {https://ui.adsabs.harvard.edu/abs/2021MNRAS.506.4188L}
}

@ARTICLE{McElroy2016,
       author = {{McElroy}, R.~E. and {Husemann}, B. and {Croom}, S.~M. and {Davis}, T.~A. and {Bennert}, V.~N. and {Busch}, G. and {Combes}, F. and {Eckart}, A. and {Perez-Torres}, M. and {Powell}, M. and {Scharw{\"a}chter}, J. and {Tremblay}, G.~R. and {Urrutia}, T.},
        title = "{The Close AGN Reference Survey (CARS). Mrk 1018 returns to the shadows after 30 years as a Seyfert 1}",
      journal = {\aap},
         year = 2016,
        month = sep,
       volume = {593},
          eid = {L8},
        pages = {L8},
          doi = {10.1051/0004-6361/201629102},
archivePrefix = {arXiv},
       eprint = {1609.04423},
 primaryClass = {astro-ph.GA},
       adsurl = {https://ui.adsabs.harvard.edu/abs/2016A&A...593L...8M}
}

@ARTICLE{Mehdipour2022,
       author = {{Mehdipour}, Missagh and {Kriss}, Gerard A. and {Brenneman}, Laura W. and {Costantini}, Elisa and {Kaastra}, Jelle S. and {Branduardi-Raymont}, Graziella and {Di Gesu}, Laura and {Ebrero}, Jacobo and {Mao}, Junjie},
        title = "{Changing-look Event in NGC 3516: Continuum or Obscuration Variability?}",
      journal = {\apj},
         year = 2022,
        month = jan,
       volume = {925},
       number = {1},
          eid = {84},
        pages = {84},
          doi = {10.3847/1538-4357/ac42ca},
archivePrefix = {arXiv},
       eprint = {2112.06297},
 primaryClass = {astro-ph.HE},
       adsurl = {https://ui.adsabs.harvard.edu/abs/2022ApJ...925...84M}
}

@ARTICLE{Tripathi2022,
       author = {{Tripathi}, Prakash and {Dewangan}, Gulab Chand},
        title = "{Thermal Comptonization in a Changing Corona in the Changing-look Active Galaxy NGC 1566}",
      journal = {\apj},
         year = 2022,
        month = may,
       volume = {930},
       number = {2},
          eid = {117},
        pages = {117},
          doi = {10.3847/1538-4357/ac610f},
archivePrefix = {arXiv},
       eprint = {2203.09958},
 primaryClass = {astro-ph.HE},
       adsurl = {https://ui.adsabs.harvard.edu/abs/2022ApJ...930..117T}
}

@BOOK{Krolik1999,
       author = {{Krolik}, Julian H.},
        title = "{Active galactic nuclei : from the central black hole to the galactic environment}",
         year = 1999,
       adsurl = {https://ui.adsabs.harvard.edu/abs/1999agnc.book.....K}
}

@ARTICLE{Malkan1982,
       author = {{Malkan}, M.~A. and {Sargent}, W.~L.~W.},
        title = "{The ultraviolet excess of Seyfert 1 galaxies and quasars.}",
      journal = {\apj},
         year = 1982,
        month = mar,
       volume = {254},
        pages = {22-37},
          doi = {10.1086/159701},
       adsurl = {https://ui.adsabs.harvard.edu/abs/1982ApJ...254...22M}
}

@ARTICLE{Mehdipour2023,
       author = {{Mehdipour}, Missagh and {Kriss}, Gerard A. and {Kaastra}, Jelle S. and {Costantini}, Elisa and {Mao}, Junjie},
        title = "{Dimming of Continuum Captured in Mrk 841: New Clues on the Nature of the Soft X-Ray Excess}",
      journal = {\apjl},
         year = 2023,
        month = jul,
       volume = {952},
       number = {1},
          eid = {L5},
        pages = {L5},
          doi = {10.3847/2041-8213/ace053},
archivePrefix = {arXiv},
       eprint = {2306.11960},
 primaryClass = {astro-ph.HE},
       adsurl = {https://ui.adsabs.harvard.edu/abs/2023ApJ...952L...5M}
}

@ARTICLE{Sobolewska2011,
       author = {{Sobolewska}, Ma{\l}gorzata A. and {Siemiginowska}, Aneta and {Gierli{\'n}ski}, Marek},
        title = "{Simulated spectral states of active galactic nuclei and observational predictions}",
      journal = {\mnras},
         year = 2011,
        month = may,
       volume = {413},
       number = {3},
        pages = {2259-2268},
          doi = {10.1111/j.1365-2966.2011.18302.x},
archivePrefix = {arXiv},
       eprint = {1102.0798},
 primaryClass = {astro-ph.GA},
       adsurl = {https://ui.adsabs.harvard.edu/abs/2011MNRAS.413.2259S}
}

@ARTICLE{Merloni2003,
       author = {{Merloni}, Andrea and {Heinz}, Sebastian and {di Matteo}, Tiziana},
        title = "{A Fundamental Plane of black hole activity}",
      journal = {\mnras},
         year = 2003,
        month = nov,
       volume = {345},
       number = {4},
        pages = {1057-1076},
          doi = {10.1046/j.1365-2966.2003.07017.x},
archivePrefix = {arXiv},
       eprint = {astro-ph/0305261},
 primaryClass = {astro-ph},
       adsurl = {https://ui.adsabs.harvard.edu/abs/2003MNRAS.345.1057M}
}

@BOOK{Netzer2013,
       author = {{Netzer}, Hagai},
        title = "{The Physics and Evolution of Active Galactic Nuclei}",
         year = 2013,
       adsurl = {https://ui.adsabs.harvard.edu/abs/2013peag.book.....N}
}

@ARTICLE{McHardy2006,
       author = {{McHardy}, I.~M. and {Koerding}, E. and {Knigge}, C. and {Uttley}, P. and {Fender}, R.~P.},
        title = "{Active galactic nuclei as scaled-up Galactic black holes}",
      journal = {\nat},
         year = 2006,
        month = dec,
       volume = {444},
       number = {7120},
        pages = {730-732},
          doi = {10.1038/nature05389},
archivePrefix = {arXiv},
       eprint = {astro-ph/0612273},
 primaryClass = {astro-ph},
       adsurl = {https://ui.adsabs.harvard.edu/abs/2006Natur.444..730M}
}

@ARTICLE{Layek2024,
       author = {{Layek}, Narendranath and {Nandi}, Prantik and {Naik}, Sachindra and {Kumari}, Neeraj and {Jana}, Arghajit and {Chhotaray}, Birendra},
        title = "{Long-term X-ray temporal and spectral study of a Seyfert galaxy Mrk 6}",
      journal = {\mnras},
         year = 2024,
        month = mar,
       volume = {528},
       number = {3},
        pages = {5269-5285},
          doi = {10.1093/mnras/stae299},
archivePrefix = {arXiv},
       eprint = {2401.16780},
 primaryClass = {astro-ph.HE},
       adsurl = {https://ui.adsabs.harvard.edu/abs/2024MNRAS.528.5269L}
}

@ARTICLE{Petrucci2013,
       author = {{Petrucci}, P. -O. and {Paltani}, S. and {Malzac}, J. and {Kaastra}, J.~S. and {Cappi}, M. and {Ponti}, G. and {De Marco}, B. and {Kriss}, G.~A. and {Steenbrugge}, K.~C. and {Bianchi}, S. and {Branduardi-Raymont}, G. and {Mehdipour}, M. and {Costantini}, E. and {Dadina}, M. and {Lubi{\'n}ski}, P.},
        title = "{Multiwavelength campaign on Mrk 509. XII. Broad band spectral analysis}",
      journal = {\aap},
         year = 2013,
        month = jan,
       volume = {549},
          eid = {A73},
        pages = {A73},
          doi = {10.1051/0004-6361/201219956},
archivePrefix = {arXiv},
       eprint = {1209.6438},
 primaryClass = {astro-ph.HE},
       adsurl = {https://ui.adsabs.harvard.edu/abs/2013A&A...549A..73P}
}

@ARTICLE{Lu2025,
       author = {{Lu}, Kai-Xing and {Li}, Yan-Rong and {Wu}, Qingwen and {Ho}, Luis C. and {Zhang}, Zhi-Xiang and {Feng}, Hai-Cheng and {Li}, Sha-Sha and {Chen}, Yong-Jie and {Sun}, Mouyuan and {Shu}, Xinwen and {Guo}, Wei-Jian and {Cheng}, Cheng and {Wang}, Jian-Guo and {Kim}, Dongchan and {Wang}, Jian-Min and {Bai}, Jin-Ming},
        title = "{A Short-lived Rejuvenation during the Decades-long Changing-look Transition in the Nucleus of Mrk 1018}",
      journal = {\apjs},
         year = 2025,
        month = feb,
       volume = {276},
       number = {2},
          eid = {51},
        pages = {51},
          doi = {10.3847/1538-4365/ad9a5a},
archivePrefix = {arXiv},
       eprint = {2411.18917},
 primaryClass = {astro-ph.GA},
       adsurl = {https://ui.adsabs.harvard.edu/abs/2025ApJS..276...51L}
}

@ARTICLE{Veronese2024,
       author = {{Veronese}, S. and {Vignali}, C. and {Severgnini}, P. and {Matzeu}, G.~A. and {Cignoni}, M.},
        title = "{Interpreting the long-term variability of the changing-look AGN Mrk 1018}",
      journal = {\aap},
         year = 2024,
        month = mar,
       volume = {683},
          eid = {A131},
        pages = {A131},
          doi = {10.1051/0004-6361/202348098},
archivePrefix = {arXiv},
       eprint = {2312.07663},
 primaryClass = {astro-ph.HE},
       adsurl = {https://ui.adsabs.harvard.edu/abs/2024A&A...683A.131V}
}

@ARTICLE{Gierlinski2004,
       author = {{Gierli{\'n}ski}, Marek and {Done}, Chris},
        title = "{Is the soft excess in active galactic nuclei real?}",
      journal = {\mnras},
         year = 2004,
        month = mar,
       volume = {349},
       number = {1},
        pages = {L7-L11},
          doi = {10.1111/j.1365-2966.2004.07687.x},
archivePrefix = {arXiv},
       eprint = {astro-ph/0312271},
 primaryClass = {astro-ph},
       adsurl = {https://ui.adsabs.harvard.edu/abs/2004MNRAS.349L...7G}
}

@ARTICLE{Done2007,
       author = {{Done}, Chris and {Gierli{\'n}ski}, Marek and {Kubota}, Aya},
        title = "{Modelling the behaviour of accretion flows in X-ray binaries. Everything you always wanted to know about accretion but were afraid to ask}",
      journal = {\aapr},
         year = 2007,
        month = dec,
       volume = {15},
       number = {1},
        pages = {1-66},
          doi = {10.1007/s00159-007-0006-1},
archivePrefix = {arXiv},
       eprint = {0708.0148},
 primaryClass = {astro-ph},
       adsurl = {https://ui.adsabs.harvard.edu/abs/2007A&ARv..15....1D}
}

@ARTICLE{Sikora2007,
       author = {{Sikora}, Marek and {Stawarz}, {\L}ukasz and {Lasota}, Jean-Pierre},
        title = "{Radio Loudness of Active Galactic Nuclei: Observational Facts and Theoretical Implications}",
      journal = {\apj},
         year = 2007,
        month = apr,
       volume = {658},
       number = {2},
        pages = {815-828},
          doi = {10.1086/511972},
archivePrefix = {arXiv},
       eprint = {astro-ph/0604095},
 primaryClass = {astro-ph},
       adsurl = {https://ui.adsabs.harvard.edu/abs/2007ApJ...658..815S}
}

@ARTICLE{Yan2020,
       author = {{Yan}, Zhen and {Xie}, Fu-Guo and {Zhang}, Wenda},
        title = "{Coronal Properties of Black Hole X-Ray Binaries in the Hard State as Seen by NuSTAR and Swift}",
      journal = {\apjl},
         year = 2020,
        month = jan,
       volume = {889},
       number = {1},
          eid = {L18},
        pages = {L18},
          doi = {10.3847/2041-8213/ab665e},
archivePrefix = {arXiv},
       eprint = {1912.12145},
 primaryClass = {astro-ph.HE},
       adsurl = {https://ui.adsabs.harvard.edu/abs/2020ApJ...889L..18Y}
}

@ARTICLE{Reis2013,
       author = {{Reis}, R.~C. and {Miller}, J.~M.},
        title = "{On the Size and Location of the X-Ray Emitting Coronae around Black Holes}",
      journal = {\apjl},
         year = 2013,
        month = may,
       volume = {769},
       number = {1},
          eid = {L7},
        pages = {L7},
          doi = {10.1088/2041-8205/769/1/L7},
archivePrefix = {arXiv},
       eprint = {1304.4947},
 primaryClass = {astro-ph.HE},
       adsurl = {https://ui.adsabs.harvard.edu/abs/2013ApJ...769L...7R}
}

@ARTICLE{Vasudevan2014,
       author = {{Vasudevan}, Ranjan V. and {Mushotzky}, Richard F. and {Reynolds}, Christopher S. and {Fabian}, Andrew C. and {Lohfink}, Anne M. and {Zoghbi}, Abderahmen and {Gallo}, Luigi C. and {Walton}, Dominic},
        title = "{The Hard X-Ray Perspective on the Soft X-Ray Excess}",
      journal = {\apj},
         year = 2014,
        month = apr,
       volume = {785},
       number = {1},
          eid = {30},
        pages = {30},
          doi = {10.1088/0004-637X/785/1/30},
archivePrefix = {arXiv},
       eprint = {1402.3591},
 primaryClass = {astro-ph.HE},
       adsurl = {https://ui.adsabs.harvard.edu/abs/2014ApJ...785...30V}
}

@ARTICLE{Petrucci2018,
       author = {{Petrucci}, P. -O. and {Ursini}, F. and {De Rosa}, A. and {Bianchi}, S. and {Cappi}, M. and {Matt}, G. and {Dadina}, M. and {Malzac}, J.},
        title = "{Testing warm Comptonization models for the origin of the soft X-ray excess in AGNs}",
      journal = {\aap},
         year = 2018,
        month = mar,
       volume = {611},
          eid = {A59},
        pages = {A59},
          doi = {10.1051/0004-6361/201731580},
archivePrefix = {arXiv},
       eprint = {1710.04940},
 primaryClass = {astro-ph.HE},
       adsurl = {https://ui.adsabs.harvard.edu/abs/2018A&A...611A..59P}
}

@ARTICLE{Crummy2006,
       author = {{Crummy}, J. and {Fabian}, A.~C. and {Gallo}, L. and {Ross}, R.~R.},
        title = "{An explanation for the soft X-ray excess in active galactic nuclei}",
      journal = {\mnras},
         year = 2006,
        month = feb,
       volume = {365},
       number = {4},
        pages = {1067-1081},
          doi = {10.1111/j.1365-2966.2005.09844.x},
archivePrefix = {arXiv},
       eprint = {astro-ph/0511457},
 primaryClass = {astro-ph},
       adsurl = {https://ui.adsabs.harvard.edu/abs/2006MNRAS.365.1067C}
}

@ARTICLE{She2018,
       author = {{She}, Rui and {Ho}, Luis C. and {Feng}, Hua and {Cui}, Can},
        title = "{Chandra Survey of Nearby Galaxies: Testing the Accretion Model for Low-luminosity AGNs}",
      journal = {\apj},
         year = 2018,
        month = jun,
       volume = {859},
       number = {2},
          eid = {152},
        pages = {152},
          doi = {10.3847/1538-4357/aabfe7},
archivePrefix = {arXiv},
       eprint = {1804.07482},
 primaryClass = {astro-ph.HE},
       adsurl = {https://ui.adsabs.harvard.edu/abs/2018ApJ...859..152S}
}

@ARTICLE{Zdziarski2014,
       author = {{Zdziarski}, Andrzej A. and {Pjanka}, Patryk and {Sikora}, Marek and {Stawarz}, {\L}ukasz},
        title = "{Jet contributions to the broad-band spectrum of Cyg X-1 in the hard state}",
      journal = {\mnras},
         year = 2014,
        month = aug,
       volume = {442},
       number = {4},
        pages = {3243-3255},
          doi = {10.1093/mnras/stu1009},
archivePrefix = {arXiv},
       eprint = {1403.4768},
 primaryClass = {astro-ph.HE},
       adsurl = {https://ui.adsabs.harvard.edu/abs/2014MNRAS.442.3243Z}
}

@ARTICLE{Yuan2014,
       author = {{Yuan}, Feng and {Narayan}, Ramesh},
        title = "{Hot Accretion Flows Around Black Holes}",
      journal = {\araa},
         year = 2014,
        month = aug,
       volume = {52},
        pages = {529-588},
          doi = {10.1146/annurev-astro-082812-141003},
archivePrefix = {arXiv},
       eprint = {1401.0586},
 primaryClass = {astro-ph.HE},
       adsurl = {https://ui.adsabs.harvard.edu/abs/2014ARA&A..52..529Y}
}

@ARTICLE{Nandi2023,
       author = {{Nandi}, Prantik and {Chatterjee}, Arka and {Jana}, Arghajit and {Chakrabarti}, Sandip K. and {Naik}, Sachindra and {Safi-Harb}, Samar and {Chang}, Hsiang-Kuang and {Heyl}, Jeremy},
        title = "{Survey of Bare Active Galactic Nuclei in the Local Universe (z < 0.2). I. On the Origin of Soft Excess}",
      journal = {\apjs},
         year = 2023,
        month = nov,
       volume = {269},
       number = {1},
          eid = {15},
        pages = {15},
          doi = {10.3847/1538-4365/acf4f9},
archivePrefix = {arXiv},
       eprint = {2308.08528},
 primaryClass = {astro-ph.GA},
       adsurl = {https://ui.adsabs.harvard.edu/abs/2023ApJS..269...15N}
}

@ARTICLE{Gupta2024,
       author = {{Gupta}, Kriti K. and {Ricci}, Claudio and {Temple}, Matthew J. and {Tortosa}, Alessia and {Koss}, Michael J. and {Assef}, Roberto J. and {Bauer}, Franz E. and {Mushotzy}, Richard and {Ricci}, Federica and {Ueda}, Yoshihiro and {Rojas}, Alejandra F. and {Trakhtenbrot}, Benny and {Chang}, Chin-Shin and {Oh}, Kyuseok and {Li}, Ruancun and {Kawamuro}, Taiki and {Diaz}, Yaherlyn and {Powell}, Meredith C. and {Stern}, Daniel and {Megan Urry}, C. and {Harrison}, Fiona and {Cenko}, Brad},
        title = "{BASS: XLIII. Optical, UV, and X-ray emission properties of unobscured Swift/BAT active galactic nuclei}",
      journal = {\aap},
         year = 2024,
        month = nov,
       volume = {691},
          eid = {A203},
        pages = {A203},
          doi = {10.1051/0004-6361/202450567},
archivePrefix = {arXiv},
       eprint = {2409.12239},
 primaryClass = {astro-ph.GA},
       adsurl = {https://ui.adsabs.harvard.edu/abs/2024A&A...691A.203G}
}

@ARTICLE{AJ2025cl,
       author = {{Jana}, A. and {Ricci}, C. and {Temple}, M.~J. and {Chang}, H. -K. and {Shablovinskaya}, E. and {Trakhtenbrot}, B. and {Diaz}, Y. and {Ilic}, D. and {Nandi}, P. and {Koss}, M.},
        title = "{Investigating changing-look active galactic nuclei with long-term optical and X-ray observations}",
      journal = {\aap},
         year = 2025,
        month = jan,
       volume = {693},
          eid = {A35},
        pages = {A35},
          doi = {10.1051/0004-6361/202451058},
archivePrefix = {arXiv},
       eprint = {2411.08676},
 primaryClass = {astro-ph.GA},
       adsurl = {https://ui.adsabs.harvard.edu/abs/2025A&A...693A..35J}
}

@ARTICLE{Diaz2023,
       author = {{Diaz}, Y. and {Hern{\`a}ndez-Garc{\'\i}a}, L. and {Ar{\'e}valo}, P. and {L{\'o}pez-Navas}, E. and {Ricci}, C. and {Koss}, M. and {Gonzalez-Martin}, O. and {Balokovi{\'c}}, M. and {Osorio-Clavijo}, N. and {Garc{\'\i}a}, J.~A. and {Malizia}, A.},
        title = "{Constraining the X-ray reflection in low accretion-rate active galactic nuclei using XMM-Newton, NuSTAR, and Swift}",
      journal = {\aap},
         year = 2023,
        month = jan,
       volume = {669},
          eid = {A114},
        pages = {A114},
          doi = {10.1051/0004-6361/202244678},
archivePrefix = {arXiv},
       eprint = {2210.15376},
 primaryClass = {astro-ph.HE},
       adsurl = {https://ui.adsabs.harvard.edu/abs/2023A&A...669A.114D}
}

@ARTICLE{Yang2015,
       author = {{Yang}, Qi-Xiang and {Xie}, Fu-Guo and {Yuan}, Feng and {Zdziarski}, Andrzej A. and {Gierli{\'n}ski}, Marek and {Ho}, Luis C. and {Yu}, Zhaolong},
        title = "{Correlation between the photon index and X-ray luminosity of black hole X-ray binaries and active galactic nuclei: observations and interpretation}",
      journal = {\mnras},
         year = 2015,
        month = feb,
       volume = {447},
       number = {2},
        pages = {1692-1704},
          doi = {10.1093/mnras/stu2571},
archivePrefix = {arXiv},
       eprint = {1412.1358},
 primaryClass = {astro-ph.HE},
       adsurl = {https://ui.adsabs.harvard.edu/abs/2015MNRAS.447.1692Y}
}

@ARTICLE{AJ2022a,
       author = {{Jana}, Arghajit and {Ricci}, Claudio and {Naik}, Sachindra and {Tanimoto}, Atsushi and {Kumari}, Neeraj and {Chang}, Hsiang-Kuang and {Nandi}, Prantik and {Chatterjee}, Arka and {Safi-Harb}, Samar},
        title = "{Absorption variability of the highly obscured active galactic nucleus NGC 4507}",
      journal = {\mnras},
         year = 2022,
        month = jun,
       volume = {512},
       number = {4},
        pages = {5942-5959},
          doi = {10.1093/mnras/stac799},
archivePrefix = {arXiv},
       eprint = {2203.10550},
 primaryClass = {astro-ph.HE},
       adsurl = {https://ui.adsabs.harvard.edu/abs/2022MNRAS.512.5942J}
}

@ARTICLE{Risaliti2005,
       author = {{Risaliti}, G. and {Elvis}, M. and {Fabbiano}, G. and {Baldi}, A. and {Zezas}, A.},
        title = "{Rapid Compton-thick/Compton-thin Transitions in the Seyfert 2 Galaxy NGC 1365}",
      journal = {\apjl},
         year = 2005,
        month = apr,
       volume = {623},
       number = {2},
        pages = {L93-L96},
          doi = {10.1086/430252},
archivePrefix = {arXiv},
       eprint = {astro-ph/0503351},
 primaryClass = {astro-ph},
       adsurl = {https://ui.adsabs.harvard.edu/abs/2005ApJ...623L..93R}
}

@ARTICLE{Ricci2020,
       author = {{Ricci}, C. and {Kara}, E. and {Loewenstein}, M. and {Trakhtenbrot}, B. and
         {Arcavi}, I. and {Remillard}, R. and {Fabian}, A.~C. and
         {Gendreau}, K.~C. and {Arzoumanian}, Z. and {Li}, R. and {Ho}, L.~C. and
         {MacLeod}, C.~L. and {Cackett}, E. and {Altamirano}, D. and {Gand
        hi}, P. and {Kosec}, P. and {Pasham}, D. and {Steiner}, J. and
         {Chan}, C. -H.},
        title = "{The Destruction and Recreation of the X-Ray Corona in a Changing-look Active Galactic Nucleus}",
      journal = {\apjl},
         year = 2020,
        month = jul,
       volume = {898},
       number = {1},
          eid = {L1},
        pages = {L1},
          doi = {10.3847/2041-8213/ab91a1},
archivePrefix = {arXiv},
       eprint = {2007.07275},
 primaryClass = {astro-ph.HE},
       adsurl = {https://ui.adsabs.harvard.edu/abs/2020ApJ...898L...1R}
}

@ARTICLE{Stern2018,
       author = {{Stern}, Daniel and {McKernan}, Barry and {Graham}, Matthew J. and {Ford}, K.~E.~S. and {Ross}, Nicholas P. and {Meisner}, Aaron M. and {Assef}, Roberto J. and {Balokovi{\'c}}, Mislav and {Brightman}, Murray and {Dey}, Arjun and {Drake}, Andrew and {Djorgovski}, S.~G. and {Eisenhardt}, Peter and {Jun}, Hyunsung D.},
        title = "{A Mid-IR Selected Changing-look Quasar and Physical Scenarios for Abrupt AGN Fading}",
      journal = {\apj},
         year = 2018,
        month = sep,
       volume = {864},
       number = {1},
          eid = {27},
        pages = {27},
          doi = {10.3847/1538-4357/aac726},
archivePrefix = {arXiv},
       eprint = {1805.06920},
 primaryClass = {astro-ph.GA},
       adsurl = {https://ui.adsabs.harvard.edu/abs/2018ApJ...864...27S}
}

@ARTICLE{Magdziarz1998,
       author = {{Magdziarz}, Pawel and {Blaes}, Omer M. and {Zdziarski}, Andrzej A. and
         {Johnson}, W. Neil and {Smith}, David A.},
        title = "{A spectral decomposition of the variable optical, ultraviolet and X-ray continuum of NGC 5548}",
      journal = {\mnras},
         year = 1998,
        month = nov,
       volume = {301},
       number = {1},
        pages = {179-192},
          doi = {10.1046/j.1365-8711.1998.02015.x},
       adsurl = {https://ui.adsabs.harvard.edu/abs/1998MNRAS.301..179M}
}

@ARTICLE{Singh1985,
       author = {{Singh}, K.~P. and {Garmire}, G.~P. and {Nousek}, J.},
        title = "{Observations of Soft X-Ray Spectra from a Seyfert 1 and a Narrow Emission-Line Galaxy}",
      journal = {\apj},
         year = 1985,
        month = oct,
       volume = {297},
        pages = {633},
          doi = {10.1086/163560},
       adsurl = {https://ui.adsabs.harvard.edu/abs/1985ApJ...297..633S}
}

@ARTICLE{Arnaud1985,
       author = {{Arnaud}, K.~A. and {Branduardi-Raymont}, G. and {Culhane}, J.~L. and
         {Fabian}, A.~C. and {Hazard}, C. and {McGlynn}, T.~A. and
         {Shafer}, R.~A. and {Tennant}, A.~F. and {Ward}, M.~J.},
        title = "{EXOSAT observations of a strong soft X-ray excess in MKN 841.}",
      journal = {\mnras},
         year = 1985,
        month = nov,
       volume = {217},
        pages = {105-113},
          doi = {10.1093/mnras/217.1.105},
       adsurl = {https://ui.adsabs.harvard.edu/abs/1985MNRAS.217..105A}
}

@ARTICLE{Done2012,
       author = {{Done}, Chris and {Davis}, S.~W. and {Jin}, C. and {Blaes}, O. and
         {Ward}, M.},
        title = "{Intrinsic disc emission and the soft X-ray excess in active galactic nuclei}",
      journal = {\mnras},
         year = 2012,
        month = mar,
       volume = {420},
       number = {3},
        pages = {1848-1860},
          doi = {10.1111/j.1365-2966.2011.19779.x},
archivePrefix = {arXiv},
       eprint = {1107.5429},
 primaryClass = {astro-ph.HE},
       adsurl = {https://ui.adsabs.harvard.edu/abs/2012MNRAS.420.1848D}
}

@ARTICLE{Garcia2019,
       author = {{Garc{\'\i}a}, Javier A. and {Kara}, Erin and {Walton}, Dominic and
         {Beuchert}, Tobias and {Dauser}, Thomas and {Gatuzz}, Efrain and
         {Balokovic}, Mislav and {Steiner}, James F. and {Tombesi}, Francesco and
         {Connors}, Riley M.~T. and {Kallman}, Timothy R. and
         {Harrison}, Fiona A. and {Fabian}, Andrew and {Wilms}, J{\"o}rn and
         {Stern}, Daniel and {Lanz}, Lauranne and {Ricci}, Claudio and
         {Ballantyne}, David R.},
        title = "{Implications of the Warm Corona and Relativistic Reflection Models for the Soft Excess in Mrk 509}",
      journal = {\apj},
         year = 2019,
        month = jan,
       volume = {871},
       number = {1},
          eid = {88},
        pages = {88},
          doi = {10.3847/1538-4357/aaf739},
archivePrefix = {arXiv},
       eprint = {1812.03194},
 primaryClass = {astro-ph.HE},
       adsurl = {https://ui.adsabs.harvard.edu/abs/2019ApJ...871...88G}
}

@ARTICLE{ST80,
       author = {{Sunyaev}, R.~A. and {Titarchuk}, L.~G.},
        title = "{Comptonization of X-rays in plasma clouds. Typical radiation spectra.}",
      journal = {\aap},
         year = 1980,
        month = jun,
       volume = {500},
        pages = {167-184},
       adsurl = {https://ui.adsabs.harvard.edu/abs/1980A&A....86..121S}
}

@ARTICLE{Dauser2014,
       author = {{Dauser}, T. and {Garcia}, J. and {Parker}, M.~L. and {Fabian}, A.~C. and
         {Wilms}, J.},
        title = "{The role of the reflection fraction in constraining black hole spin.}",
      journal = {\mnras},
         year = 2014,
        month = oct,
       volume = {444},
        pages = {L100-L104},
          doi = {10.1093/mnrasl/slu125},
archivePrefix = {arXiv},
       eprint = {1408.2347},
 primaryClass = {astro-ph.HE},
       adsurl = {https://ui.adsabs.harvard.edu/abs/2014MNRAS.444L.100D}
}

@ARTICLE{Garcia2010,
       author = {{Garc{\'\i}a}, J. and {Kallman}, T.~R.},
        title = "{X-ray Reflected Spectra from Accretion Disk Models. I. Constant Density Atmospheres}",
      journal = {\apj},
         year = 2010,
        month = aug,
       volume = {718},
       number = {2},
        pages = {695-706},
          doi = {10.1088/0004-637X/718/2/695},
archivePrefix = {arXiv},
       eprint = {1006.0485},
 primaryClass = {astro-ph.HE},
       adsurl = {https://ui.adsabs.harvard.edu/abs/2010ApJ...718..695G}
}

@ARTICLE{RM06,
       author = {{Remillard}, Ronald A. and {McClintock}, Jeffrey E.},
        title = "{X-Ray Properties of Black-Hole Binaries}",
      journal = {\araa},
         year = 2006,
        month = sep,
       volume = {44},
       number = {1},
        pages = {49-92},
          doi = {10.1146/annurev.astro.44.051905.092532},
archivePrefix = {arXiv},
       eprint = {astro-ph/0606352},
 primaryClass = {astro-ph},
       adsurl = {https://ui.adsabs.harvard.edu/abs/2006ARA&A..44...49R}
}

@ARTICLE{HM1991,
       author = {{Haardt}, F. and {Maraschi}, L.},
        title = "{A Two-Phase Model for the X-Ray Emission from Seyfert Galaxies}",
      journal = {\apjl},
         year = 1991,
        month = oct,
       volume = {380},
        pages = {L51},
          doi = {10.1086/186171},
       adsurl = {https://ui.adsabs.harvard.edu/abs/1991ApJ...380L..51H}
}

@ARTICLE{Noda2018,
       author = {{Noda}, Hirofumi and {Done}, Chris},
        title = "{Explaining changing-look AGN with state transition triggered by rapid mass accretion rate drop}",
      journal = {\mnras},
         year = 2018,
        month = nov,
       volume = {480},
       number = {3},
        pages = {3898-3906},
          doi = {10.1093/mnras/sty2032},
archivePrefix = {arXiv},
       eprint = {1805.07873},
 primaryClass = {astro-ph.GA},
       adsurl = {https://ui.adsabs.harvard.edu/abs/2018MNRAS.480.3898N}
}

@ARTICLE{Emmanoulopoulos2012,
       author = {{Emmanoulopoulos}, D. and {Papadakis}, I.~E. and {McHardy}, I.~M. and {Ar{\'e}valo}, P. and {Calvelo}, D.~E. and {Uttley}, P.},
        title = "{The 'harder when brighter' X-ray behaviour of the low-luminosity active galactic nucleus NGC 7213}",
      journal = {\mnras},
         year = 2012,
        month = aug,
       volume = {424},
       number = {2},
        pages = {1327-1334},
          doi = {10.1111/j.1365-2966.2012.21316.x},
archivePrefix = {arXiv},
       eprint = {1205.3524},
 primaryClass = {astro-ph.CO},
       adsurl = {https://ui.adsabs.harvard.edu/abs/2012MNRAS.424.1327E}
}

@ARTICLE{Brightman2013,
       author = {{Brightman}, M. and {Silverman}, J.~D. and {Mainieri}, V. and {Ueda}, Y. and {Schramm}, M. and {Matsuoka}, K. and {Nagao}, T. and {Steinhardt}, C. and {Kartaltepe}, J. and {Sanders}, D.~B. and {Treister}, E. and {Shemmer}, O. and {Brandt}, W.~N. and {Brusa}, M. and {Comastri}, A. and {Ho}, L.~C. and {Lanzuisi}, G. and {Lusso}, E. and {Nandra}, K. and {Salvato}, M. and {Zamorani}, G. and {Akiyama}, M. and {Alexander}, D.~M. and {Bongiorno}, A. and {Capak}, P. and {Civano}, F. and {Del Moro}, A. and {Doi}, A. and {Elvis}, M. and {Hasinger}, G. and {Laird}, E.~S. and {Masters}, D. and {Mignoli}, M. and {Ohta}, K. and {Schawinski}, K. and {Taniguchi}, Y.},
        title = "{A statistical relation between the X-ray spectral index and Eddington ratio of active galactic nuclei in deep surveys}",
      journal = {\mnras},
         year = 2013,
        month = aug,
       volume = {433},
       number = {3},
        pages = {2485-2496},
          doi = {10.1093/mnras/stt920},
archivePrefix = {arXiv},
       eprint = {1305.3917},
 primaryClass = {astro-ph.HE},
       adsurl = {https://ui.adsabs.harvard.edu/abs/2013MNRAS.433.2485B}
}

@ARTICLE{Esin1997,
       author = {{Esin}, Ann A. and {McClintock}, Jeffrey E. and {Narayan}, Ramesh},
        title = "{Advection-Dominated Accretion and the Spectral States of Black Hole X-Ray Binaries: Application to Nova Muscae 1991}",
      journal = {\apj},
         year = 1997,
        month = nov,
       volume = {489},
       number = {2},
        pages = {865-889},
          doi = {10.1086/304829},
archivePrefix = {arXiv},
       eprint = {astro-ph/9705237},
 primaryClass = {astro-ph},
       adsurl = {https://ui.adsabs.harvard.edu/abs/1997ApJ...489..865E}
}

@ARTICLE{Chen2025b,
       author = {{Chen}, Shi-Jiang and {Buchner}, Johannes and {Liu}, Teng and {Hagen}, Scott and {Waddell}, Sophia G.~H. and {Nandra}, Kirpal and {Salvato}, Mara and {Igo}, Zsofi and {Aydar}, Catarina and {Merloni}, Andrea and {Ni}, Qingling and {Kang}, Jia-Lai and {Cai}, Zhen-Yi and {Wang}, Jun-Xian and {Li}, Ruancun and {Ramos-Ceja}, Miriam E. and {Sanders}, Jeremy and {Georgakakis}, Antonis and {Zhang}, Yi},
        title = "{The Average Soft X-ray Spectra of eROSITA Active Galactic Nuclei}",
      journal = {arXiv e-prints},
         year = 2025,
        month = jun,
          eid = {arXiv:2506.17150},
        pages = {arXiv:2506.17150},
          doi = {10.48550/arXiv.2506.17150},
archivePrefix = {arXiv},
       eprint = {2506.17150},
 primaryClass = {astro-ph.HE},
       adsurl = {https://ui.adsabs.harvard.edu/abs/2025arXiv250617150C}
}

@ARTICLE{Svoboda2017,
       author = {{Svoboda}, J. and {Guainazzi}, M. and {Merloni}, A.},
        title = "{AGN spectral states from simultaneous UV and X-ray observations by XMM-Newton}",
      journal = {\aap},
         year = 2017,
        month = jul,
       volume = {603},
          eid = {A127},
        pages = {A127},
          doi = {10.1051/0004-6361/201630181},
archivePrefix = {arXiv},
       eprint = {1704.07268},
 primaryClass = {astro-ph.GA},
       adsurl = {https://ui.adsabs.harvard.edu/abs/2017A&A...603A.127S}
}

@ARTICLE{Gupta2025,
       author = {{Gupta}, Kriti Kamal and {Ricci}, Claudio and {Tortosa}, Alessia and {Temple}, Matthew J. and {Koss}, Michael J. and {Trakhtenbrot}, Benny and {Bauer}, Franz E. and {Treister}, Ezequiel and {Mushotzky}, Richard and {Kammoun}, Elias and {Papadakis}, Iossif and {Oh}, Kyuseok and {Rojas}, Alejandra and {Chang}, Chin-Shin and {Diaz}, Yaherlyn and {Jana}, Arghajit and {Kakkad}, Darshan and {del Moral-Castro}, Ignacio and {Peca}, Alessandro and {Powell}, Meredith C. and {Stern}, Daniel and {Urry}, C. Megan and {Harrison}, Fiona},
        title = "{BASS. LIII. The Eddington Ratio as the Primary Regulator of the Fraction of X-Ray Emission in Active Galactic Nuclei}",
      journal = {\apj},
         year = 2025,
        month = sep,
       volume = {990},
       number = {1},
          eid = {86},
        pages = {86},
          doi = {10.3847/1538-4357/adf0f8},
archivePrefix = {arXiv},
       eprint = {2507.12541},
 primaryClass = {astro-ph.GA},
       adsurl = {https://ui.adsabs.harvard.edu/abs/2025ApJ...990...86G}
}

@ARTICLE{Ricci2017apjs,
       author = {{Ricci}, C. and {Trakhtenbrot}, B. and {Koss}, M.~J. and {Ueda}, Y. and {Del Vecchio}, I. and {Treister}, E. and {Schawinski}, K. and {Paltani}, S. and {Oh}, K. and {Lamperti}, I. and {Berney}, S. and {Gandhi}, P. and {Ichikawa}, K. and {Bauer}, F.~E. and {Ho}, L.~C. and {Asmus}, D. and {Beckmann}, V. and {Soldi}, S. and {Balokovi{\'c}}, M. and {Gehrels}, N. and {Markwardt}, C.~B.},
        title = "{BAT AGN Spectroscopic Survey. V. X-Ray Properties of the Swift/BAT 70-month AGN Catalog}",
      journal = {\apjs},
         year = 2017,
        month = dec,
       volume = {233},
       number = {2},
          eid = {17},
        pages = {17},
          doi = {10.3847/1538-4365/aa96ad},
archivePrefix = {arXiv},
       eprint = {1709.03989},
 primaryClass = {astro-ph.HE},
       adsurl = {https://ui.adsabs.harvard.edu/abs/2017ApJS..233...17R}
}

@ARTICLE{Gupta2021,
       author = {{Gupta}, K.~K. and {Ricci}, C. and {Tortosa}, A. and {Ueda}, Y. and {Kawamuro}, T. and {Koss}, M. and {Trakhtenbrot}, B. and {Oh}, K. and {Bauer}, F.~E. and {Ricci}, F. and {Privon}, G.~C. and {Zappacosta}, L. and {Stern}, D. and {Kakkad}, D. and {Piconcelli}, E. and {Veilleux}, S. and {Mushotzky}, R. and {Caglar}, T. and {Ichikawa}, K. and {Elagali}, A. and {Powell}, M.~C. and {Urry}, C.~M. and {Harrison}, F.},
        title = "{BAT AGN Spectroscopic Survey XXVII: scattered X-Ray radiation in obscured active galactic nuclei}",
      journal = {\mnras},
         year = 2021,
        month = jun,
       volume = {504},
       number = {1},
        pages = {428-443},
          doi = {10.1093/mnras/stab839},
archivePrefix = {arXiv},
       eprint = {2103.10543},
 primaryClass = {astro-ph.HE},
       adsurl = {https://ui.adsabs.harvard.edu/abs/2021MNRAS.504..428G}
}

@ARTICLE{Oh2022,
       author = {{Oh}, Kyuseok and {Koss}, Michael J. and {Ueda}, Yoshihiro and {Stern}, Daniel and {Ricci}, Claudio and {Trakhtenbrot}, Benny and {Powell}, Meredith C. and {den Brok}, Jakob S. and {Lamperti}, Isabella and {Mushotzky}, Richard and {Ricci}, Federica and {B{\"a}r}, Rudolf E. and {Rojas}, Alejandra F. and {Ichikawa}, Kohei and {Riffel}, Rog{\'e}rio and {Treister}, Ezequiel and {Harrison}, Fiona and {Urry}, C. Megan and {Bauer}, Franz E. and {Schawinski}, Kevin},
        title = "{BASS. XXIV. The BASS DR2 Spectroscopic Line Measurements and AGN Demographics}",
      journal = {\apjs},
         year = 2022,
        month = jul,
       volume = {261},
       number = {1},
          eid = {4},
        pages = {4},
          doi = {10.3847/1538-4365/ac5b68},
archivePrefix = {arXiv},
       eprint = {2203.00017},
 primaryClass = {astro-ph.GA},
       adsurl = {https://ui.adsabs.harvard.edu/abs/2022ApJS..261....4O}
}

@ARTICLE{Shemmer2006,
       author = {{Shemmer}, Ohad and {Brandt}, W.~N. and {Netzer}, Hagai and {Maiolino}, Roberto and {Kaspi}, Shai},
        title = "{The Hard X-Ray Spectral Slope as an Accretion Rate Indicator in Radio-quiet Active Galactic Nuclei}",
      journal = {\apjl},
         year = 2006,
        month = jul,
       volume = {646},
       number = {1},
        pages = {L29-L32},
          doi = {10.1086/506911},
archivePrefix = {arXiv},
       eprint = {astro-ph/0606389},
 primaryClass = {astro-ph},
       adsurl = {https://ui.adsabs.harvard.edu/abs/2006ApJ...646L..29S}
}

@INPROCEEDINGS{Reynolds2023axis,
       author = {{Reynolds}, Christopher S. and {Kara}, Erin A. and {Mushotzky}, Richard F. and {Ptak}, Andrew and {Koss}, Michael J. and {Williams}, Brian J. and {Allen}, Steven W. and {Bauer}, Franz E. and {Bautz}, Marshall and {Bogadhee}, Arash and {Burdge}, Kevin B. and {Cappelluti}, Nico and {Cenko}, Brad and {Chartas}, George and {Chan}, Kai-Wing and {Corrales}, L{\'\i}a. and {Daylan}, Tansu and {Falcone}, Abraham D. and {Foord}, Adi and {Grant}, Catherine E. and {Habouzit}, M{\'e}lanie and {Haggard}, Daryl and {Herrmann}, Sven and {Hodges-Kluck}, Edmund and {Kargaltsev}, Oleg and {King}, George W. and {Kounkel}, Marina and {Lopez}, Laura A. and {Marchesi}, Stefano and {McDonald}, Michael and {Meyer}, Eileen and {Miller}, Eric D. and {Nynka}, Melania and {Okajima}, Takashi and {Pacucci}, Fabio and {Russell}, Helen R. and {Safi-Harb}, Samar and {Strassun}, Keivan G. and {Trindade Falc{\~a}o}, Anna and {Walker}, Stephen A. and {Wilms}, Joern and {Yukita}, Mihoko and {Zhang}, William W.},
        title = "{Overview of the advanced x-ray imaging satellite (AXIS)}",
    booktitle = {UV, X-Ray, and Gamma-Ray Space Instrumentation for Astronomy XXIII},
         year = 2023,
       editor = {{Siegmund}, Oswald H. and {Hoadley}, Keri},
       series = {Society of Photo-Optical Instrumentation Engineers (SPIE) Conference Series},
       volume = {12678},
        month = oct,
          eid = {126781E},
        pages = {126781E},
          doi = {10.1117/12.2677468},
archivePrefix = {arXiv},
       eprint = {2311.00780},
 primaryClass = {astro-ph.IM},
       adsurl = {https://ui.adsabs.harvard.edu/abs/2023SPIE12678E..1ER}
}

@INPROCEEDINGS{Paul2022polix,
       author = {{Paul}, Biswajit},
        title = "{The X-ray Polarimetry Satellite XPoSat}",
    booktitle = {44th COSPAR Scientific Assembly. Held 16-24 July},
         year = 2022,
       volume = {44},
        month = jul,
        pages = {1853},
       adsurl = {https://ui.adsabs.harvard.edu/abs/2022cosp...44.1853P}
}

@ARTICLE{Radhakrishna2025xspect,
       author = {{V}, Radhakrishna and {Tyagi}, Anurag and {Vadodariya}, Koushal and {Agrawal}, Vivek K and {Chatterjee}, Rwitika and {C}, Ramadevi M and {Jayasurya}, Kiran M and {Kumar} and {S}, Vaishali and {Tadepalli}, Srikar P and {K}, Sreedatta Reddy and {Garg}, Lokesh K and {Sharma}, Nidhi and {Justin}, Evangelin L},
        title = "{X-Ray spectroscopy and timing (XSPECT) experiment on XPoSat -- instrument configuration and science prospects}",
      journal = {arXiv e-prints},
         year = 2025,
        month = may,
          eid = {arXiv:2505.20061},
        pages = {arXiv:2505.20061},
          doi = {10.48550/arXiv.2505.20061},
archivePrefix = {arXiv},
       eprint = {2505.20061},
 primaryClass = {astro-ph.IM},
       adsurl = {https://ui.adsabs.harvard.edu/abs/2025arXiv250520061V}
}

@ARTICLE{Cruise2025newAthena,
       author = {{Cruise}, Mike and {Guainazzi}, Matteo and {Aird}, James and {Carrera}, Francisco J. and {Costantini}, Elisa and {Corrales}, Lia and {Dauser}, Thomas and {Eckert}, Dominique and {Gastaldello}, Fabio and {Matsumoto}, Hironori and {Osten}, Rachel and {Petrucci}, Pierre-Olivier and {Porquet}, Delphine and {Pratt}, Gabriel W. and {Rea}, Nanda and {Reiprich}, Thomas H. and {Simionescu}, Aurora and {Spiga}, Daniele and {Troja}, Eleonora},
        title = "{The NewAthena mission concept in the context of the next decade of X-ray astronomy}",
      journal = {Nature Astronomy},
         year = 2025,
        month = jan,
       volume = {9},
        pages = {36-44},
          doi = {10.1038/s41550-024-02416-3},
archivePrefix = {arXiv},
       eprint = {2501.03100},
 primaryClass = {astro-ph.IM},
       adsurl = {https://ui.adsabs.harvard.edu/abs/2025NatAs...9...36C}
}

@ARTICLE{Trakhtenbrot2017,
       author = {{Trakhtenbrot}, Benny and {Ricci}, Claudio and {Koss}, Michael J. and {Schawinski}, Kevin and {Mushotzky}, Richard and {Ueda}, Yoshihiro and {Veilleux}, Sylvain and {Lamperti}, Isabella and {Oh}, Kyuseok and {Treister}, Ezequiel and {Stern}, Daniel and {Harrison}, Fiona and {Balokovi{\'c}}, Mislav and {Gehrels}, Neil},
        title = "{BAT AGN Spectroscopic Survey (BASS) - VI. The {\ensuremath{\Gamma}}$_{X}$-L/L$_{Edd}$ relation}",
      journal = {\mnras},
         year = 2017,
        month = sep,
       volume = {470},
       number = {1},
        pages = {800-814},
          doi = {10.1093/mnras/stx1117},
archivePrefix = {arXiv},
       eprint = {1705.01550},
 primaryClass = {astro-ph.GA},
       adsurl = {https://ui.adsabs.harvard.edu/abs/2017MNRAS.470..800T}
}

@ARTICLE{Beuchert2017,
       author = {{Beuchert}, T. and {Markowitz}, A.~G. and {Dauser}, T. and {Garc{\'\i}a}, J.~A. and {Keck}, M.~L. and {Wilms}, J. and {Kadler}, M. and {Brenneman}, L.~W. and {Zdziarski}, A.~A.},
        title = "{A Suzaku, NuSTAR, and XMM-Newton view on variable absorption and relativistic reflection in NGC 4151}",
      journal = {\aap},
         year = 2017,
        month = jul,
       volume = {603},
          eid = {A50},
        pages = {A50},
          doi = {10.1051/0004-6361/201630293},
archivePrefix = {arXiv},
       eprint = {1703.10856},
 primaryClass = {astro-ph.HE},
       adsurl = {https://ui.adsabs.harvard.edu/abs/2017A&A...603A..50B}
}

@BOOK{Frank2002,
       author = {{Frank}, Juhan and {King}, Andrew and {Raine}, Derek J.},
        title = "{Accretion Power in Astrophysics: Third Edition}",
         year = 2002,
   publishers = "{Cambridge University Press}",
       adsurl = {https://ui.adsabs.harvard.edu/abs/2002apa..book.....F}
}

@ARTICLE{Feigelson1985,
       author = {{Feigelson}, E.~D. and {Nelson}, P.~I.},
        title = "{Statistical methods for astronomical data with upper limits. I. Univariate distributions.}",
      journal = {\apj},
         year = 1985,
        month = jun,
       volume = {293},
        pages = {192-206},
          doi = {10.1086/163225},
       adsurl = {https://ui.adsabs.harvard.edu/abs/1985ApJ...293..192F}
}

@ARTICLE{Shimizu2017,
       author = {{Shimizu}, T. Taro and {Mushotzky}, Richard F. and {Mel{\'e}ndez}, Marcio and {Koss}, Michael J. and {Barger}, Amy J. and {Cowie}, Lennox L.},
        title = "{Herschel far-infrared photometry of the Swift Burst Alert Telescope active galactic nuclei sample of the local universe - III. Global star-forming properties and the lack of a connection to nuclear activity}",
      journal = {\mnras},
         year = 2017,
        month = apr,
       volume = {466},
       number = {3},
        pages = {3161-3183},
          doi = {10.1093/mnras/stw3268},
archivePrefix = {arXiv},
       eprint = {1612.03941},
 primaryClass = {astro-ph.GA},
       adsurl = {https://ui.adsabs.harvard.edu/abs/2017MNRAS.466.3161S}
}

@ARTICLE{Burrows2005,
       author = {{Burrows}, David N. and {Hill}, J.~E. and {Nousek}, J.~A. and {Kennea}, J.~A. and {Wells}, A. and {Osborne}, J.~P. and {Abbey}, A.~F. and {Beardmore}, A. and {Mukerjee}, K. and {Short}, A.~D.~T. and {Chincarini}, G. and {Campana}, S. and {Citterio}, O. and {Moretti}, A. and {Pagani}, C. and {Tagliaferri}, G. and {Giommi}, P. and {Capalbi}, M. and {Tamburelli}, F. and {Angelini}, L. and {Cusumano}, G. and {Br{\"a}uninger}, H.~W. and {Burkert}, W. and {Hartner}, G.~D.},
        title = "{The Swift X-Ray Telescope}",
      journal = {\ssr},
         year = 2005,
        month = oct,
       volume = {120},
       number = {3-4},
        pages = {165-195},
          doi = {10.1007/s11214-005-5097-2},
archivePrefix = {arXiv},
       eprint = {astro-ph/0508071},
 primaryClass = {astro-ph},
       adsurl = {https://ui.adsabs.harvard.edu/abs/2005SSRv..120..165B}
}

@ARTICLE{Gruber1999,
       author = {{Gruber}, D.~E. and {Matteson}, J.~L. and {Peterson}, L.~E. and {Jung}, G.~V.},
        title = "{The Spectrum of Diffuse Cosmic Hard X-Rays Measured with HEAO 1}",
      journal = {\apj},
         year = 1999,
        month = jul,
       volume = {520},
       number = {1},
        pages = {124-129},
          doi = {10.1086/307450},
archivePrefix = {arXiv},
       eprint = {astro-ph/9903492},
 primaryClass = {astro-ph},
       adsurl = {https://ui.adsabs.harvard.edu/abs/1999ApJ...520..124G}
}

@article{sksurv2020,
  author  = {Sebastian P{\"o}lsterl},
  title   = {scikit-survival: A Library for Time-to-Event Analysis Built on Top of scikit-learn},
  journal = {Journal of Machine Learning Research},
  year    = {2020},
  volume  = {21},
  number  = {212},
  pages   = {1-6},
  url     = {http://jmlr.org/papers/v21/20-729.html}
}

@ARTICLE{Mehdipour2022n55,
       author = {{Mehdipour}, Missagh and {Kriss}, Gerard A. and {Costantini}, Elisa and {Gu}, Liyi and {Kaastra}, Jelle S. and {Landt}, Hermine and {Mao}, Junjie},
        title = "{10 Yr Transformation of the Obscuring Wind in NGC 5548}",
      journal = {\apjl},
         year = 2022,
        month = aug,
       volume = {934},
       number = {2},
          eid = {L24},
        pages = {L24},
          doi = {10.3847/2041-8213/ac822f},
archivePrefix = {arXiv},
       eprint = {2207.09464},
 primaryClass = {astro-ph.HE},
       adsurl = {https://ui.adsabs.harvard.edu/abs/2022ApJ...934L..24M}
}

@ARTICLE{Edelson2017,
       author = {{Edelson}, R. and {Gelbord}, J. and {Cackett}, E. and {Connolly}, S. and {Done}, C. and {Fausnaugh}, M. and {Gardner}, E. and {Gehrels}, N. and {Goad}, M. and {Horne}, K. and {McHardy}, I. and {Peterson}, B.~M. and {Vaughan}, S. and {Vestergaard}, M. and {Breeveld}, A. and {Barth}, A.~J. and {Bentz}, M. and {Bottorff}, M. and {Brandt}, W.~N. and {Crawford}, S.~M. and {Dalla Bont{\`a}}, E. and {Emmanoulopoulos}, D. and {Evans}, P. and {Figuera Jaimes}, R. and {Filippenko}, A.~V. and {Ferland}, G. and {Grupe}, D. and {Joner}, M. and {Kennea}, J. and {Korista}, K.~T. and {Krimm}, H.~A. and {Kriss}, G. and {Leonard}, D.~C. and {Mathur}, S. and {Netzer}, H. and {Nousek}, J. and {Page}, K. and {Romero-Colmenero}, E. and {Siegel}, M. and {Starkey}, D.~A. and {Treu}, T. and {Vogler}, H.~A. and {Winkler}, H. and {Zheng}, W.},
        title = "{Swift Monitoring of NGC 4151: Evidence for a Second X-Ray/UV Reprocessing}",
      journal = {\apj},
         year = 2017,
        month = may,
       volume = {840},
       number = {1},
          eid = {41},
        pages = {41},
          doi = {10.3847/1538-4357/aa6890},
archivePrefix = {arXiv},
       eprint = {1703.06901},
 primaryClass = {astro-ph.HE},
       adsurl = {https://ui.adsabs.harvard.edu/abs/2017ApJ...840...41E}
}

@ARTICLE{Xiang2022,
       author = {{Xiang}, X. and {Ballantyne}, D.~R. and {Bianchi}, S. and {De Rosa}, A. and {Matt}, G. and {Middei}, R. and {Petrucci}, P. -O. and {R{\'o}{\.z}a{\'n}ska}, A. and {Ursini}, F.},
        title = "{REXCOR: a model of the X-ray spectrum of active galactic nuclei that combines ionized reflection and a warm corona}",
      journal = {\mnras},
         year = 2022,
        month = sep,
       volume = {515},
       number = {1},
        pages = {353-368},
          doi = {10.1093/mnras/stac1646},
archivePrefix = {arXiv},
       eprint = {2206.06825},
 primaryClass = {astro-ph.HE},
       adsurl = {https://ui.adsabs.harvard.edu/abs/2022MNRAS.515..353X}
}

@ARTICLE{Palit2024,
       author = {{Palit}, B. and {R{\'o}{\.z}a{\'n}ska}, A. and {Petrucci}, P.~O. and {Gronkiewicz}, D. and {Barnier}, S. and {Bianchi}, S. and {Ballantyne}, D.~R. and {Gianolli}, V.~E. and {Middei}, R. and {Belmont}, R. and {Ursini}, F.},
        title = "{X-ray view of dissipative warm corona in active galactic nuclei}",
      journal = {\aap},
         year = 2024,
        month = oct,
       volume = {690},
          eid = {A308},
        pages = {A308},
          doi = {10.1051/0004-6361/202450111},
archivePrefix = {arXiv},
       eprint = {2406.14378},
 primaryClass = {astro-ph.HE},
       adsurl = {https://ui.adsabs.harvard.edu/abs/2024A&A...690A.308P}
}

@ARTICLE{Lawther2023,
       author = {{Lawther}, D. and {Vestergaard}, M. and {Raimundo}, S. and {Koay}, J.~Y. and {Peterson}, B.~M. and {Fan}, X. and {Grupe}, D. and {Mathur}, S.},
        title = "{Flares in the changing look AGN Mrk 590 - I. The UV response to X-ray outbursts suggests a more complex reprocessing geometry than a standard disc}",
      journal = {\mnras},
         year = 2023,
        month = mar,
       volume = {519},
       number = {3},
        pages = {3903-3922},
          doi = {10.1093/mnras/stac3515},
archivePrefix = {arXiv},
       eprint = {2211.14371},
 primaryClass = {astro-ph.HE},
       adsurl = {https://ui.adsabs.harvard.edu/abs/2023MNRAS.519.3903L}
}

@ARTICLE{Xu2024,
       author = {{Xu}, D.~W. and {Komossa}, S. and {Grupe}, D. and {Wang}, J. and {Xin}, L.~P. and {Han}, X.~H. and {Wei}, J.~Y. and {Bai}, J.~Y. and {Bon}, E. and {Cangemi}, F. and {Cordier}, B. and {Dennefeld}, M. and {Gallo}, L.~C. and {Kollatschny}, W. and {Kong}, De-Feng and {Ochmann}, M.~W. and {Qiu}, Y.~L. and {Schartel}, N.},
        title = "{Changing-Look Narrow-Line Seyfert 1 Galaxies, their Detection with SVOM, and the Case of NGC 1566}",
      journal = {Universe},
         year = 2024,
        month = jan,
       volume = {10},
       number = {2},
          eid = {61},
        pages = {61},
          doi = {10.3390/universe10020061},
archivePrefix = {arXiv},
       eprint = {2401.10600},
 primaryClass = {astro-ph.GA},
       adsurl = {https://ui.adsabs.harvard.edu/abs/2024Univ...10...61X}
}

@ARTICLE{Wilkins2016,
       author = {{Wilkins}, D.~R. and {Cackett}, E.~M. and {Fabian}, A.~C. and {Reynolds}, C.~S.},
        title = "{Towards modelling X-ray reverberation in AGN: piecing together the extended corona}",
      journal = {\mnras},
         year = 2016,
        month = may,
       volume = {458},
       number = {1},
        pages = {200-225},
          doi = {10.1093/mnras/stw276},
archivePrefix = {arXiv},
       eprint = {1602.00022},
 primaryClass = {astro-ph.HE},
       adsurl = {https://ui.adsabs.harvard.edu/abs/2016MNRAS.458..200W}
}

@ARTICLE{Middei2020,
       author = {{Middei}, R. and {Petrucci}, P. -O. and {Bianchi}, S. and {Ursini}, F. and {Cappi}, M. and {Clavel}, M. and {De Rosa}, A. and {Marinucci}, A. and {Matt}, G. and {Tortosa}, A.},
        title = "{The soft excess of the NLS1 galaxy Mrk 359 studied with an XMM-Newton-NuSTAR monitoring campaign}",
      journal = {\aap},
         year = 2020,
        month = aug,
       volume = {640},
          eid = {A99},
        pages = {A99},
          doi = {10.1051/0004-6361/202038112},
archivePrefix = {arXiv},
       eprint = {2006.09005},
 primaryClass = {astro-ph.HE},
       adsurl = {https://ui.adsabs.harvard.edu/abs/2020A&A...640A..99M}
}

@ARTICLE{Kubota2018,
       author = {{Kubota}, Aya and {Done}, Chris},
        title = "{A physical model of the broad-band continuum of AGN and its implications for the UV/X relation and optical variability}",
      journal = {\mnras},
         year = 2018,
        month = oct,
       volume = {480},
       number = {1},
        pages = {1247-1262},
          doi = {10.1093/mnras/sty1890},
archivePrefix = {arXiv},
       eprint = {1804.00171},
 primaryClass = {astro-ph.HE},
       adsurl = {https://ui.adsabs.harvard.edu/abs/2018MNRAS.480.1247K}
}

@article{Chen2025,
doi = {10.3847/1538-4357/ada035},
url = {https://dx.doi.org/10.3847/1538-4357/ada035},
year = {2025},
month = {jan},
publisher = {The American Astronomical Society},
volume = {980},
number = {1},
pages = {23},
author = {Chen, Shi-Jiang and Wang, Jun-Xian and Kang, Jia-Lai and Kang, Wen-Yong and Sou, Hao and Liu, Teng and Cai, Zhen-Yi and Su, Zhen-Bo},
title = {A UV to X-Ray View of Soft Excess in Type 1 Active Galactic Nuclei. I. Sample Selection and Spectral Profile},
journal = {The Astrophysical Journal},
}

@ARTICLE{Kollatschny2020,
       author = {{Kollatschny}, W. and {Grupe}, D. and {Parker}, M.~L. and {Ochmann}, M.~W. and {Schartel}, N. and {Herwig}, E. and {Komossa}, S. and {Romero-Colmenero}, E. and {Santos-Lleo}, M.},
        title = "{Optical and X-ray discovery of the changing-look AGN IRAS 23226-3843 showing extremely broad and double-peaked Balmer profiles}",
      journal = {\aap},
         year = 2020,
        month = jun,
       volume = {638},
          eid = {A91},
        pages = {A91},
          doi = {10.1051/0004-6361/202037897},
archivePrefix = {arXiv},
       eprint = {2004.01711},
 primaryClass = {astro-ph.GA},
       adsurl = {https://ui.adsabs.harvard.edu/abs/2020A&A...638A..91K}
}

@ARTICLE{Kollatschny2023,
       author = {{Kollatschny}, W. and {Grupe}, D. and {Parker}, M.~L. and {Ochmann}, M.~W. and {Schartel}, N. and {Romero-Colmenero}, E. and {Winkler}, H. and {Komossa}, S. and {Famula}, P. and {Probst}, M.~A. and {Santos-Lleo}, M.},
        title = "{The outburst of the changing-look AGN IRAS 23226-3843 in 2019}",
      journal = {\aap},
         year = 2023,
        month = feb,
       volume = {670},
          eid = {A103},
        pages = {A103},
          doi = {10.1051/0004-6361/202244786},
archivePrefix = {arXiv},
       eprint = {2212.07270},
 primaryClass = {astro-ph.HE},
       adsurl = {https://ui.adsabs.harvard.edu/abs/2023A&A...670A.103K}
}

@ARTICLE{Ai2020,
       author = {{Ai}, Yanli and {Dou}, Liming and {Yang}, Chenwei and {Sun}, Luming and {Xie}, Fu-Guo and {Yao}, Su and {Wu}, Xue-Bing and {Wang}, Tinggui and {Shu}, Xinwen and {Jiang}, Ning},
        title = "{X-Ray Spectral Shape Variation in Changing-look Seyfert Galaxy SDSS J155258+273728}",
      journal = {\apjl},
         year = 2020,
        month = feb,
       volume = {890},
       number = {2},
          eid = {L29},
        pages = {L29},
          doi = {10.3847/2041-8213/ab7306},
       adsurl = {https://ui.adsabs.harvard.edu/abs/2020ApJ...890L..29A}
}

@ARTICLE{Gallo2019,
       author = {{Gallo}, L.~C. and {Gonzalez}, A.~G. and {Waddell}, S.~G.~H. and {Ehler}, H.~J.~S. and {Wilkins}, D.~R. and {Longinotti}, A.~L. and {Grupe}, D. and {Komossa}, S. and {Kriss}, G.~A. and {Pinto}, C. and {Tripathi}, S. and {Fabian}, A.~C. and {Krongold}, Y. and {Mathur}, S. and {Parker}, M.~L. and {Pradhan}, A.},
        title = "{Evidence for an emerging disc wind and collimated outflow during an X-ray flare in the narrow-line Seyfert 1 galaxy Mrk 335}",
      journal = {\mnras},
         year = 2019,
        month = apr,
       volume = {484},
       number = {3},
        pages = {4287-4297},
          doi = {10.1093/mnras/stz274},
archivePrefix = {arXiv},
       eprint = {1901.07899},
 primaryClass = {astro-ph.HE},
       adsurl = {https://ui.adsabs.harvard.edu/abs/2019MNRAS.484.4287G}
}

@ARTICLE{Ruan2019,
       author = {{Ruan}, John J. and {Anderson}, Scott F. and {Eracleous}, Michael and {Green}, Paul J. and {Haggard}, Daryl and {MacLeod}, Chelsea L. and {Runnoe}, Jessie C. and {Sobolewska}, Malgosia A.},
        title = "{The Analogous Structure of Accretion Flows in Supermassive and Stellar Mass Black Holes: New Insights from Faded Changing-look Quasars}",
      journal = {\apj},
         year = 2019,
        month = sep,
       volume = {883},
       number = {1},
          eid = {76},
        pages = {76},
          doi = {10.3847/1538-4357/ab3c1a},
archivePrefix = {arXiv},
       eprint = {1903.02553},
 primaryClass = {astro-ph.HE},
       adsurl = {https://ui.adsabs.harvard.edu/abs/2019ApJ...883...76R}
}

@ARTICLE{Kang2025,
       author = {{Kang}, Jia-Lai and {Done}, Chris and {Hagen}, Scott and {Temple}, Matthew J. and {Silverman}, John D. and {Li}, Junyao and {Liu}, Teng},
        title = "{Systematic collapse of the accretion disc in AGN confirmed by UV photometry and broad-line spectra}",
      journal = {\mnras},
         year = 2025,
        month = mar,
       volume = {538},
       number = {1},
        pages = {121-131},
          doi = {10.1093/mnras/staf145},
archivePrefix = {arXiv},
       eprint = {2410.06730},
 primaryClass = {astro-ph.HE},
       adsurl = {https://ui.adsabs.harvard.edu/abs/2025MNRAS.538..121K}
}

@ARTICLE{Hagen2024,
       author = {{Hagen}, Scott and {Done}, Chris and {Silverman}, John D. and {Li}, Junyao and {Liu}, Teng and {Ren}, Wenke and {Buchner}, Johannes and {Merloni}, Andrea and {Nagao}, Tohru and {Salvato}, Mara},
        title = "{Systematic collapse of the accretion disc across the supermassive black hole population}",
      journal = {\mnras},
         year = 2024,
        month = nov,
       volume = {534},
       number = {3},
        pages = {2803-2818},
          doi = {10.1093/mnras/stae2272},
archivePrefix = {arXiv},
       eprint = {2406.06674},
 primaryClass = {astro-ph.HE},
       adsurl = {https://ui.adsabs.harvard.edu/abs/2024MNRAS.534.2803H}
}

@ARTICLE{Ding2024,
       author = {{Ding}, Yuanze and {Garc{\i}a}, Javier A. and {Kallman}, Timothy R. and {Mendoza}, Claudio and {Bautista}, Manuel and {Harrison}, Fiona A. and {Tomsick}, John A. and {Dong}, Jameson},
        title = "{Next-generation Accretion Disk Reflection Model: High-density Plasma Effects}",
      journal = {\apj},
         year = 2024,
        month = oct,
       volume = {974},
       number = {2},
          eid = {280},
        pages = {280},
          doi = {10.3847/1538-4357/ad76a1},
archivePrefix = {arXiv},
       eprint = {2409.00253},
 primaryClass = {astro-ph.HE},
       adsurl = {https://ui.adsabs.harvard.edu/abs/2024ApJ...974..280D}
}

@ARTICLE{Panda2024,
       author = {{Panda}, Swayamtrupta and {{\'S}niegowska}, Marzena},
        title = "{Changing-look Active Galactic Nuclei. I. Tracking the Transition on the Main Sequence of Quasars}",
      journal = {\apjs},
         year = 2024,
        month = may,
       volume = {272},
       number = {1},
          eid = {13},
        pages = {13},
          doi = {10.3847/1538-4365/ad344f},
archivePrefix = {arXiv},
       eprint = {2206.10056},
 primaryClass = {astro-ph.HE},
       adsurl = {https://ui.adsabs.harvard.edu/abs/2024ApJS..272...13P}
}

@ARTICLE{Trakhtenbrot2019,
       author = {{Trakhtenbrot}, Benny and {Arcavi}, Iair and {MacLeod}, Chelsea L. and {Ricci}, Claudio and {Kara}, Erin and {Graham}, Melissa L. and {Stern}, Daniel and {Harrison}, Fiona A. and {Burke}, Jamison and {Hiramatsu}, Daichi and {Hosseinzadeh}, Griffin and {Howell}, D. Andrew and {Smartt}, Stephen J. and {Rest}, Armin and {Prieto}, Jose L. and {Shappee}, Benjamin J. and {Holoien}, Thomas W. -S. and {Bersier}, David and {Filippenko}, Alexei V. and {Brink}, Thomas G. and {Zheng}, WeiKang and {Li}, Ruancun and {Remillard}, Ronald A. and {Loewenstein}, Michael},
        title = "{1ES 1927+654: An AGN Caught Changing Look on a Timescale of Months}",
      journal = {\apj},
         year = 2019,
        month = sep,
       volume = {883},
       number = {1},
          eid = {94},
        pages = {94},
          doi = {10.3847/1538-4357/ab39e4},
archivePrefix = {arXiv},
       eprint = {1903.11084},
 primaryClass = {astro-ph.GA},
       adsurl = {https://ui.adsabs.harvard.edu/abs/2019ApJ...883...94T}
}

@ARTICLE{Risaliti2009,
       author = {{Risaliti}, G. and {Salvati}, M. and {Elvis}, M. and {Fabbiano}, G. and {Baldi}, A. and {Bianchi}, S. and {Braito}, V. and {Guainazzi}, M. and {Matt}, G. and {Miniutti}, G. and {Reeves}, J. and {Soria}, R. and {Zezas}, A.},
        title = "{The XMM-Newton long look of NGC 1365: uncovering of the obscured X-ray source}",
      journal = {\mnras},
         year = 2009,
        month = feb,
       volume = {393},
       number = {1},
        pages = {L1-L5},
          doi = {10.1111/j.1745-3933.2008.00580.x},
archivePrefix = {arXiv},
       eprint = {0811.1594},
 primaryClass = {astro-ph},
       adsurl = {https://ui.adsabs.harvard.edu/abs/2009MNRAS.393L...1R}
}

@ARTICLE{Laha2022,
       author = {{Laha}, Sibasish and {Meyer}, Eileen and {Roychowdhury}, Agniva and {Becerra Gonzalez}, Josefa and {Acosta-Pulido}, J.~A. and {Thapa}, Aditya and {Ghosh}, Ritesh and {Behar}, Ehud and {Gallo}, Luigi C. and {Kriss}, Gerard A. and {Panessa}, Francesca and {Bianchi}, Stefano and {La Franca}, Fabio and {Scepi}, Nicolas and {Begelman}, Mitchell C. and {Longinotti}, Anna Lia and {Lusso}, Elisabeta and {Oates}, Samantha and {Nicholl}, Matt and {Cenko}, S. Bradley},
        title = "{A Radio, Optical, UV, and X-Ray View of the Enigmatic Changing-look Active Galactic Nucleus 1ES 1927+654 from Its Pre- to Postflare States}",
      journal = {\apj},
         year = 2022,
        month = may,
       volume = {931},
       number = {1},
          eid = {5},
        pages = {5},
          doi = {10.3847/1538-4357/ac63aa},
archivePrefix = {arXiv},
       eprint = {2203.07446},
 primaryClass = {astro-ph.HE},
       adsurl = {https://ui.adsabs.harvard.edu/abs/2022ApJ...931....5L}
}

@ARTICLE{Fukazawa2009,
       author = {{Fukazawa}, Yasushi and {Mizuno}, Tsunefumi and {Watanabe}, Shin and {Kokubun}, Motohide and {Takahashi}, Hiromitsu and {Kawano}, Naomi and {Nishino}, Sho and {Sasada}, Mahito and {Shirai}, Hirohisa and {Takahashi}, Takuya and {Umeki}, Yudai and {Yamasaki}, Tomonori and {Yasuda}, Tomonori and {Bamba}, Aya and {Ohno}, Masanori and {Takahashi}, Tadayuki and {Ushio}, Masayoshi and {Enoto}, Teruaki and {Kitaguchi}, Takao and {Makishima}, Kazuo and {Nakazawa}, Kazuhiro and {Uehara}, Yuichi and {Yamada}, Shin'ya and {Yuasa}, Takayuki and {Isobe}, Naoki and {Kawaharada}, Madoka and {Tanaka}, Takaaki and {Tashiro}, Makoto S. and {Terada}, Yukikatsu and {Yamaoka}, Kazutaka},
        title = "{Modeling and Reproducibility of Suzaku HXD PIN/GSO Background}",
      journal = {\pasj},
         year = 2009,
        month = jan,
       volume = {61},
        pages = {S17},
          doi = {10.1093/pasj/61.sp1.S17},
       adsurl = {https://ui.adsabs.harvard.edu/abs/2009PASJ...61S..17F}
}

@ARTICLE{Takahashi2007,
       author = {{Takahashi}, Tadayuki and {Abe}, Keiichi and {Endo}, Manabu and {Endo}, Yasuhiko and {Ezoe}, Yuuichiro and {Fukazawa}, Yasushi and {Hamaya}, Masahito and {Hirakuri}, Shinya and {Hong}, Soojing and {Horii}, Michihiro and {Inoue}, Hokuto and {Isobe}, Naoki and {Itoh}, Takeshi and {Iyomoto}, Naoko and {Kamae}, Tuneyoshi and {Kasama}, Daisuke and {Kataoka}, Jun and {Kato}, Hiroshi and {Kawaharada}, Madoka and {Kawano}, Naomi and {Kawashima}, Kengo and {Kawasoe}, Satoshi and {Kishishita}, Tetsuichi and {Kitaguchi}, Takao and {Kobayashi}, Yoshihito and {Kokubun}, Motohide and {Kotoku}, Jun'ichi and {Kouda}, Manabu and {Kubota}, Aya and {Kuroda}, Yoshikatsu and {Madejski}, Greg and {Makishima}, Kazuo and {Masukawa}, Kazunori and {Matsumoto}, Yukari and {Mitani}, Takefumi and {Miyawaki}, Ryohei and {Mizuno}, Tsunefumi and {Mori}, Kunishiro and {Mori}, Masanori and {Murashima}, Mio and {Murakami}, Toshio and {Nakazawa}, Kazuhiro and {Niko}, Hisako and {Nomachi}, Masaharu and {Okada}, Yuu and {Ohno}, Masanori and {Oonuki}, Kousuke and {Ota}, Naomi and {Ozawa}, Hideki and {Sato}, Goro and {Shinoda}, Shingo and {Sugiho}, Masahiko and {Suzuki}, Masaya and {Taguchi}, Koji and {Takahashi}, Hiromitsu and {Takahashi}, Isao and {Takeda}, Shin'ichiro and {Tamura}, Ken-Ichi and {Tamura}, Takayuki and {Tanaka}, Takaaki and {Tanihata}, Chiharu and {Tashiro}, Makoto and {Terada}, Yukikatsu and {Tominaga}, Shin'ya and {Uchiyama}, Yasunobu and {Watanabe}, Shin and {Yamaoka}, Kazutaka and {Yanagida}, Takayuki and {Yonetoku}, Daisuke},
        title = "{Hard X-Ray Detector (HXD) on Board Suzaku}",
      journal = {\pasj},
         year = 2007,
        month = jan,
       volume = {59},
        pages = {35-51},
          doi = {10.1093/pasj/59.sp1.S35},
archivePrefix = {arXiv},
       eprint = {astro-ph/0611232},
 primaryClass = {astro-ph},
       adsurl = {https://ui.adsabs.harvard.edu/abs/2007PASJ...59S..35T}
}

@ARTICLE{Koyama2007,
       author = {{Koyama}, Katsuji and {Tsunemi}, Hiroshi and {Dotani}, Tadayasu and {Bautz}, Mark W. and {Hayashida}, Kiyoshi and {Tsuru}, Takeshi Go and {Matsumoto}, Hironori and {Ogawara}, Yoshiaki and {Ricker}, George R. and {Doty}, John and {Kissel}, Steven E. and {Foster}, Rick and {Nakajima}, Hiroshi and {Yamaguchi}, Hiroya and {Mori}, Hideyuki and {Sakano}, Masaaki and {Hamaguchi}, Kenji and {Nishiuchi}, Mamiko and {Miyata}, Emi and {Torii}, Ken'ichi and {Namiki}, Masaaki and {Katsuda}, Satoru and {Matsuura}, Daisuke and {Miyauchi}, Tomofumi and {Anabuki}, Naohisa and {Tawa}, Noriaki and {Ozaki}, Masanobu and {Murakami}, Hiroshi and {Maeda}, Yoshitomo and {Ichikawa}, Yoshinori and {Prigozhin}, Gregory Y. and {Boughan}, Edward A. and {Lamarr}, Beverly and {Miller}, Eric D. and {Burke}, Barry E. and {Gregory}, James A. and {Pillsbury}, Allen and {Bamba}, Aya and {Hiraga}, Junko S. and {Senda}, Atsushi and {Katayama}, Haruyoshi and {Kitamoto}, Shunji and {Tsujimoto}, Masahiro and {Kohmura}, Takayoshi and {Tsuboi}, Yohko and {Awaki}, Hisamitsu},
        title = "{X-Ray Imaging Spectrometer (XIS) on Board Suzaku}",
      journal = {\pasj},
         year = 2007,
        month = jan,
       volume = {59},
        pages = {23-33},
          doi = {10.1093/pasj/59.sp1.S23},
       adsurl = {https://ui.adsabs.harvard.edu/abs/2007PASJ...59S..23K}
}

@ARTICLE{Palit2025,
       author = {{Palit}, Biswaraj and {{\'S}niegowska}, Marzena and {Markowitz}, Alex and {R{\'o}{\.z}a{\'n}ska}, Agata and {Trakhtenbrot}, Benny and {Farah}, Joseph and {Howell}, Andy},
        title = "{Markarian 590: The AGN Awakens}",
      journal = {arXiv e-prints},
         year = 2025,
        month = jan,
          eid = {arXiv:2501.07225},
        pages = {arXiv:2501.07225},
          doi = {10.48550/arXiv.2501.07225},
archivePrefix = {arXiv},
       eprint = {2501.07225},
 primaryClass = {astro-ph.HE},
       adsurl = {https://ui.adsabs.harvard.edu/abs/2025arXiv250107225P}
}

@ARTICLE{Rivers2012,
       author = {{Rivers}, Elizabeth and {Markowitz}, Alex and {Duro}, Refiz and {Rothschild}, Richard},
        title = "{A Suzaku Observation of Mkn 590 Reveals a Vanishing Soft Excess}",
      journal = {\apj},
         year = 2012,
        month = nov,
       volume = {759},
       number = {1},
          eid = {63},
        pages = {63},
          doi = {10.1088/0004-637X/759/1/63},
archivePrefix = {arXiv},
       eprint = {1210.3330},
 primaryClass = {astro-ph.HE},
       adsurl = {https://ui.adsabs.harvard.edu/abs/2012ApJ...759...63R}
}

@ARTICLE{Ghosh2022,
       author = {{Ghosh}, Ritesh and {Laha}, Sibasish and {Deshmukh}, Kunal and {Bhalerao}, Varun and {Dewangan}, Gulab C. and {Chatterjee}, Ritaban},
        title = "{The Origin of the Vanishing Soft X-Ray Excess in the Changing-look Active Galactic Nucleus Mrk 590}",
      journal = {\apj},
         year = 2022,
        month = sep,
       volume = {937},
       number = {1},
          eid = {31},
        pages = {31},
          doi = {10.3847/1538-4357/ac887e},
archivePrefix = {arXiv},
       eprint = {2208.04067},
 primaryClass = {astro-ph.HE},
       adsurl = {https://ui.adsabs.harvard.edu/abs/2022ApJ...937...31G}
}

@ARTICLE{Koss2022,
       author = {{Koss}, Michael J. and {Ricci}, Claudio and {Trakhtenbrot}, Benny and {Oh}, Kyuseok and {den Brok}, Jakob S. and {Mej{\'\i}a-Restrepo}, Julian E. and {Stern}, Daniel and {Privon}, George C. and {Treister}, Ezequiel and {Powell}, Meredith C. and {Mushotzky}, Richard and {Bauer}, Franz E. and {Ananna}, Tonima T. and {Balokovi{\'c}}, Mislav and {B{\"a}r}, Rudolf E. and {Becker}, George and {Bessiere}, Patricia and {Burtscher}, Leonard and {Caglar}, Turgay and {Congiu}, Enrico and {Evans}, Phil and {Harrison}, Fiona and {Heida}, Marianne and {Ichikawa}, Kohei and {Kamraj}, Nikita and {Lamperti}, Isabella and {Pacucci}, Fabio and {Ricci}, Federica and {Riffel}, Rog{\'e}rio and {Rojas}, Alejandra F. and {Schawinski}, Kevin and {Temple}, Matthew J. and {Urry}, C. Megan and {Veilleux}, Sylvain and {Williams}, Jonathan},
        title = "{BASS. XXII. The BASS DR2 AGN Catalog and Data}",
      journal = {\apjs},
         year = 2022,
        month = jul,
       volume = {261},
       number = {1},
          eid = {2},
        pages = {2},
          doi = {10.3847/1538-4365/ac6c05},
archivePrefix = {arXiv},
       eprint = {2207.12432},
 primaryClass = {astro-ph.GA},
       adsurl = {https://ui.adsabs.harvard.edu/abs/2022ApJS..261....2K}
}

@ARTICLE{koss2022b,
       author = {{Koss}, Michael J. and {Trakhtenbrot}, Benny and {Ricci}, Claudio and {Oh}, Kyuseok and {Bauer}, Franz E. and {Stern}, Daniel and {Caglar}, Turgay and {den Brok}, Jakob S. and {Mushotzky}, Richard and {Ricci}, Federica and {Mej{\'\i}a-Restrepo}, Julian E. and {Lamperti}, Isabella and {Treister}, Ezequiel and {B{\"a}r}, Rudolf E. and {Harrison}, Fiona and {Powell}, Meredith C. and {Privon}, George C. and {Riffel}, Rog{\'e}rio and {Rojas}, Alejandra F. and {Schawinski}, Kevin and {Urry}, C. Megan},
        title = "{BASS. XXVI. DR2 Host Galaxy Stellar Velocity Dispersions}",
      journal = {\apjs},
         year = 2022,
        month = jul,
       volume = {261},
       number = {1},
          eid = {6},
        pages = {6},
          doi = {10.3847/1538-4365/ac650b},
archivePrefix = {arXiv},
       eprint = {2207.12435},
 primaryClass = {astro-ph.GA},
       adsurl = {https://ui.adsabs.harvard.edu/abs/2022ApJS..261....6K}
}

@ARTICLE{Vasudevan2007,
       author = {{Vasudevan}, R.~V. and {Fabian}, A.~C.},
        title = "{Piecing together the X-ray background: bolometric corrections for active galactic nuclei}",
      journal = {\mnras},
         year = 2007,
        month = nov,
       volume = {381},
       number = {3},
        pages = {1235-1251},
          doi = {10.1111/j.1365-2966.2007.12328.x},
archivePrefix = {arXiv},
       eprint = {0708.4308},
 primaryClass = {astro-ph},
       adsurl = {https://ui.adsabs.harvard.edu/abs/2007MNRAS.381.1235V}
}

@ARTICLE{Rees1988,
       author = {{Rees}, Martin J.},
        title = "{Tidal disruption of stars by black holes of {}10$^{6}$-{}10$^{8}$ solar masses in nearby galaxies}",
      journal = {\nat},
         year = 1988,
        month = jun,
       volume = {333},
       number = {6173},
        pages = {523-528},
          doi = {10.1038/333523a0},
       adsurl = {https://ui.adsabs.harvard.edu/abs/1988Natur.333..523R}
}

@ARTICLE{Evans2009,
       author = {{Evans}, P.~A. and {Beardmore}, A.~P. and {Page}, K.~L. and
         {Osborne}, J.~P. and {O'Brien}, P.~T. and {Willingale}, R. and
         {Starling}, R.~L.~C. and {Burrows}, D.~N. and {Godet}, O. and
         {Vetere}, L. and {Racusin}, J. and {Goad}, M.~R. and {Wiersema}, K. and
         {Angelini}, L. and {Capalbi}, M. and {Chincarini}, G. and
         {Gehrels}, N. and {Kennea}, J.~A. and {Margutti}, R. and
         {Morris}, D.~C. and {Mountford}, C.~J. and {Pagani}, C. and
         {Perri}, M. and {Romano}, P. and {Tanvir}, N.},
        title = "{Methods and results of an automatic analysis of a complete sample of Swift-XRT observations of GRBs}",
      journal = {\mnras},
         year = 2009,
        month = aug,
       volume = {397},
       number = {3},
        pages = {1177-1201},
          doi = {10.1111/j.1365-2966.2009.14913.x},
archivePrefix = {arXiv},
       eprint = {0812.3662},
 primaryClass = {astro-ph},
       adsurl = {https://ui.adsabs.harvard.edu/abs/2009MNRAS.397.1177E}
}

@ARTICLE{Harrison2013,
       author = {{Harrison}, Fiona A. and {Craig}, William W. and {Christensen}, Finn E. and
         {Hailey}, Charles J. and {Zhang}, William W. and {Boggs}, Steven E. and
         {Stern}, Daniel and {Cook}, W. Rick and {Forster}, Karl and
         {Giommi}, Paolo and {Grefenstette}, Brian W. and {Kim}, Yunjin and
         {Kitaguchi}, Takao and {Koglin}, Jason E. and {Madsen}, Kristin K. and
         {Mao}, Peter H. and {Miyasaka}, Hiromasa and {Mori}, Kaya and
         {Perri}, Matteo and {Pivovaroff}, Michael J. and {Puccetti}, Simonetta and
         {Rana}, Vikram R. and {Westergaard}, Niels J. and {Willis}, Jason and
         {Zoglauer}, Andreas and {An}, Hongjun and {Bachetti}, Matteo and
         {Barri{\`e}re}, Nicolas M. and {Bellm}, Eric C. and {Bhalerao}, Varun and
         {Brejnholt}, Nicolai F. and {Fuerst}, Felix and {Liebe}, Carl C. and
         {Markwardt}, Craig B. and {Nynka}, Melania and {Vogel}, Julia K. and
         {Walton}, Dominic J. and {Wik}, Daniel R. and {Alexander}, David M. and
         {Cominsky}, Lynn R. and {Hornschemeier}, Ann E. and {Hornstrup}, Allan and
         {Kaspi}, Victoria M. and {Madejski}, Greg M. and {Matt}, Giorgio and
         {Molendi}, Silvano and {Smith}, David M. and {Tomsick}, John A. and
         {Ajello}, Marco and {Ballantyne}, David R. and {Balokovi{\'c}}, Mislav and
         {Barret}, Didier and {Bauer}, Franz E. and {Blandford}, Roger D. and
         {Brandt}, W. Niel and {Brenneman}, Laura W. and {Chiang}, James and
         {Chakrabarty}, Deepto and {Chenevez}, Jerome and {Comastri}, Andrea and
         {Dufour}, Francois and {Elvis}, Martin and {Fabian}, Andrew C. and
         {Farrah}, Duncan and {Fryer}, Chris L. and {Gotthelf}, Eric V. and
         {Grindlay}, Jonathan E. and {Helfand}, David J. and {Krivonos}, Roman and
         {Meier}, David L. and {Miller}, Jon M. and {Natalucci}, Lorenzo and
         {Ogle}, Patrick and {Ofek}, Eran O. and {Ptak}, Andrew and
         {Reynolds}, Stephen P. and {Rigby}, Jane R. and
         {Tagliaferri}, Gianpiero and {Thorsett}, Stephen E. and
         {Treister}, Ezequiel and {Urry}, C. Megan},
        title = "{The Nuclear Spectroscopic Telescope Array (NuSTAR) High-energy X-Ray Mission}",
      journal = {\apj},
         year = 2013,
        month = jun,
       volume = {770},
       number = {2},
          eid = {103},
        pages = {103},
          doi = {10.1088/0004-637X/770/2/103},
archivePrefix = {arXiv},
       eprint = {1301.7307},
 primaryClass = {astro-ph.IM},
       adsurl = {https://ui.adsabs.harvard.edu/abs/2013ApJ...770..103H}
}

@ARTICLE{Denney2014,
       author = {{Denney}, K.~D. and {De Rosa}, G. and {Croxall}, K. and {Gupta}, A. and
         {Bentz}, M.~C. and {Fausnaugh}, M.~M. and {Grier}, C.~J. and
         {Martini}, P. and {Mathur}, S. and {Peterson}, B.~M. and
         {Pogge}, R.~W. and {Shappee}, B.~J.},
        title = "{The Typecasting of Active Galactic Nuclei: Mrk 590 no Longer Fits the Role}",
      journal = {\apj},
         year = 2014,
        month = dec,
       volume = {796},
       number = {2},
          eid = {134},
        pages = {134},
          doi = {10.1088/0004-637X/796/2/134},
archivePrefix = {arXiv},
       eprint = {1404.4879},
 primaryClass = {astro-ph.GA},
       adsurl = {https://ui.adsabs.harvard.edu/abs/2014ApJ...796..134D}
}

@ARTICLE{Shappee2014,
       author = {{Shappee}, B.~J. and {Prieto}, J.~L. and {Grupe}, D. and
         {Kochanek}, C.~S. and {Stanek}, K.~Z. and {De Rosa}, G. and
         {Mathur}, S. and {Zu}, Y. and {Peterson}, B.~M. and {Pogge}, R.~W. and
         {Komossa}, S. and {Im}, M. and {Jencson}, J. and {Holoien}, T.~W. -S. and
         {Basu}, U. and {Beacom}, J.~F. and {Szczygie{\l}}, D.~M. and
         {Brimacombe}, J. and {Adams}, S. and {Campillay}, A. and {Choi}, C. and
         {Contreras}, C. and {Dietrich}, M. and {Dubberley}, M. and
         {Elphick}, M. and {Foale}, S. and {Giustini}, M. and {Gonzalez}, C. and
         {Hawkins}, E. and {Howell}, D.~A. and {Hsiao}, E.~Y. and {Koss}, M. and
         {Leighly}, K.~M. and {Morrell}, N. and {Mudd}, D. and {Mullins}, D. and
         {Nugent}, J.~M. and {Parrent}, J. and {Phillips}, M.~M. and
         {Pojmanski}, G. and {Rosing}, W. and {Ross}, R. and {Sand}, D. and
         {Terndrup}, D.~M. and {Valenti}, S. and {Walker}, Z. and {Yoon}, Y.},
        title = "{The Man behind the Curtain: X-Rays Drive the UV through NIR Variability in the 2013 Active Galactic Nucleus Outburst in NGC 2617}",
      journal = {\apj},
         year = 2014,
        month = jun,
       volume = {788},
       number = {1},
          eid = {48},
        pages = {48},
          doi = {10.1088/0004-637X/788/1/48},
archivePrefix = {arXiv},
       eprint = {1310.2241},
 primaryClass = {astro-ph.HE},
       adsurl = {https://ui.adsabs.harvard.edu/abs/2014ApJ...788...48S}
}

@ARTICLE{Merloni2015,
       author = {{Merloni}, A. and {Dwelly}, T. and {Salvato}, M. and {Georgakakis}, A. and
         {Greiner}, J. and {Krumpe}, M. and {Nandra}, K. and {Ponti}, G. and
         {Rau}, A.},
        title = "{A tidal disruption flare in a massive galaxy? Implications for the fuelling mechanisms of nuclear black holes}",
      journal = {\mnras},
         year = 2015,
        month = sep,
       volume = {452},
       number = {1},
        pages = {69-87},
          doi = {10.1093/mnras/stv1095},
archivePrefix = {arXiv},
       eprint = {1503.04870},
 primaryClass = {astro-ph.HE},
       adsurl = {https://ui.adsabs.harvard.edu/abs/2015MNRAS.452...69M}
}

@ARTICLE{MacLeod2016,
       author = {{MacLeod}, Chelsea L. and {Ross}, Nicholas P. and {Lawrence}, Andy and
         {Goad}, Mike and {Horne}, Keith and {Burgett}, William and
         {Chambers}, Ken C. and {Flewelling}, Heather and {Hodapp}, Klaus and
         {Kaiser}, Nick and {Magnier}, Eugene and {Wainscoat}, Richard and
         {Waters}, Christopher},
        title = "{A systematic search for changing-look quasars in SDSS}",
      journal = {\mnras},
         year = 2016,
        month = mar,
       volume = {457},
       number = {1},
        pages = {389-404},
          doi = {10.1093/mnras/stv2997},
archivePrefix = {arXiv},
       eprint = {1509.08393},
 primaryClass = {astro-ph.GA},
       adsurl = {https://ui.adsabs.harvard.edu/abs/2016MNRAS.457..389M}
}

@ARTICLE{Cohen1986,
       author = {{Cohen}, Ross D. and {Rudy}, Richard J. and {Puetter}, R.~C. and
         {Ake}, T.~B. and {Foltz}, Craig B.},
        title = "{Variability of Markarian 1018: Seyfert 1.9 to Seyfert 1}",
      journal = {\apj},
         year = 1986,
        month = dec,
       volume = {311},
        pages = {135},
          doi = {10.1086/164758},
       adsurl = {https://ui.adsabs.harvard.edu/abs/1986ApJ...311..135C}
}

@ARTICLE{Jansen2001,
       author = {{Jansen}, F. and {Lumb}, D. and {Altieri}, B. and {Clavel}, J. and
         {Ehle}, M. and {Erd}, C. and {Gabriel}, C. and {Guainazzi}, M. and
         {Gondoin}, P. and {Much}, R. and {Munoz}, R. and {Santos}, M. and
         {Schartel}, N. and {Texier}, D. and {Vacanti}, G.},
        title = "{XMM-Newton observatory. I. The spacecraft and operations}",
      journal = {\aap},
         year = 2001,
        month = jan,
       volume = {365},
        pages = {L1-L6},
          doi = {10.1051/0004-6361:20000036},
       adsurl = {https://ui.adsabs.harvard.edu/abs/2001A&A...365L...1J}
}

@ARTICLE{Ross2005,
       author = {{Ross}, R.~R. and {Fabian}, A.~C.},
        title = "{A comprehensive range of X-ray ionized-reflection models}",
      journal = {\mnras},
         year = 2005,
        month = mar,
       volume = {358},
       number = {1},
        pages = {211-216},
          doi = {10.1111/j.1365-2966.2005.08797.x},
archivePrefix = {arXiv},
       eprint = {astro-ph/0501116},
 primaryClass = {astro-ph},
       adsurl = {https://ui.adsabs.harvard.edu/abs/2005MNRAS.358..211R}
}

@ARTICLE{Walton2013,
       author = {{Walton}, D.~J. and {Nardini}, E. and {Fabian}, A.~C. and
         {Gallo}, L.~C. and {Reis}, R.~C.},
        title = "{Suzaku observations of `bare' active galactic nuclei}",
      journal = {\mnras},
         year = 2013,
        month = feb,
       volume = {428},
       number = {4},
        pages = {2901-2920},
          doi = {10.1093/mnras/sts227},
archivePrefix = {arXiv},
       eprint = {1210.4593},
 primaryClass = {astro-ph.HE},
       adsurl = {https://ui.adsabs.harvard.edu/abs/2013MNRAS.428.2901W}
}

@ARTICLE{Ochmann2020,
       author = {{Ochmann}, M.~W. and {Kollatschny}, W. and {Zetzl}, M.},
        title = "{Spectral changes and BLR kinematics of eruptive changing-look AGN}",
      journal = {Contributions of the Astronomical Observatory Skalnate Pleso},
         year = 2020,
        month = jan,
       volume = {50},
       number = {1},
        pages = {318-327},
          doi = {10.31577/caosp.2020.50.1.318},
       adsurl = {https://ui.adsabs.harvard.edu/abs/2020CoSka..50..318O}
}

@ARTICLE{Parker2019,
       author = {{Parker}, M.~L. and {Schartel}, N. and {Grupe}, D. and {Komossa}, S. and
         {Harrison}, F. and {Kollatschny}, W. and {Mikula}, R. and
         {Santos-Lle{\'o}}, M. and {Tom{\'a}s}, L.},
        title = "{X-ray spectra reveal the reawakening of the repeat changing-look AGN NGC 1566}",
      journal = {\mnras},
         year = 2019,
        month = feb,
       volume = {483},
       number = {1},
        pages = {L88-L92},
          doi = {10.1093/mnrasl/sly224},
archivePrefix = {arXiv},
       eprint = {1811.10289},
 primaryClass = {astro-ph.HE},
       adsurl = {https://ui.adsabs.harvard.edu/abs/2019MNRAS.483L..88P}
}

@ARTICLE{Ricci2018,
       author = {{Ricci}, C. and {Ho}, L.~C. and {Fabian}, A.~C. and {Trakhtenbrot}, B. and {Koss}, M.~J. and {Ueda}, Y. and {Lohfink}, A. and {Shimizu}, T. and {Bauer}, F.~E. and {Mushotzky}, R. and {Schawinski}, K. and {Paltani}, S. and {Lamperti}, I. and {Treister}, E. and {Oh}, K.},
        title = "{BAT AGN Spectroscopic Survey - XII. The relation between coronal properties of active galactic nuclei and the Eddington ratio}",
      journal = {\mnras},
         year = 2018,
        month = oct,
       volume = {480},
       number = {2},
        pages = {1819-1830},
          doi = {10.1093/mnras/sty1879},
archivePrefix = {arXiv},
       eprint = {1809.04076},
 primaryClass = {astro-ph.HE},
       adsurl = {https://ui.adsabs.harvard.edu/abs/2018MNRAS.480.1819R}
}

@ARTICLE{Oknyansky2019,
       author = {{Oknyansky}, V.~L. and {Winkler}, H. and {Tsygankov}, S.~S. and
         {Lipunov}, V.~M. and {Gorbovskoy}, E.~S. and {van Wyk}, F. and
         {Buckley}, D.~A.~H. and {Tyurina}, N.~V.},
        title = "{New changing look case in NGC 1566}",
      journal = {\mnras},
         year = 2019,
        month = feb,
       volume = {483},
       number = {1},
        pages = {558-564},
          doi = {10.1093/mnras/sty3133},
archivePrefix = {arXiv},
       eprint = {1811.06926},
 primaryClass = {astro-ph.GA},
       adsurl = {https://ui.adsabs.harvard.edu/abs/2019MNRAS.483..558O}
}

@ARTICLE{SS73,
       author = {{Shakura}, N.~I. and {Sunyaev}, R.~A.},
        title = "{Reprint of 1973A\&A....24..337S. Black holes in binary systems. Observational appearance.}",
      journal = {\aap},
         year = 1973,
        month = jun,
       volume = {500},
        pages = {33-51},
       adsurl = {https://ui.adsabs.harvard.edu/abs/1973A&A....24..337S}
}

@ARTICLE{Oknyansky2023ATel,
       author = {{Oknyansky}, V.~L. and {Tsygankov}, S.~S. and {Dodin}, A.~S. and {Tatarnikov}, A.~M. and {Metlov}, V.~G. and {Burlak}, M.~A. and {Ikonnikova}, N.~P. and {Belinski}, A.~A. and {Shatsky}, N.~I. and {Chelouche}, D. and {Figaredo}, C. Sobrino and {Kaspi}, S. and {Brotherton}, M. and {Pu}, D.},
        title = "{The Changing Look AGN NGC 2617 is in the deepest low state}",
      journal = {The Astronomer's Telegram},
         year = 2023,
        month = nov,
       volume = {16324},
        pages = {1},
       adsurl = {https://ui.adsabs.harvard.edu/abs/2023ATel16324....1O}
}

@ARTICLE{Fender2004,
       author = {{Fender}, Rob and {Belloni}, Tomaso},
        title = "{GRS 1915+105 and the Disc-Jet Coupling in Accreting Black Hole Systems}",
      journal = {\araa},
         year = 2004,
        month = sep,
       volume = {42},
       number = {1},
        pages = {317-364},
          doi = {10.1146/annurev.astro.42.053102.134031},
archivePrefix = {arXiv},
       eprint = {astro-ph/0406483},
 primaryClass = {astro-ph},
       adsurl = {https://ui.adsabs.harvard.edu/abs/2004ARA&A..42..317F}
}

@INPROCEEDINGS{Arnaud1996,
       author = {{Arnaud}, K.~A.},
        title = "{XSPEC: The First Ten Years}",
    booktitle = {Astronomical Data Analysis Software and Systems V},
         year = 1996,
       editor = {{Jacoby}, George H. and {Barnes}, Jeannette},
       series = {Astronomical Society of the Pacific Conference Series},
       volume = {101},
        month = jan,
        pages = {17},
       adsurl = {https://ui.adsabs.harvard.edu/abs/1996ASPC..101...17A}
}

@ARTICLE{Maccarone2003,
       author = {{Maccarone}, T.~J.},
        title = "{Do X-ray binary spectral state transition luminosities vary?}",
      journal = {\aap},
         year = 2003,
        month = oct,
       volume = {409},
        pages = {697-706},
          doi = {10.1051/0004-6361:20031146},
archivePrefix = {arXiv},
       eprint = {astro-ph/0308036},
 primaryClass = {astro-ph},
       adsurl = {https://ui.adsabs.harvard.edu/abs/2003A&A...409..697M}
}

@ARTICLE{CT95,
       author = {{Chakrabarti}, Sandip and {Titarchuk}, Lev G.},
        title = "{Spectral Properties of Accretion Disks around Galactic and Extragalactic Black Holes}",
      journal = {\apj},
         year = 1995,
        month = dec,
       volume = {455},
        pages = {623},
          doi = {10.1086/176610},
archivePrefix = {arXiv},
       eprint = {astro-ph/9510005},
 primaryClass = {astro-ph},
       adsurl = {https://ui.adsabs.harvard.edu/abs/1995ApJ...455..623C}
}

@ARTICLE{Gu2009,
       author = {{Gu}, Minfeng and {Cao}, Xinwu},
        title = "{The anticorrelation between the hard X-ray photon index and the Eddington ratio in low-luminosity active galactic nuclei}",
      journal = {\mnras},
         year = 2009,
        month = oct,
       volume = {399},
       number = {1},
        pages = {349-356},
          doi = {10.1111/j.1365-2966.2009.15277.x},
archivePrefix = {arXiv},
       eprint = {0906.3560},
 primaryClass = {astro-ph.GA},
       adsurl = {https://ui.adsabs.harvard.edu/abs/2009MNRAS.399..349G}
}

@ARTICLE{Mitchell2023,
       author = {{Mitchell}, Jake A.~J. and {Done}, Chris and {Ward}, Martin J. and {Kynoch}, Daniel and {Hagen}, Scott and {Lusso}, Elisabeta and {Landt}, Hermine},
        title = "{The SOUX AGN sample: optical/UV/X-ray SEDs and the nature of the disc}",
      journal = {\mnras},
         year = 2023,
        month = sep,
       volume = {524},
       number = {2},
        pages = {1796-1825},
          doi = {10.1093/mnras/stad1830},
archivePrefix = {arXiv},
       eprint = {2210.11977},
 primaryClass = {astro-ph.GA},
       adsurl = {https://ui.adsabs.harvard.edu/abs/2023MNRAS.524.1796M}
}

@ARTICLE{Trump2011,
       author = {{Trump}, Jonathan R. and {Impey}, Christopher D. and {Kelly}, Brandon C. and {Civano}, Francesca and {Gabor}, Jared M. and {Diamond-Stanic}, Aleksandar M. and {Merloni}, Andrea and {Urry}, C. Megan and {Hao}, Heng and {Jahnke}, Knud and {Nagao}, Tohru and {Taniguchi}, Yoshi and {Koekemoer}, Anton M. and {Lanzuisi}, Giorgio and {Liu}, Charles and {Mainieri}, Vincenzo and {Salvato}, Mara and {Scoville}, Nick Z.},
        title = "{Accretion Rate and the Physical Nature of Unobscured Active Galaxies}",
      journal = {\apj},
         year = 2011,
        month = may,
       volume = {733},
       number = {1},
          eid = {60},
        pages = {60},
          doi = {10.1088/0004-637X/733/1/60},
archivePrefix = {arXiv},
       eprint = {1103.0276},
 primaryClass = {astro-ph.CO},
       adsurl = {https://ui.adsabs.harvard.edu/abs/2011ApJ...733...60T}
}

@ARTICLE{Dauser2016,
       author = {{Dauser}, T. and {Garc{\'\i}a}, J. and {Walton}, D.~J. and
         {Eikmann}, W. and {Kallman}, T. and {McClintock}, J. and {Wilms}, J.},
        title = "{Normalizing a relativistic model of X-ray reflection. Definition of the reflection fraction and its implementation in relxill}",
      journal = {\aap},
         year = 2016,
        month = may,
       volume = {590},
          eid = {A76},
        pages = {A76},
          doi = {10.1051/0004-6361/201628135},
archivePrefix = {arXiv},
       eprint = {1601.03771},
 primaryClass = {astro-ph.HE},
       adsurl = {https://ui.adsabs.harvard.edu/abs/2016A&A...590A..76D}
}

@ARTICLE{AJ2021,
       author = {{Jana}, Arghajit and {Kumari}, Neeraj and {Nandi}, Prantik and {Naik}, Sachindra and {Chatterjee}, Arka and {Jaisawal}, Gaurava K. and {Hayasaki}, Kimitake and {Ricci}, Claudio},
        title = "{Broad-band X-ray observations of the 2018 outburst of the changing-look active galactic nucleus NGC 1566}",
      journal = {\mnras},
         year = 2021,
        month = oct,
       volume = {507},
       number = {1},
        pages = {687-703},
          doi = {10.1093/mnras/stab2155},
archivePrefix = {arXiv},
       eprint = {2107.11127},
 primaryClass = {astro-ph.HE},
       adsurl = {https://ui.adsabs.harvard.edu/abs/2021MNRAS.507..687J}
}

@ARTICLE{Kawamuro2013,
       author = {{Kawamuro}, Taiki and {Ueda}, Yoshihiro and {Tazaki}, Fumie and {Terashima}, Yuichi},
        title = "{Broadband X-Ray Spectra of Two Low-luminosity Active Galactic Nuclei NGC 1566 and NGC 4941 Observed with Suzaku}",
      journal = {\apj},
         year = 2013,
        month = jun,
       volume = {770},
       number = {2},
          eid = {157},
        pages = {157},
          doi = {10.1088/0004-637X/770/2/157},
archivePrefix = {arXiv},
       eprint = {1305.2912},
 primaryClass = {astro-ph.GA},
       adsurl = {https://ui.adsabs.harvard.edu/abs/2013ApJ...770..157K}
}

@ARTICLE{Temple2023,
       author = {{Temple}, Matthew J. and {Ricci}, Claudio and {Koss}, Michael J. and {Trakhtenbrot}, Benny and {Bauer}, Franz E. and {Mushotzky}, Richard and {Rojas}, Alejandra F. and {Caglar}, Turgay and {Harrison}, Fiona and {Oh}, Kyuseok and {Padilla Gonzalez}, Estefania and {Powell}, Meredith C. and {Ricci}, Federica and {Riffel}, Rog{\'e}rio and {Stern}, Daniel and {Urry}, C. Megan},
        title = "{BASS XXXIX: Swift-BAT AGN with changing-look optical spectra}",
      journal = {\mnras},
         year = 2023,
        month = jan,
       volume = {518},
       number = {2},
        pages = {2938-2953},
          doi = {10.1093/mnras/stac3279},
archivePrefix = {arXiv},
       eprint = {2211.04478},
 primaryClass = {astro-ph.HE},
       adsurl = {https://ui.adsabs.harvard.edu/abs/2023MNRAS.518.2938T}
}

@ARTICLE{Boissay2016,
       author = {{Boissay}, Rozenn and {Ricci}, Claudio and {Paltani}, St{\'e}phane},
        title = "{A hard X-ray view of the soft excess in AGN}",
      journal = {\aap},
         year = 2016,
        month = apr,
       volume = {588},
          eid = {A70},
        pages = {A70},
          doi = {10.1051/0004-6361/201526982},
archivePrefix = {arXiv},
       eprint = {1511.08168},
 primaryClass = {astro-ph.HE},
       adsurl = {https://ui.adsabs.harvard.edu/abs/2016A&A...588A..70B}
}

@ARTICLE{Giustini2017,
       author = {{Giustini}, M. and {Costantini}, E. and {De Marco}, B. and {Svoboda}, J. and {Motta}, S.~E. and {Proga}, D. and {Saxton}, R. and {Ferrigno}, C. and {Longinotti}, A.~L. and {Miniutti}, G. and {Grupe}, D. and {Mathur}, S. and {Shappee}, B.~J. and {Prieto}, J.~L. and {Stanek}, K.},
        title = "{Direct probe of the inner accretion flow around the supermassive black hole in NGC 2617}",
      journal = {\aap},
         year = 2017,
        month = jan,
       volume = {597},
          eid = {A66},
        pages = {A66},
          doi = {10.1051/0004-6361/201628686},
archivePrefix = {arXiv},
       eprint = {1608.00233},
 primaryClass = {astro-ph.GA},
       adsurl = {https://ui.adsabs.harvard.edu/abs/2017A&A...597A..66G}
}

@ARTICLE{Oknyansky2017,
       author = {{Oknyansky}, V.~L. and {Gaskell}, C.~M. and {Huseynov}, N.~A. and {Lipunov}, V.~M. and {Shatsky}, N.~I. and {Tsygankov}, S.~S. and {Gorbovskoy}, E.~S. and {Mikailov}, Kh. M. and {Tatarnikov}, A.~M. and {Buckley}, D.~A.~H. and {Metlov}, V.~G. and {Nadzhip}, A.~E. and {Kuznetsov}, A.~S. and {Balanutza}, P.~V. and {Burlak}, M.~A. and {Galazutdinov}, G.~A. and {Artamonov}, B.~P. and {Salmanov}, I.~R. and {Malanchev}, K.~L. and {Oknyansky}, R.~S.},
        title = "{The curtain remains open: NGC 2617 continues in a high state}",
      journal = {\mnras},
         year = 2017,
        month = may,
       volume = {467},
       number = {2},
        pages = {1496-1504},
          doi = {10.1093/mnras/stx149},
archivePrefix = {arXiv},
       eprint = {1701.05042},
 primaryClass = {astro-ph.HE},
       adsurl = {https://ui.adsabs.harvard.edu/abs/2017MNRAS.467.1496O}
}

@ARTICLE{Feng2021,
       author = {{Feng}, Hai-Cheng and {Liu}, H.~T. and {Bai}, J.~M. and {Yang}, Zi-Xu and {Hu}, Chen and {Li}, Sha-Sha and {Yang}, Sen and {Lu}, Kai-Xing and {Xiao}, Ming},
        title = "{Velocity-resolved Reverberation Mapping of Changing-look AGN NGC 2617}",
      journal = {\apj},
         year = 2021,
        month = may,
       volume = {912},
       number = {2},
          eid = {92},
        pages = {92},
          doi = {10.3847/1538-4357/abefe0},
archivePrefix = {arXiv},
       eprint = {2103.03508},
 primaryClass = {astro-ph.GA},
       adsurl = {https://ui.adsabs.harvard.edu/abs/2021ApJ...912...92F}
}
%
\newpage

\appendix

\section{Observation log}
\label{sec:obs_log}

\begin{table*}
\caption{Observation log}
\label{tab:log}
\begin{tabular}{ccccccccc}
\hline
\hline
Epochs & UT date & Soft X-ray & Obs ID  & Exp. & Hard X-ray & Obs ID & Exp. & Type \\
 & (yyyy-mm-dd) & instrument & & (ks) & instrument &  & (ks) &  \\
\hline
        ~ & ~ & ~ & ~ & ~ & ~ & ~ & ~ & ~ \\
\multicolumn{9}{c}{NGC\,1566} \\ 
        ~ & ~ & ~ & ~ & ~ & ~ & ~ & ~ & ~ \\
E1 & 2012-05-19 & Suzaku/XIS & 707002010 & 73 & Suzaku/HXD & 707002010 & 73 & 1.9 \\  
        E2 & 2015-11-05 & XMM & 763500201 & 92 & -- & -- & -- & 1.9 \\  
        E3 & 2018-06-26 & XMM & 800840201 & 94 & NuSTAR & 80301601002 & 57 & 1 \\  
        E4 & 2018-10-04 & XMM & 820530401 & 108 & NuSTAR & 80401601002 & 75 & 1 \\  
        E5 & 2019-06-05 & XMM & 840800401 & 94 & NuSTAR & 80502606002 & 57 & 1 \\  
        E6 & 2019-08-08 & Swift/XRT & 88910001 & 2 & NuSTAR & 60501031002 & 59 & 1 \\  
        E7 & 2019-08-11 & XMM & 851980101 & 18 & -- & -- & -- & 1 \\  
        E8 & 2019-08-18 & Swift/XRT & 88910002 & 2 & NuSTAR & 60501031004 & 77 & 1 \\  
        E9 & 2019-08-21 & Swift/XRT & 88910003 & 2 & NuSTAR & 60501031006 & 86 & 1 \\  \hline
        ~ & ~ & ~ & ~ & ~ & ~ & ~ & ~ & ~ \\
        \multicolumn{9}{c}{NGC\,2617} \\
        ~ & ~ & ~ & ~ & ~ & ~ & ~ & ~ & ~ \\ 
        E1 & 2013-04-27 & XMM & 701981601 & 35 & -- & -- & -- & 1 \\  
        E2 & 2013-05-24 & XMM & 701981901 & 66 & -- & -- & -- & 1 \\   \hline
        ~ & ~ & ~ & ~ & ~ & ~ & ~ & ~ & ~ \\  
        \multicolumn{9}{c}{Mrk\,590} \\
        ~ & ~ & ~ & ~ & ~ & ~ & ~ & ~ & ~ \\ 
        E1 & 2002-01-01 & XMM & 109130301 & 11 & -- & -- & -- & 1 \\  
        E2 & 2004-07-04 & XMM & 201020201 & 113 & -- & -- & -- & 1 \\  
        E3 & 2011-01-23 & Suzaku/XIS & 705043010 & 62 & Suzaku/HXD & 705043010 & 62 & 1 \\  
        E4 & 2011-01-26 & Suzaku/XIS & 705043020 & 41 & Suzaku/HXD & 705043020 & 41 & 1 \\  
        E5 & 2016-02-05 & Swift/XRT & 80903001 & 7 & NuSTAR & 60160095002 & 21 & 2 \\  
        E6 & 2016-12-02 & Swift/XRT & 88014001 & 2 & NuSTAR & 90201043002 & 51 & 1 \\  
        E7 & 2018-10-07 & Swift/XRT & 94095012 & 2 & NuSTAR & 80402610002 & 21 & 1 \\  
        E8 & 2019-08-31 & Swift/XRT & 11542001 & 5 & NuSTAR & 80502630002 & 68 & 1 \\  
        E9 & 2020-01-21 & Swift/XRT & 13172002 & 5 & NuSTAR & 80502630004 & 50 & 1 \\  
        E10 & 2020-07-04 & XMM & 865470201 & 27 & -- & -- & -- & 1 \\  
        E11 & 2021-01-03 & XMM & 865470301 & 27 & -- & -- & -- & 1 \\  
        E12 & 2021-01-10 & Swift/XRT & 95662033 & 10 & NuSTAR & 80502630006 & 42 & 1 \\  
        E13 & 2021-08-18 & Swift/XRT & 89297002 & 2 & NuSTAR & 60761012002 & 19 & 1 \\  
        E14 & 2021-12-22 & Swift/XRT & 89297004 & 7 & NuSTAR & 80602604002 & 53 & 1 \\  
        E15 & 2022-01-24 & XMM & 870840201 & 26 & ~ & ~ & ~ & 1 \\  
        E16 & 2022-07-28 & XMM & 912400101 & 27 & ~ & ~ & ~ & 1 \\  
        E17 & 2023-02-06 & XMM & 870840301 & 38 & NuSTAR & 80602604004 & 41 & 1 \\ \hline
                ~ & ~ & ~ & ~ & ~ & ~ & ~ & ~ & ~ \\
\multicolumn{9}{c}{Mrk\,1018} \\
~ & ~ & ~ & ~ & ~ & ~ & ~ & ~ & ~ \\  
        E1 & 2005-01-15 & XMM & 201090201 & 12 & -- & -- & -- & 1 \\  
        E2 & 2008-08-07 & XMM & 554920301 & 18 & -- & -- & -- & 1 \\  
        E3 & 2009-07-03 & Suzaku/XIS & 704044010 & 44 & Suzaku/HXD & 704044010 & 44 & 1 \\  
        E4 & 2016-02-10 & Swift/XRT & 80898001 & 4 & NuSTAR & 60160087002 & 22 & 2 \\  
        E5 & 2018-03-05 & Swift/XRT & -- & -- & NuSTAR & 60301022003 & 43 & 2 \\  
        E6 & 2018-07-17 & Swift/XRT & 88207003 & 2 & NuSTAR & 60301022005 & 42 & 2 \\  
        E7 & 2018-07-23 & XMM & 821240201 & 75 & -- & -- & -- & 2 \\ 
        E8 & 2019-01-04 & XMM & 821240301 & 68 & -- & -- & -- & 2 \\  
        E9  & 2021-02-04 & XMM & 864350101 & 65 & -- & -- & -- & 2 \\  
        E10 & 2021-08-15 & Swift/XRT & 89296001 & 2 & NuSTAR & 60761011002 & 18 & 2 \\  \hline
~ & ~ & ~ & ~ & ~ & ~ & ~ & ~ & ~ \\
\multicolumn{9}{c}{IRAS\,23226--3843} \\
~ & ~ & ~ & ~ & ~ & ~ & ~ & ~ & ~ \\ 
        E1 & 2016-07-08 & Swift/XRT & 81303001 & 6 & NuSTAR & 60160826002 & 22 & 1.8 \\  
        E2 & 2017-05-02 & XMM & 760020101 & 16 & -- & -- & -- & 1.9 \\  
        E3 & 2017-06-11 & XMM & 760020401 & 86 & NuSTAR & 80101001002 & 97 & 1.9 \\  
        E4 & 2019-11-07 & XMM & 840800501 & 90 & NuSTAR & 80502607002 & 55 & 1 \\   
\hline \hline
\end{tabular}
\end{table*}

\section{Notes on individual sources}
\label{sec:sources}

\subsection{NGC\,1566}
\label{subsec:ngc1566}

NGC\,1566 is a nearby CSAGN (z=0.00476), that showed multiple CS transitions over the years \citep[see][for details]{AJ2025cl}.
More recently, NGC\,1566 underwent an outburst in 2018 when the source transitioned from type\,1.9 to type\,1 states \citep[e.g.,][]{Parker2019,Oknyansky2017,Oknyansky2019}. During this outburst, both primary continuum and soft-excess flux increased along with optical, ultraviolet, mid-infrared, and millimeter fluxes \citep[e.g.,][]{Oknyansky2019,Tripathi2022}. The source again moved back to type\,2 state with fluxes in all wavebands gradually decreasing \citep{Ochmann2020,Xu2024}.

The X-ray properties of NGC\,1566 has been studied extensively in the past \citep[e.g.,][]{Kawamuro2013,Parker2019,AJ2021,Tripathi2022}. Previous studies have found an evolving SE that decreases when the flux decreases, consistent with the findings of the current work \citep[e.g.,][]{Tripathi2022}. It is also found that reflection was very weak with $R_{\rm f}<0.2$ \citep[e.g.][]{AJ2021}. Modeling with \textsc{relxillcp} required an additional \textsc{blackbody} component for the SE, indicating that ionized, blurred reflection is not the origin of the SE in this source, but rather warm Comptonization. \citet{Tripathi2022} also found a similar result.

\begin{figure}
\centering
\includegraphics[width=1.1\linewidth]{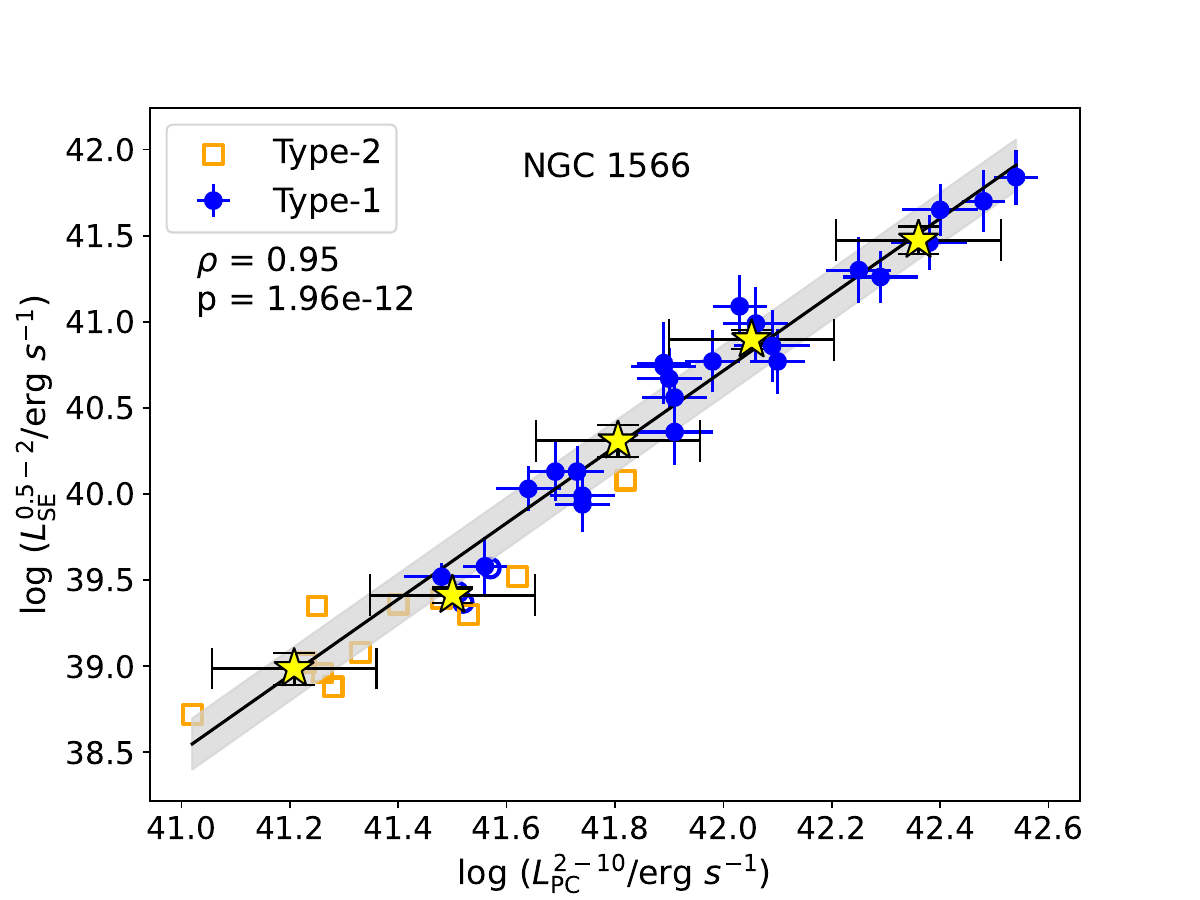}
\caption{The 0.5--2~keV soft-excess luminosity ($L_{\rm SE}^{\rm 0.5-2}$) as a function of 2--10~keV primary continuum luminosity ($L_{\rm PC}^{\rm 2-10}$). The black solid line represents the linear best-fit. The gray region marks the $1\sigma$ scatter.}
\label{fig:n15_lx_se}
\end{figure}

In the current work, we employed phenomenological models; however, our findings are consistent with previous studies that have examined the source in detail using physical models. Figure~\ref{fig:n15_lx_se} displays the relation of the SE with the continuum emission of NGC\,1566. In our study, we found $\log (L_{\rm SE}^{\rm 0.5-2}/10^{42})=(2.21 \pm 0.08)\log (L_{\rm PC}^{\rm 2-10}/10^{42}) - (1.33 \pm 0.09)$, with intrinsic scatter of 0.15\,dex ($1\sigma$). The slope is steeper compared to the entire sample. The tight correlation of the SE and PC is consistent with the warm Comptonization scenario as the origin of the SE emission in NGC\,1566.

\begin{figure}
\centering
\includegraphics[width=0.95\linewidth]{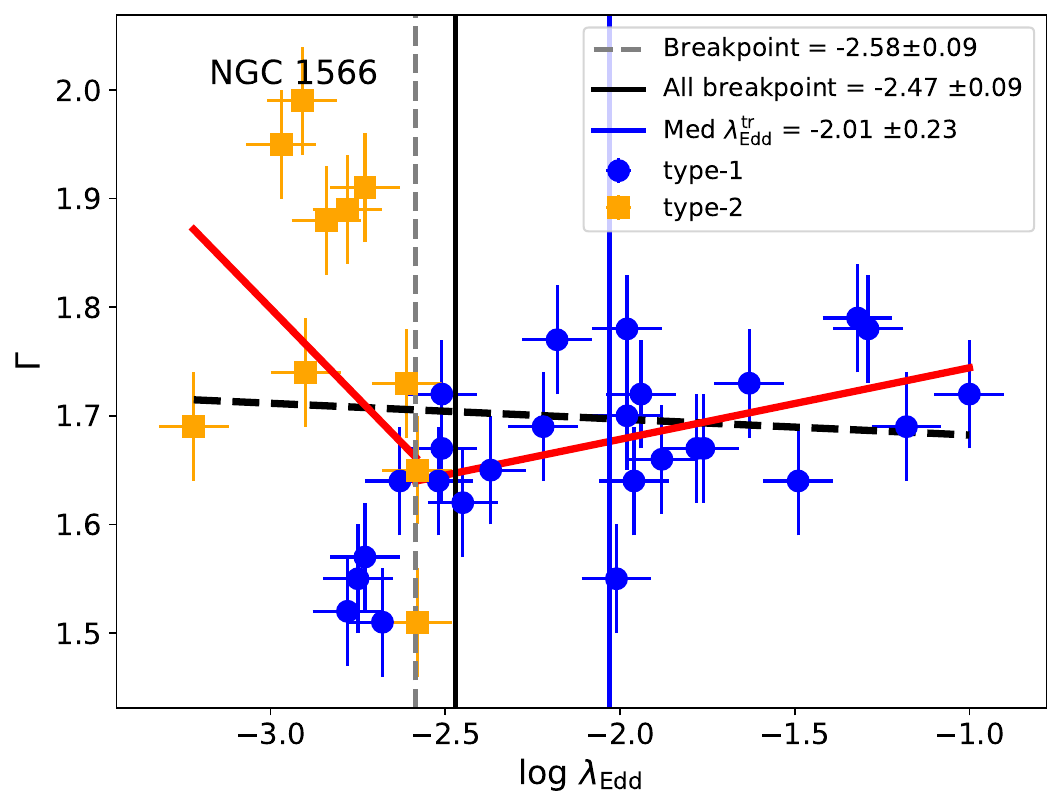}
\caption{Relation between the photon index ($\Gamma$) and the Eddington ratio ($\lambda_{\rm Edd}$). The black dashed line represents the linear best-fit. Two red lines represent the two-linear fit of the data, with a break. The vertical solid blue and solid black lines represent the median of transition Eddington ratio ($\lambda_{\rm Edd}^{\rm tr}$) and breakpoint for the entire sample, respectively. The gray dashed line represents the breakpoint for NGC\,1566.}
\label{fig:n15_ged}
\end{figure}

\begin{figure}
\centering
\includegraphics[width=1.1\linewidth]{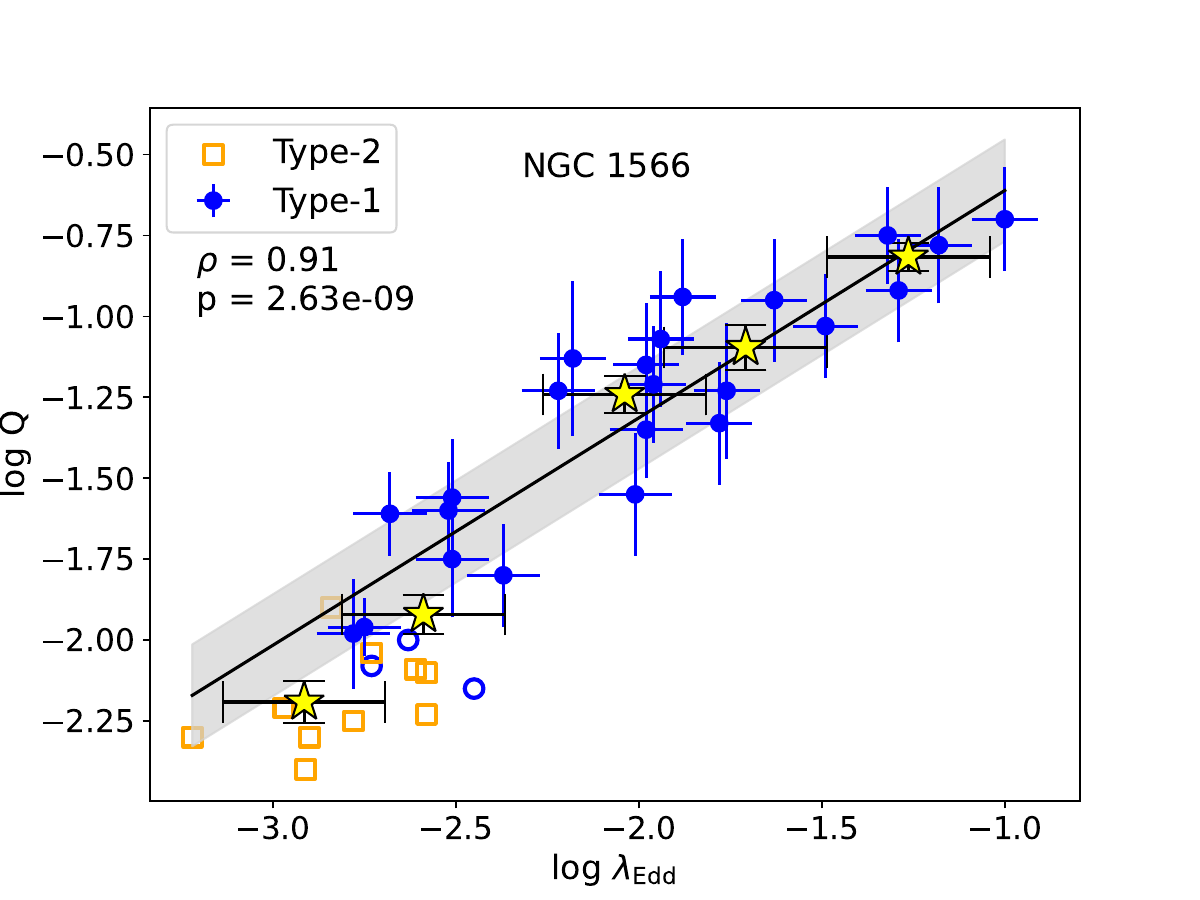}
\caption{Variation of soft-excess strength ($Q$) as a function of Eddington ratio ($\lambda_{\rm Edd}$). The black solid line represents the linear best-fit. The gray region marks the $1\sigma$ scatter.}
\label{fig:n15_qed}
\end{figure}

We show the $\Gamma-\lambda_{\rm Edd}$ relation in Figure~\ref{fig:n15_ged}. We obtained a `V'-shaped relation, with a break at $\log \lambda_{\rm Edd} = -2.58 \pm 0.09$, which is consistent with the entire sample. Figure~\ref{fig:n15_qed} shows the variation of $Q$ as a function of $\lambda_{\rm Edd}$. The positive correlation indicates that the SE strongly deepens on the accretion rate, and SE emission vanishes rapidly at low accretion state, possibly at $\log \lambda_{\rm Edd}<-2.5$. Both $Q-\lambda_{\rm Edd}$ and $\Gamma-\lambda_{\rm Edd}$ relations indicate the change of the accretion geometry at $\lambda_{\rm Edd} \sim 0.003$, below which the warm corona weakens significantly.

\begin{table*}
\centering
\caption{Broadband X-ray properties of NGC\,1566}
\small
\begin{tabular}{lcccccccccc}
\hline \hline
Epoch & $\log L_{\rm PC}^{\rm 2-10}$ & $\log L_{\rm SE}^{\rm 0.5-2}$ & $\Gamma$ & $E_{\rm cut}$ & $\log R_{\rm S}$ & $kT_{\rm BB}$ & $\log Q$ & $\log \lambda_{\rm Edd}$ & $\chi^2/{\rm dof}$ \\
      &   &    &   & (keV) &   &  (eV) &  & \\
(1) & (2) & (3) & (4) & (5) & (6) & (7) & (8) & (9) & (10) \\
\hline
E1	&$	41.06	\pm	0.01	$&$	<39.22			$&$	1.96	\pm	0.02	$&$	179	^{+	145	}_{-	109	}$&$	-0.37	\pm	0.14	$&$	120	^*		$&$	<-2.03			$&$	-2.87	\pm	0.05	$&	538/422	\\
E2	&$	41.19	\pm	0.03	$&$	39.60	\pm	0.08	$&$	1.74	\pm	0.03	$&$	200	^*				$&$	-			$&$	159	\pm	9	$&$	-1.58	\pm	0.09	$&$	-2.74	\pm	0.06	$&	1075/1000	\\
E3	&$	41.77	\pm	0.02	$&$	40.49	\pm	0.08	$&$	1.69	\pm	0.02	$&$	90	^{+	30	}_{-	27	}$&$	-1.22	\pm	0.48	$&$	134	\pm	17	$&$	-1.28	\pm	0.08	$&$	-2.13	\pm	0.05	$&	1570/1646	\\
E4	&$	42.01	\pm	0.01	$&$	40.81	\pm	0.10	$&$	1.69	\pm	0.03	$&$	142	^{+	105	}_{-	78	}$&$	-1.09	\pm	0.30	$&$	109	\pm	12	$&$	-1.20	\pm	0.10	$&$	-1.85	\pm	0.05	$&	715/663	\\
E5	&$	41.82	\pm	0.01	$&$	40.42	\pm	0.19	$&$	1.67	\pm	0.03	$&$	200	^*				$&$	-			$&$	104	\pm	39	$&$	-1.40	\pm	0.19	$&$	-2.08	\pm	0.05	$&	838/833	\\
E6	&$	41.75	\pm	0.01	$&$	40.35	\pm	0.21	$&$	1.70	\pm	0.04	$&$	174	^{+	119	}_{-	77	}$&$	-0.41	\pm	0.21	$&$	117^*$&$	-1.40	\pm	0.21	$&$	-2.15	\pm	0.05	$&	528/549	\\
E7	&$	41.81	\pm	0.01	$&$	40.39	\pm	0.22	$&$	1.69	\pm	0.05	$&$	>114					$&$	-1.00	\pm	0.22	$&$	115^*	$&$	-1.42	\pm	0.22	$&$	-2.09	\pm	0.05	$&	558/567	\\
\hline
\end{tabular}
\label{tab:n15}
\end{table*}

\subsection{NGC\,2617}
\label{subsec:ngc2617}

NGC\,2617 transitioned to type\,1 state from type\,2 state between 2003 and 2013 \citep{Shappee2014}. Since then, the source remained in type\,1 state until 2023. In October 2023, NGC\,2617 lost its broad H$\beta$ line and entered a type\,2 state \citep{Oknyansky2023ATel}. NGC\,2617 showed the presence of soft excess in 2013, when it was in the high flux state \citep{Giustini2017}. Later the SE emission decreased as it moved towards the type\,2 state.

\begin{figure}
\centering
\includegraphics[width=1.1\linewidth]{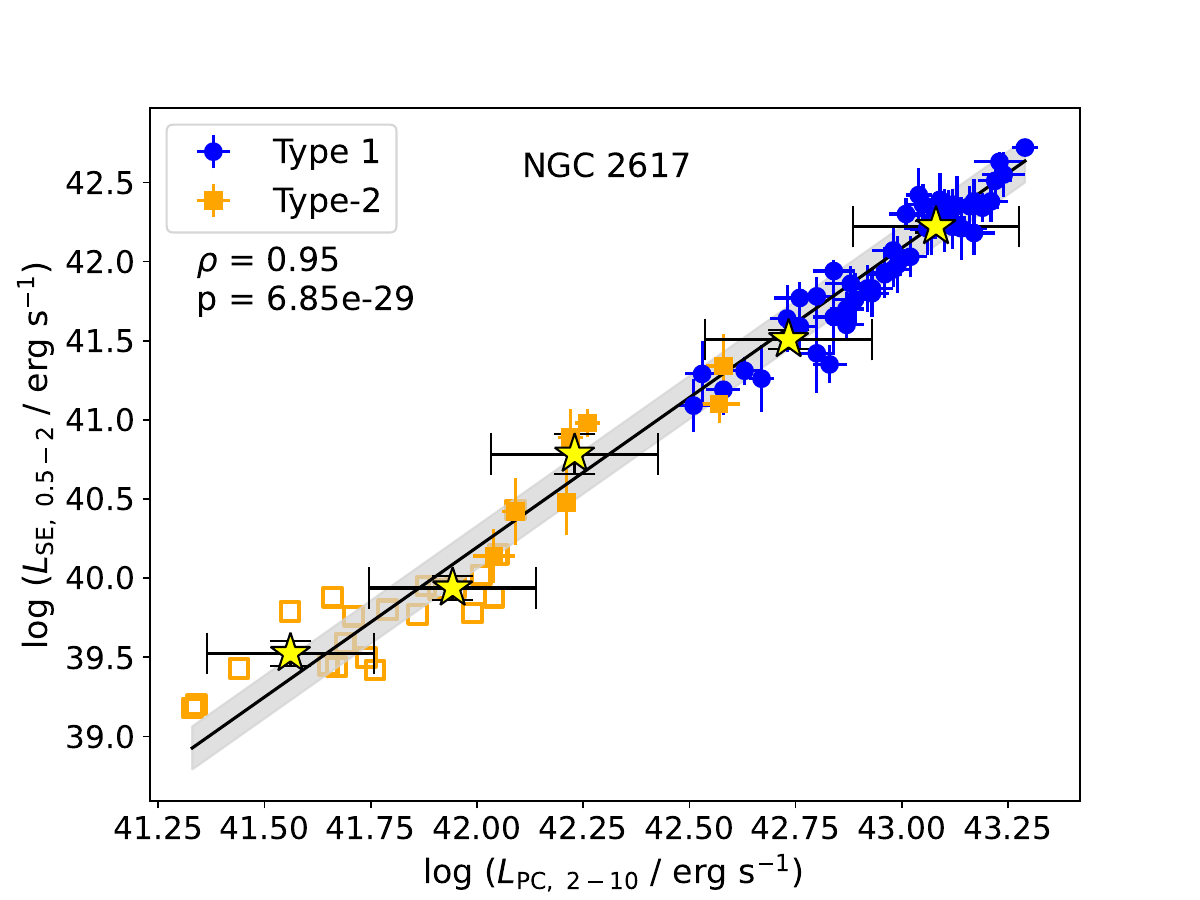}
\caption{The 0.5--2~keV soft-excess luminosity ($L_{\rm SE}^{\rm 0.5-2}$) as a function of 2--10~keV primary continuum luminosity ($L_{\rm PC}^{\rm 2-10}$). The black solid line represents the linear best-fit. The gray region marks the $1\sigma$ scatter.}
\label{fig:n26_lx_se}
\end{figure}

NGC\,2617 showed a strong positive correlation between the SE and continuum emission, with a Spearman correlation index of 0.95 with $p \ll 10^{-10}$. Figure~\ref{fig:n26_lx_se} shows the variation of the SE emission as a function of continuum emission. The linear fit between the two quantities revealed, 
$\log (L_{\rm SE}^{\rm 0.5-2}/10^{42})=(1.89 \pm 0.07) \log (L_{\rm PC}^{\rm 2-10}/10^{42})-(1.89\pm0.07)$. The observed slope is slightly steeper than that of the whole sample.

\begin{figure}
\centering
\includegraphics[width=0.95\linewidth]{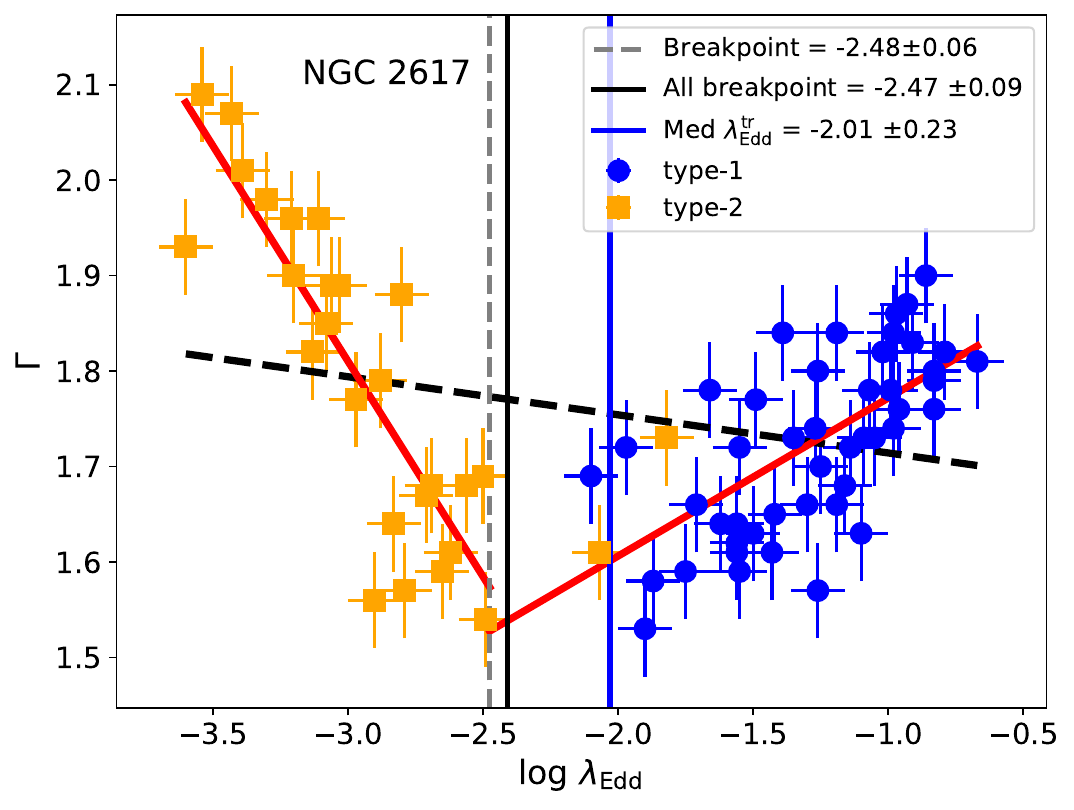}
\caption{Relation between the photon index ($\Gamma$) and the Eddington ratio ($\lambda_{\rm Edd}$). The black dashed line represents the linear best-fit. Two red lines represent the two-linear fit of the data, with a break. The vertical solid blue and solid black lines represent the median of transition Eddington ratio ($\lambda_{\rm Edd}^{\rm tr}$) and breakpoint for the entire sample, respectively. The gray dashed line represents the breakpoint for NGC\,2617.}
\label{fig:n26_ged}
\end{figure}

\begin{figure}
\centering
\includegraphics[width=1.1\linewidth]{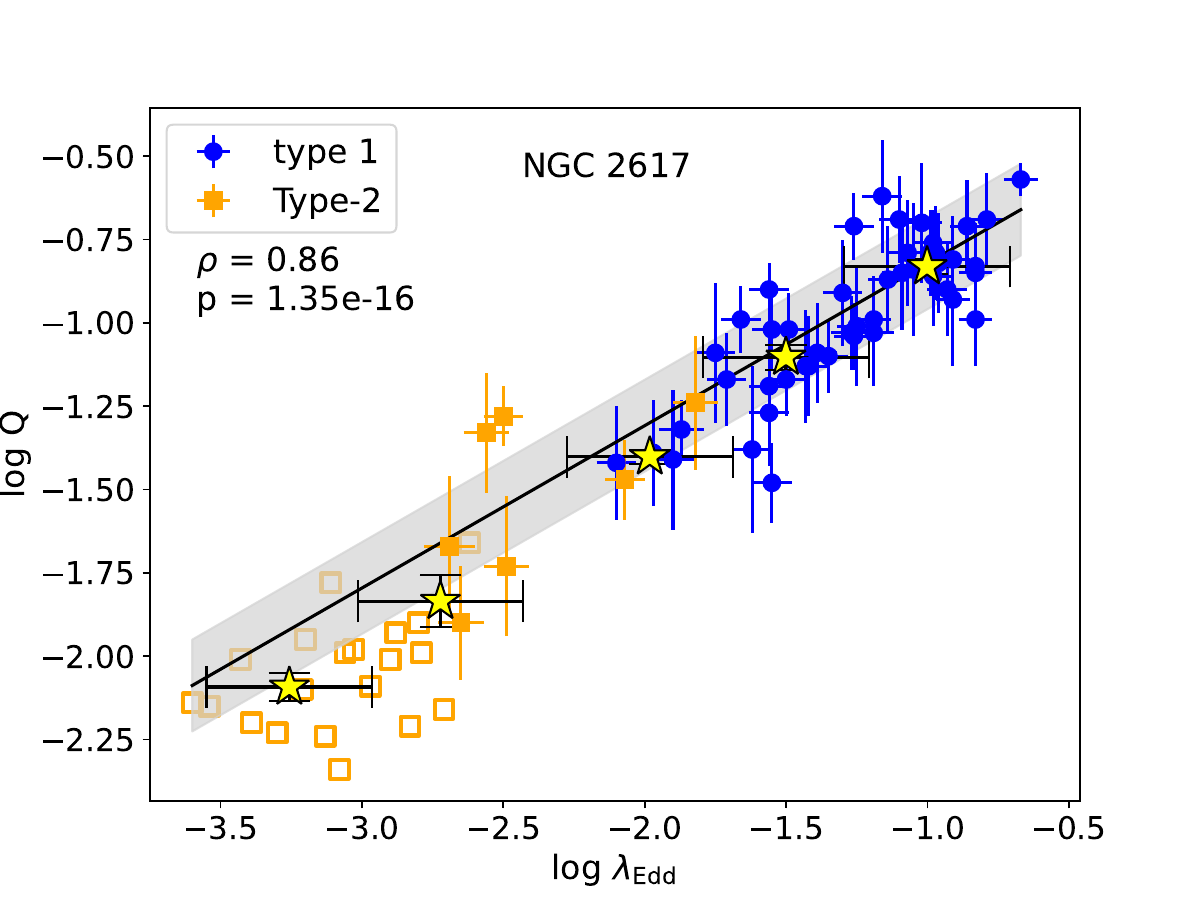}
\caption{Variation of soft-excess strength ($Q$) as a function of Eddington ratio ($\lambda_{\rm Edd}$). The black solid line represents the linear best-fit. The gray region marks the $1\sigma$ scatter.}
\label{fig:n26_qed}
\end{figure}

In Figure~\ref{fig:n26_ged}, we show the relation between the $\Gamma$ and $\lambda_{\rm Edd}$. The $\Gamma-\lambda_{\rm Edd}$ relation shows a sharp break at 
$\log \lambda_{\rm Edd}=-2.45\pm0.06$, indicating a change of geometry in the inner accretion flow. Figure~\ref{fig:n26_qed} displays the relation between the 
$Q$ and $\lambda_{\rm Edd}$, which shows strong positive correlations, with $\rho = 0.87$ and $ p \ll 10^{-10}$. The observed 
correlations among different parameters are consistent with the warm Comptonization scenario as the origin of the SE emission. The findings are also consistent with the entire sample.

We found that at $\log \lambda_{\rm Edd} \sim -2.5$, the SE emission diminishes very rapidly with decreasing accretion rate in NGC\,2617.
At the same $\lambda_{\rm Edd}$, the accretion geometry is also found to change. This suggests that the warm corona is closely linked with 
the inner accretion flow, possibly related to the inner accretion disk. Once the inner accretion disk receded, the warm corona also became weak.

\begin{table*}
\centering
\caption{Broadband X-ray properties of NGC\,2617}
\small
\begin{tabular}{lcccccccccc}
\hline \hline
Epoch & $\log L_{\rm PC}^{\rm 2-10}$ & $\log L_{\rm SE}^{\rm 0.5-2}$ & $\Gamma$ & $E_{\rm cut}$ & $\log R_{\rm S}$ & $kT_{\rm BB}$ & $\log Q$ & $\log \lambda_{\rm Edd}$ & $\chi^2/{\rm dof}$ \\
      &   &    &   & (keV) &   &  (eV) &  & \\
(1) & (2) & (3) & (4) & (5) & (6) & (7) & (8) & (9) & (10) \\
\hline
E1	&$	42.97	\pm	0.03	$&$	41.89	\pm	0.07	$&$	1.78	\pm	0.03	$&$	200	^*				$&$	-			$&$	125	\pm	8	$&$	-1.08	\pm	0.08	$&$	-1.21	\pm	0.09	$&	1596/1439	\\
E2	&$	43.21	\pm	0.03	$&$	42.19	\pm	0.11	$&$	1.96	\pm	0.03	$&$	200	^*				$&$	-			$&$	99	\pm	14	$&$	-1.02	\pm	0.11	$&$	-0.91	\pm	0.09	$&	1435/1314	\\
E3	&$	42.00	\pm	0.21	$&$	<	41.11		$&$	2.09	\pm	0.14	$&$	200	^*				$&$	-			$&$	120	^*		$&$	<	-0.89		$&$	-2.31	\pm	0.22	$&	59/63	\\
\hline
\end{tabular}
\label{tab:n26}
\end{table*}

\subsection{Mrk\,590}
\label{subsec:mrk590}

Mrk\,590 showed several CS transitions in the past four decades \citep[e.g.,][]{Denney2014,AJ2025cl}. The source entered a type\,2 state in 2006 \citep{Denney2014}, where it remained until 2014. After that, the flux slowly increased, and the source entered a type\,1 state in 2017 \citep{Oh2022}. The broadband X-ray properties of Mrk\,590 have been studied extensively in the past \citep[e.g.,][]{Rivers2012,Ghosh2022,Lawther2023}. Mrk\,590 showed a variable SE emission, where the SE generally weakens when the primary flux decreases \citep[e.g.,][]{Ghosh2022}. \cite{Rivers2012} found that the SE emission decreased by a factor of $\sim 20-30$, while the PC emission only changed about $\sim 10\%$, suggesting the SE emission is independent of the PC emission.
\citet{Ghosh2022} found that SE emission could be explained with both ionized reflection and warm Comptonization scenarios.
\cite{Palit2025} reported the emergence of the SE emission when the source transitioned to a type\,1 state, when the UV flux also increased. The authors suggested that it is linked with the warm corona, which forms or becomes strong with increasing accretion rate.

\begin{figure}
\centering
\includegraphics[width=1.1\linewidth]{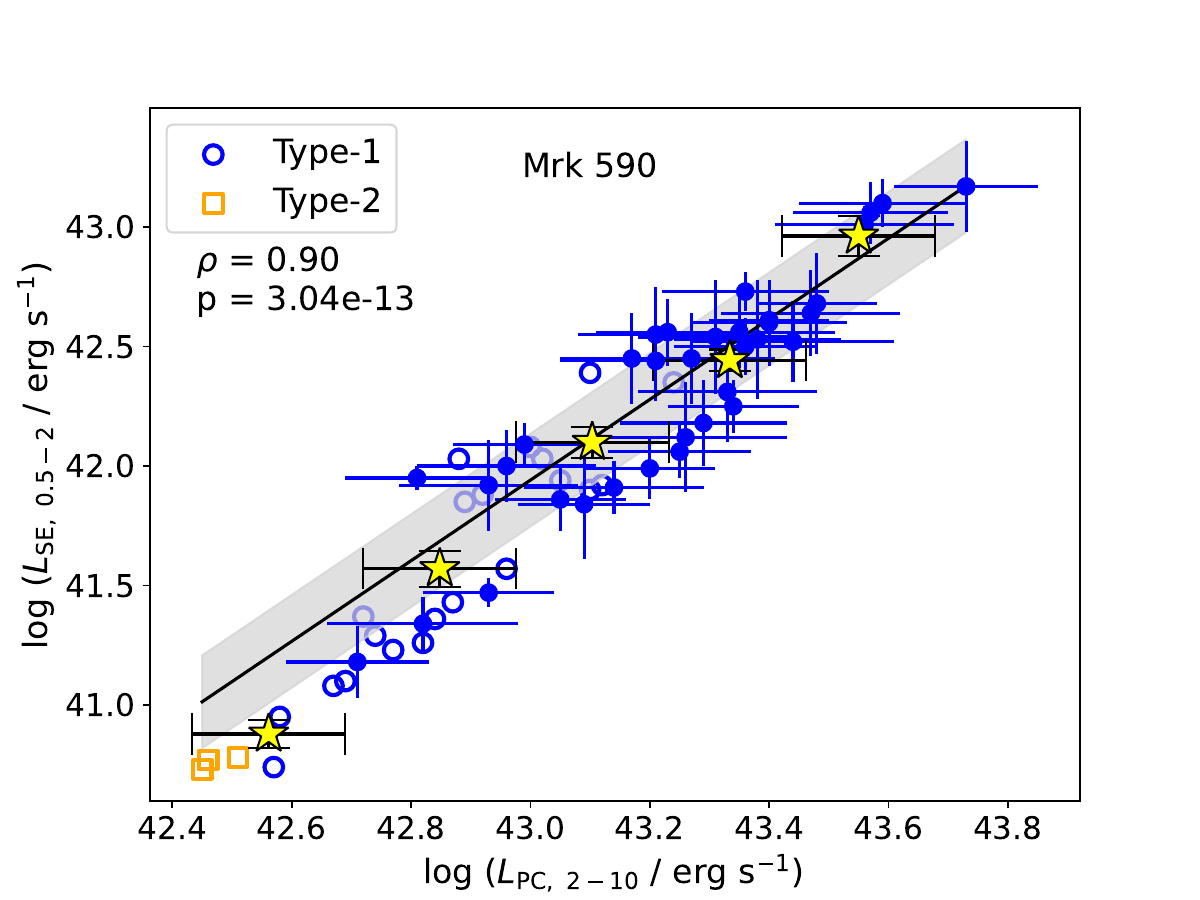}
\caption{The 0.5--2~keV soft-excess luminosity ($L_{\rm SE}^{\rm 0.5-2}$) as a function of 2--10~keV primary continuum luminosity ($L_{\rm PC}^{\rm 2-10}$). The black solid line represents the linear best-fit. The gray region marks the $1\sigma$ scatter.}
\label{fig:m59_lx_se}
\end{figure}

Figure~\ref{fig:m59_lx_se} displays the relation between the SE emission and the continuum emission. The SE emission has a strong positive correlation with the continuum emission with $\rho = 0.90$ and $ p\ll 10^{-10}$. With linear fitting, we obtained $\log (L_{\rm SE}^{\rm 0.5-2}/10^{42})=(1.69\pm0.15)\log (L_{\rm PC}^{\rm 2-10}/10^{42})- (1.88\pm0.18)$, with an intrinsic scatter of 0.19\,dex ($1\sigma$).

\begin{figure}
\centering
\includegraphics[width=0.95\linewidth]{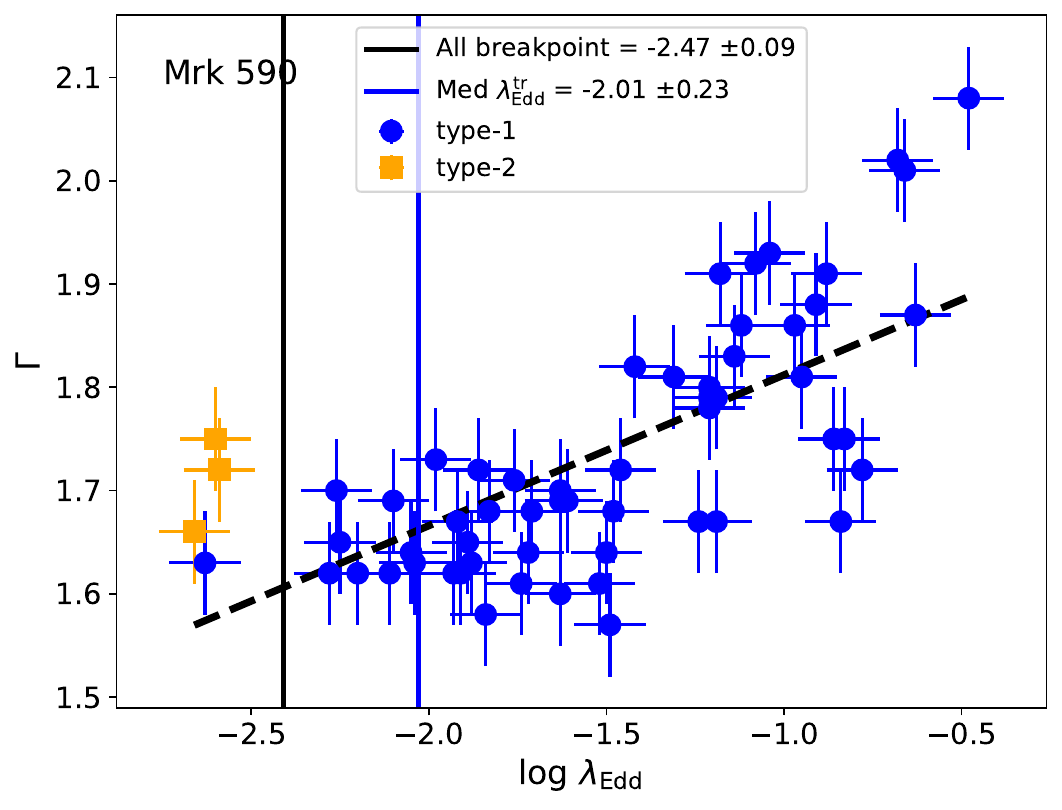}
\caption{Relation between the photon index ($\Gamma$) and the Eddington ratio ($\lambda_{\rm Edd}$). The black dashed line represents the linear best-fit. The vertical solid blue and solid black lines represent the median of transition Eddington ratio ($\lambda_{\rm Edd}^{\rm tr}$) and breakpoint for the entire sample, respectively.}
\label{fig:m59_ged}
\end{figure}

\begin{figure}
\centering
\includegraphics[width=1.1\linewidth]{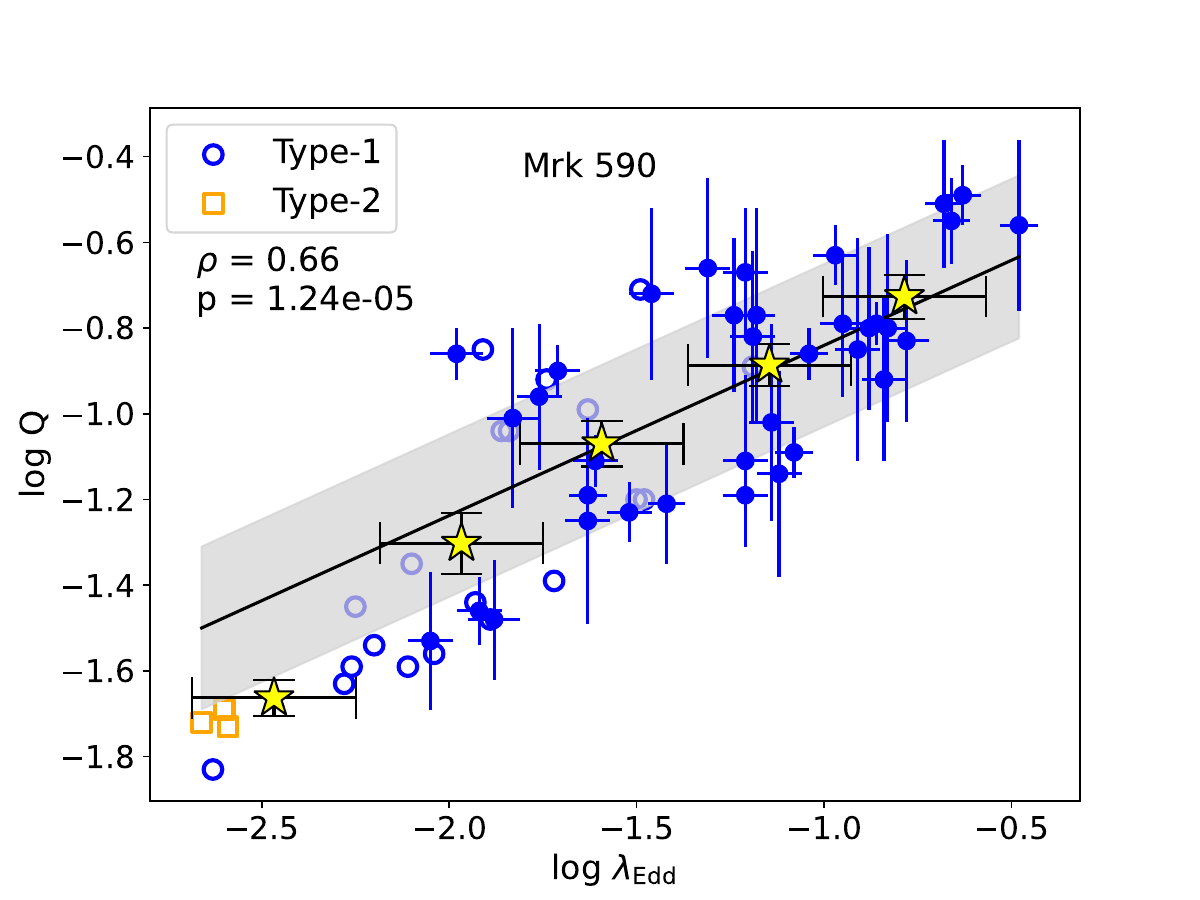}
\caption{Variation of soft-excess strength ($Q$) as a function of Eddington ratio ($\lambda_{\rm Edd}$). The black solid line represents the linear best-fit. The gray region marks the $1\sigma$ scatter.}
\label{fig:m59_qed}
\end{figure}

We show the relation between the $\Gamma$ and $\lambda_{\rm Edd}$ in Figure~\ref{fig:m59_ged}. We obtained a positive correlation between these two parameters. Figure~\ref{fig:m59_qed} shows the variation of $Q$ as a function of $\lambda_{\rm Edd}$. We obtained a positive correlation with $\rho = 0.65$ and $p < 10^{-5}$. Broadband spectral analysis revealed the reflection is very weak or absent, suggesting the reflection did not contribute to the SE emission.
The correlations we obtained among different parameters indicate that the origin of the SE emission is most probably the warm corona.

In Mrk\,590, we did not observe any change of geometry in the inner accretion flow, although there is a hint of change in slope at $\log \lambda_{\rm Edd} < -1.5$. We also obtained only an upper limit of the SE flux below $\log \lambda_{\rm Edd} \sim -2$, indicating the warm corona substantially weakens below this.

\begin{table*}
\centering
\caption{Broadband X-ray properties of Mrk\,590}
\small
\begin{tabular}{lcccccccccc}
\hline \hline
Epoch & $\log L_{\rm PC}^{\rm 2-10}$ & $\log L_{\rm SE}^{\rm 0.5-2}$ & $\Gamma$ & $E_{\rm cut}$ & $\log R_{\rm S}$ & $kT_{\rm BB}$ & $\log Q$ & $\log \lambda_{\rm Edd}$ & $\chi^2/{\rm dof}$ \\
      &   &    &   & (keV) &   &  (eV) &  & \\
(1) & (2) & (3) & (4) & (5) & (6) & (7) & (8) & (9) & (10) \\
\hline
E1	&$	42.80	\pm	0.02	$&$	41.13	\pm	0.11	$&$	1.62	\pm	0.04	$&$	200	^*				$&$	-			$&$	157	\pm	29	$&$	-1.67	\pm	0.11	$&$	-1.71	\pm	0.05	$&	157/169	\\
E2	&$	42.89	\pm	0.01	$&$	41.17	\pm	0.12	$&$	1.69	\pm	0.04	$&$	200	^*				$&$	-			$&$	129	\pm	13	$&$	-1.72	\pm	0.12	$&$	-1.61	\pm	0.05	$&	679/666	\\
E3	&$	42.96	\pm	0.01	$&$	<41.34			$&$	1.67	\pm	0.04	$&$	>65					$&$	-0.82	\pm	0.10	$&$	120	^*		$&$	<-1.62			$&$	-1.52	\pm	0.05	$&	831/754	\\
E4	&$	42.94	\pm	0.01	$&$	<41.36			$&$	1.67	\pm	0.03	$&$	>58					$&$	-0.89	\pm	0.12	$&$	120	^*		$&$	<-1.58			$&$	-1.55	\pm	0.05	$&	480/445	\\
E5	&$	42.53	\pm	0.02	$&$	<40.56			$&$	1.65	\pm	0.03	$&$	>75					$&$	-0.70	\pm	0.08	$&$	120	^*		$&$	<-1.97			$&$	-2.01	\pm	0.05	$&	435/465	\\
E6	&$	42.61	\pm	0.02	$&$	<40.66			$&$	1.72	\pm	0.04	$&$	>63					$&$	-0.57	\pm	0.07	$&$	120	^*		$&$	<-1.95			$&$	-1.92	\pm	0.05	$&	385/447	\\
E7	&$	43.12	\pm	0.01	$&$	<41.49			$&$	1.73	\pm	0.04	$&$	72	^{+	39	}_{-	24	}$&$	-2.00	\pm	0.60	$&$	120	^*		$&$	<-1.63			$&$	-1.34	\pm	0.05	$&	569/684	\\
E8	&$	43.33	\pm	0.02	$&$	41.80	\pm	0.19	$&$	1.82	\pm	0.02	$&$	60	^{+	25	}_{-	17	}$&$	-0.15	\pm	0.10	$&$	152	\pm	58	$&$	-1.53	\pm	0.19	$&$	-1.08	\pm	0.05	$&	1271/1287	\\
E9	&$	43.31	\pm	0.01	$&$	41.72	\pm	0.21	$&$	1.83	\pm	0.04	$&$	55	^{+	24	}_{-	14	}$&$	-1.13	\pm	0.15	$&$	124	\pm	36	$&$	-1.59	\pm	0.21	$&$	-1.10	\pm	0.05	$&	1160/1211	\\
E10	&$	42.93	\pm	0.01	$&$	41.33	\pm	0.25	$&$	1.75	\pm	0.03	$&$	200	^*				$&$	-			$&$	166	\pm	30	$&$	-1.60	\pm	0.25	$&$	-1.56	\pm	0.05	$&	630/628	\\
E11	&$	43.03	\pm	0.01	$&$	41.45	\pm	0.21	$&$	1.64	\pm	0.02	$&$	200	^*				$&$	-			$&$	153	\pm	18	$&$	-1.58	\pm	0.21	$&$	-1.44	\pm	0.05	$&	659/678	\\
E12	&$	43.01	\pm	0.01	$&$	<41.14			$&$	1.62	\pm	0.02	$&$	66	^{+	20	}_{-	16	}$&$	-0.80	\pm	0.10	$&$	120	^*		$&$	<-1.87			$&$	-1.47	\pm	0.05	$&	947/1001	\\
E13	&$	43.32	\pm	0.01	$&$	<41.68			$&$	1.91	\pm	0.02	$&$	59	^{+	21	}_{-	11	}$&$	-0.14	\pm	0.11	$&$	120	^*		$&$	<-1.64			$&$	-1.09	\pm	0.05	$&	694/761	\\
E14	&$	42.91	\pm	0.02	$&$	<41.33			$&$	1.73	\pm	0.04	$&$	>54					$&$	-0.64	\pm	0.14	$&$	120	^*		$&$	<-1.58			$&$	-1.58	\pm	0.05	$&	894/947	\\
E15	&$	42.48	\pm	0.02	$&$	<40.97			$&$	1.76	\pm	0.03	$&$	200	^*				$&$	-			$&$	120	^*		$&$	<-1.51			$&$	-2.07	\pm	0.05	$&	348/342	\\
E16	&$	42.91	\pm	0.02	$&$	41.25	\pm	0.19	$&$	1.68	\pm	0.05	$&$	200	^*				$&$	-			$&$	154	\pm	18	$&$	-1.66	\pm	0.19	$&$	-1.58	\pm	0.05	$&	580/593	\\
E17	&$	42.85	\pm	0.02	$&$	41.11	\pm	0.17	$&$	1.61	\pm	0.02	$&$	>88					$&$	-1.06	\pm	0.13	$&$	152	\pm	12	$&$	-1.74	\pm	0.17	$&$	-1.65	\pm	0.05	$&	1104/1141	\\
\hline
\end{tabular}
\label{tab:m59}
\end{table*}

\subsection{Mrk\,1018}
\label{subsec:mrk1018}

Mrk\,1018 is one of the first AGNs that showed CS transitions \citep{Cohen1986}. The source transitioned to type\,1 state in 1984, where it remained until 2008 \citep{McElroy2016}. The source moved to type\,1.9 state in January 2015 \citep{McElroy2016}. In 2020, the source showed a full CS transition when it briefly entered a type\,1 state and moved back to type\,1.8 state \citep{Lu2025}.

The broadband X-ray properties of Mrk\,1018 have been studied extensively in the past \citep[e.g.,][]{Noda2018,Lyu2021,Veronese2024}, which agrees with our findings. We find that the reflection is very weak or absent in the broadband X-ray spectroscopy. This agrees with the work of \cite{Veronese2024}, who found that no reflection component is needed to model the hard X-ray spectra. We also found evolving SE emission, which is also reported in the previous studies \citep[e.g.,][]{Noda2018}. These suggest that warm coronal emission contributes to the SE emission in Mrk\,1018.

\begin{figure}
\centering
\includegraphics[width=1.1\linewidth]{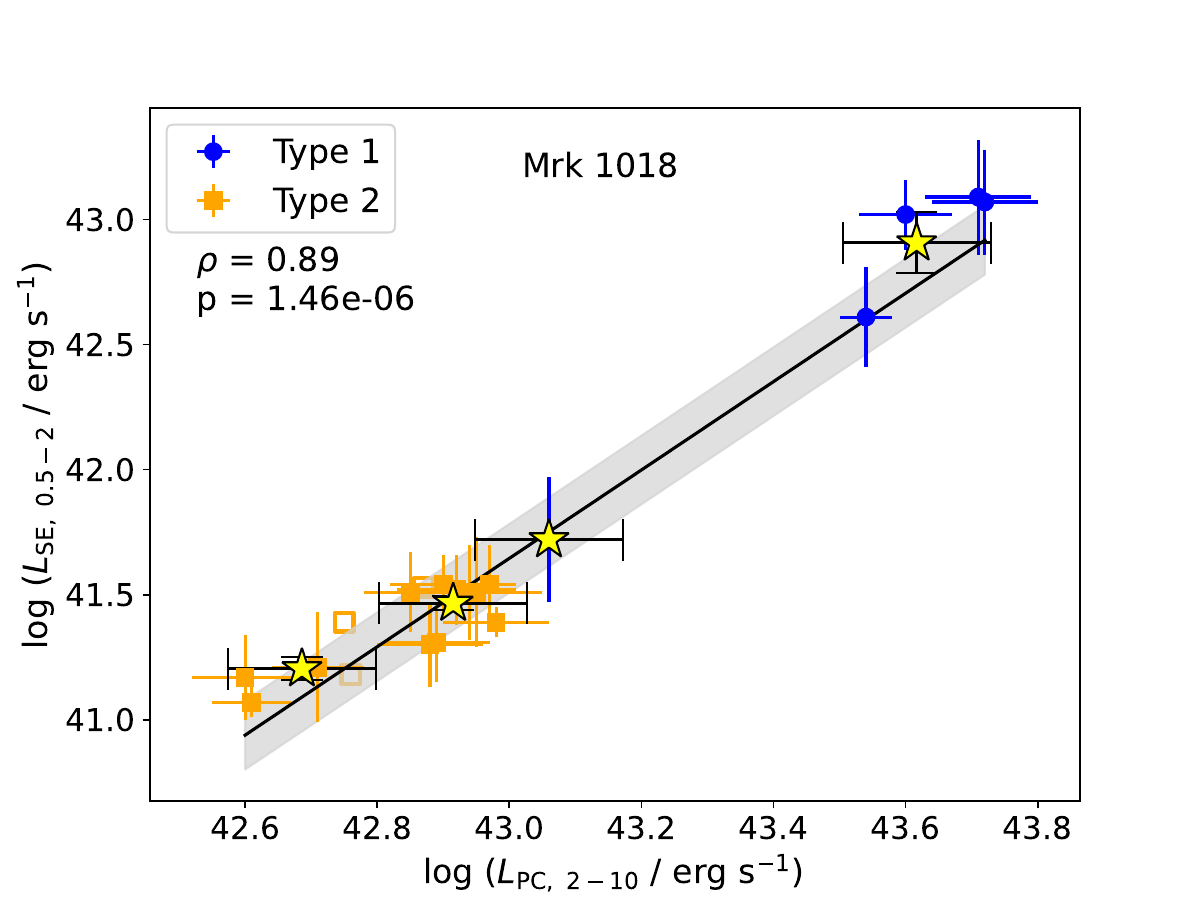}
\caption{The 0.5--2~keV soft-excess luminosity ($L_{\rm SE}^{\rm 0.5-2}$) as a function of 2--10~keV primary continuum luminosity ($L_{\rm PC}^{\rm 2-10}$). The black solid line represents the linear best-fit. The gray region marks the $1\sigma$ scatter.}
\label{fig:m10_lx_se}
\end{figure}

\begin{figure}
\centering
\includegraphics[width=0.95\linewidth]{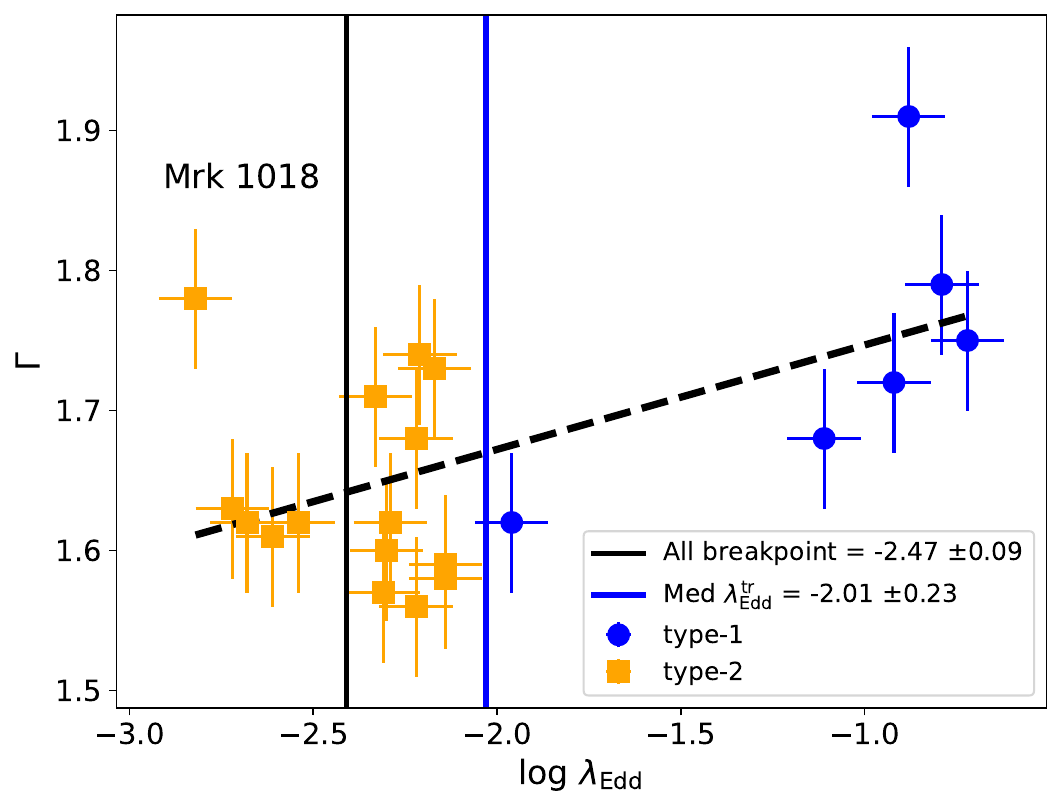}
\caption{Relation between the photon index ($\Gamma$) and the Eddington ratio ($\lambda_{\rm Edd}$). The black dashed line represents the linear best-fit. The vertical solid blue and solid black lines represent the median of transition Eddington ratio ($\lambda_{\rm Edd}^{\rm tr}$) and breakpoint for the entire sample, respectively.}
\label{fig:m10_ged}
\end{figure}

\begin{figure}
\centering
\includegraphics[width=1.1\linewidth]{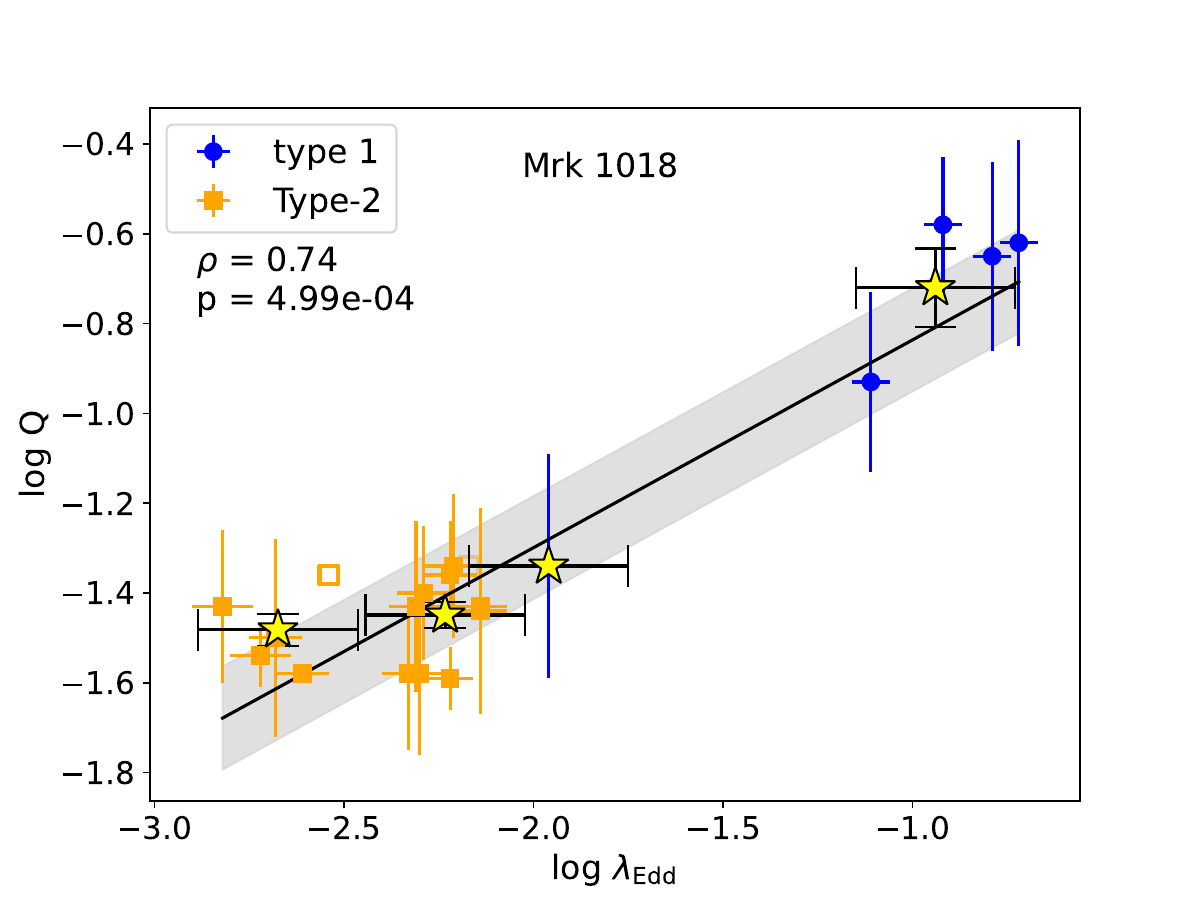}
\caption{Variation of soft-excess strength ($Q$) as a function of Eddington ratio ($\lambda_{\rm Edd}$). The black solid line represents the linear best-fit. The gray region marks the $1\sigma$ scatter.}
\label{fig:m10_qed}
\end{figure}

Figure~\ref{fig:m10_lx_se} displays the relation between the SE and continuum emission of Mrk\,1018. Strong positive relation ($\rho = 0.89$ and $ p < 10^{-5}$.) suggests a common physical origin of the SE and continuum emission. In Figure~\ref{fig:m10_ged}, we show the relation between $\Gamma$ and $\lambda_{\rm Edd}$. We obtained a positive correlation between these two parameters, with a hint of break around $\log \lambda_{\rm Edd}\sim -2.5$. Figure~\ref{fig:m10_qed} shows the variation of $Q$ with $\lambda_{\rm Edd}$. We obtained $\rho = 0.71$ and $p < 10^{-5}$, indicating $Q$ strongly linked with the accretion rate.

In Mrk\,1018, we could only probe up to $\log \lambda_{\rm Edd} \sim -2.7$. Low SE emission indicates that the warm corona becomes weak at this $\lambda_{\rm Edd}$, and still exists. We expect the warm corona would vanish or diminish at lower $\lambda_{\rm Edd}$. The observed relations among different parameters are consistent with the warm corona scenario of the SE.

\begin{table*}
\centering
\caption{Broadband X-ray properties of Mrk\,1018}
\small
\begin{tabular}{lcccccccccc}
\hline \hline
Epoch & $\log L_{\rm PC}^{\rm 2-10}$ & $\log L_{\rm SE}^{\rm 0.5-2}$ & $\Gamma$ & $E_{\rm cut}$ & $\log R_{\rm S}$ & $kT_{\rm BB}$ & $\log Q$ & $\log \lambda_{\rm Edd}$ & $\chi^2/{\rm dof}$ \\
      &   &    &   & (keV) &   &  (eV) &  & \\
(1) & (2) & (3) & (4) & (5) & (6) & (7) & (8) & (9) & (10) \\
\hline
E1	&$	43.57	\pm	0.03	$&$	42.61	\pm	0.16	$&$	1.84	\pm	0.04	$&$	200	^*				$&$	-			$&$	108	\pm	27	$&$	-0.96	\pm	0.16	$&$	-1.08	\pm	0.06	$&	194/217	\\
E2	&$	43.67	\pm	0.02	$&$	42.70	\pm	0.14	$&$	1.87	\pm	0.03	$&$	200	^*				$&$	-			$&$	127	\pm	10	$&$	-0.97	\pm	0.14	$&$	-0.95	\pm	0.05	$&	684/684	\\
E3	&$	43.51	\pm	0.02	$&$	42.55	\pm	0.13	$&$	1.77	\pm	0.04	$&$	>77					$&$	-			$&$	120	^*		$&$	-0.96	\pm	0.13	$&$	-1.15	\pm	0.05	$&	667/589	\\
E4	&$	42.72	\pm	0.03	$&$	<40.89			$&$	1.65	\pm	0.05	$&$	>98					$&$	-0.76	\pm	0.15	$&$	120	^*		$&$	<-1.83			$&$	-2.07	\pm	0.06	$&	165/219	\\
E5	&$	42.83	\pm	0.03	$&$	<41.35			$&$	1.62	\pm	0.04	$&$	>105					$&$	-1.11	\pm	0.18	$&$	120	^*		$&$	<-1.48			$&$	-1.95	\pm	0.06	$&	208/219	\\
E6	&$	42.75	\pm	0.02	$&$	-			$&$	1.69	\pm	0.04	$&$	>91					$&$	-1.10	\pm	0.14	$&$	-			$&$			-	$&$	-2.03	\pm	0.05	$&	166/161	\\
E7	&$	42.80	\pm	0.03	$&$	<41.52			$&$	1.63	\pm	0.05	$&$	>108					$&$	-1.07	\pm	0.13	$&$	120	^*		$&$	<-1.28			$&$	-1.98	\pm	0.06	$&	213/274	\\
E8	&$	42.91	\pm	0.02	$&$	41.51	\pm	0.08	$&$	1.64	\pm	0.03	$&$	200	^*				$&$	-			$&$	127	\pm	36	$&$	-1.40	\pm	0.08	$&$	-1.86	\pm	0.05	$&	664/637	\\
E9	&$	42.57	\pm	0.02	$&$	41.29	\pm	0.06	$&$	1.70	\pm	0.03	$&$	200	^*				$&$	-			$&$	125	\pm	35	$&$	-1.28	\pm	0.06	$&$	-2.23	\pm	0.05	$&	393/409	\\
E10	&$	42.57	\pm	0.02	$&$	41.25	\pm	0.05	$&$	1.69	\pm	0.03	$&$	200	^*				$&$	-			$&$	129	\pm	37	$&$	-1.32	\pm	0.05	$&$	-2.23	\pm	0.05	$&	416/404	\\
E11	&$	42.71	\pm	0.02	$&$	<41.62			$&$	1.62	\pm	0.04	$&$	>122					$&$	-2.08	\pm	0.34	$&$	120	^*		$&$	<-1.09			$&$	-2.08	\pm	0.05	$&	80/117	\\
\hline
\end{tabular}
\label{tab:m10}
\end{table*}

\subsection{IRAS\,23226--3843}
\label{subsec:iras23226}

\begin{figure}
\centering
\includegraphics[width=1.1\linewidth]{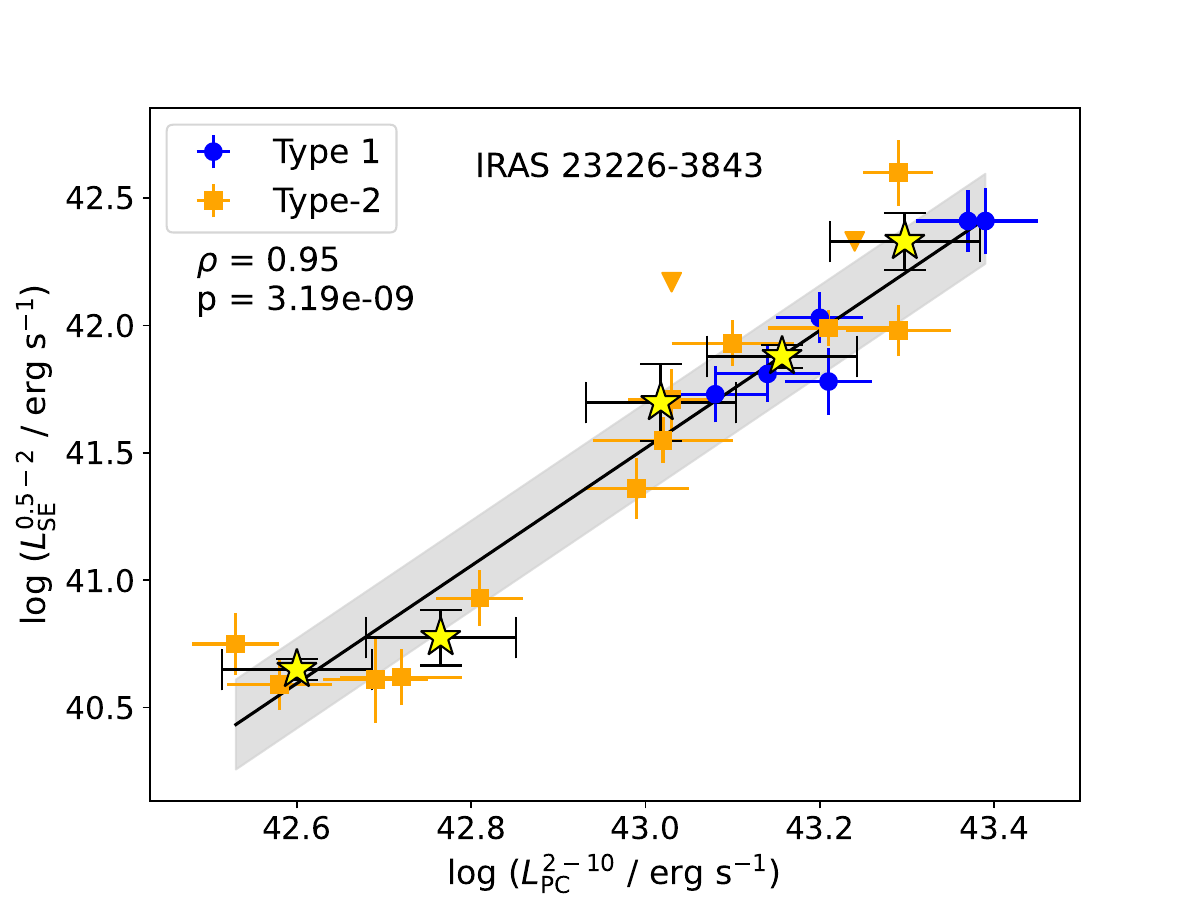}
\caption{The 0.5--2~keV soft-excess luminosity ($L_{\rm SE}^{\rm 0.5-2}$) as a function of 2--10~keV primary continuum luminosity ($L_{\rm PC}^{\rm 2-10}$). The black solid line represents the linear best-fit. The gray region marks the $1\sigma$ scatter.}
\label{fig:i22_lx_se}
\end{figure}

IRAS\,23226--3843 showed several CS transitions in the last decade \citep[e.g.,][]{Kollatschny2020,Kollatschny2023,AJ2025cl}.
We detected SE emission in the source, which evolves with the accretion rate. Figure~\ref{fig:i22_lx_se} shows the relation of the SE and continuum emission, which shows a strong positive correlation with $\rho = 0.95$ and $ p < 10^{-5}$. We find the presence of moderate reflection in the source; however, the strength of the Compton hump (reflection strength $R_{S}$) does not correlate with the $Q$. This suggests that the SE originates from a warm Corona.

\begin{figure}
\centering
\includegraphics[width=0.95\linewidth]{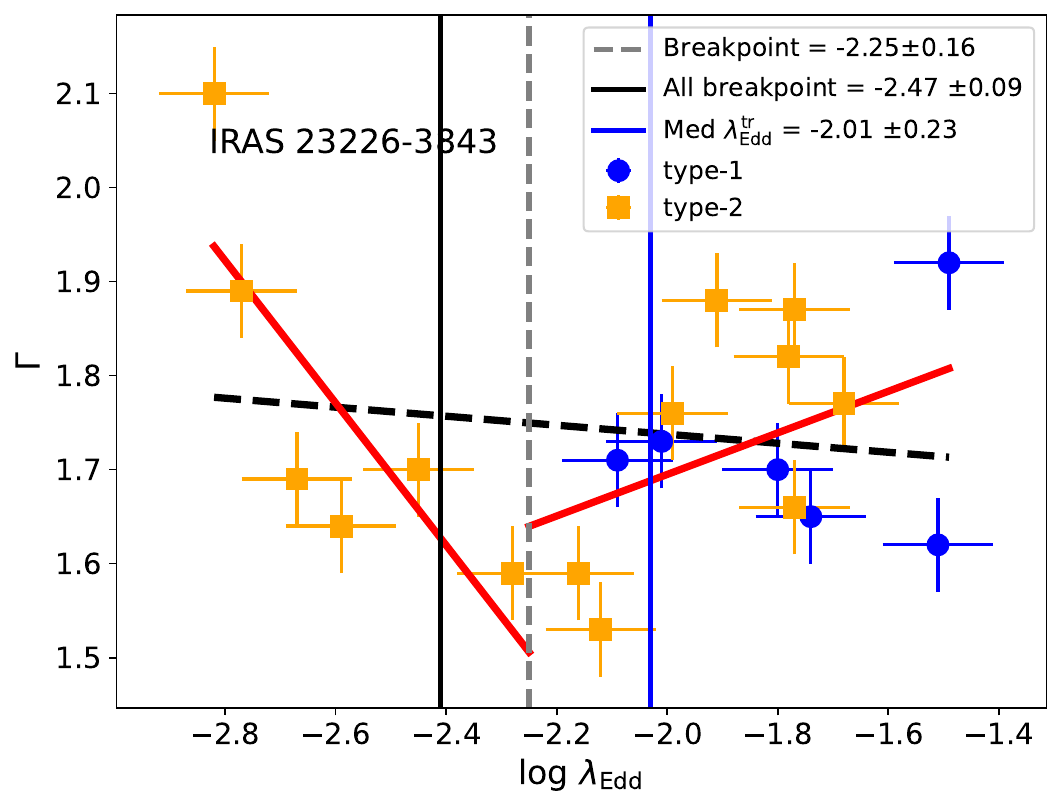}
\caption{Relation between the photon index ($\Gamma$) and the Eddington ratio ($\lambda_{\rm Edd}$). The black dashed line represents the linear best-fit. Two red lines represent the two-linear fit of the data, with a break. The vertical solid blue and solid black lines represent the median of transition Eddington ratio ($\lambda_{\rm Edd}^{\rm tr}$) and breakpoint for the entire sample, respectively. The gray dashed line represents the breakpoint for IRAS\,23226--3843.}
\label{fig:i22_ged}
\end{figure}

\begin{figure}
\centering
\includegraphics[width=1.1\linewidth]{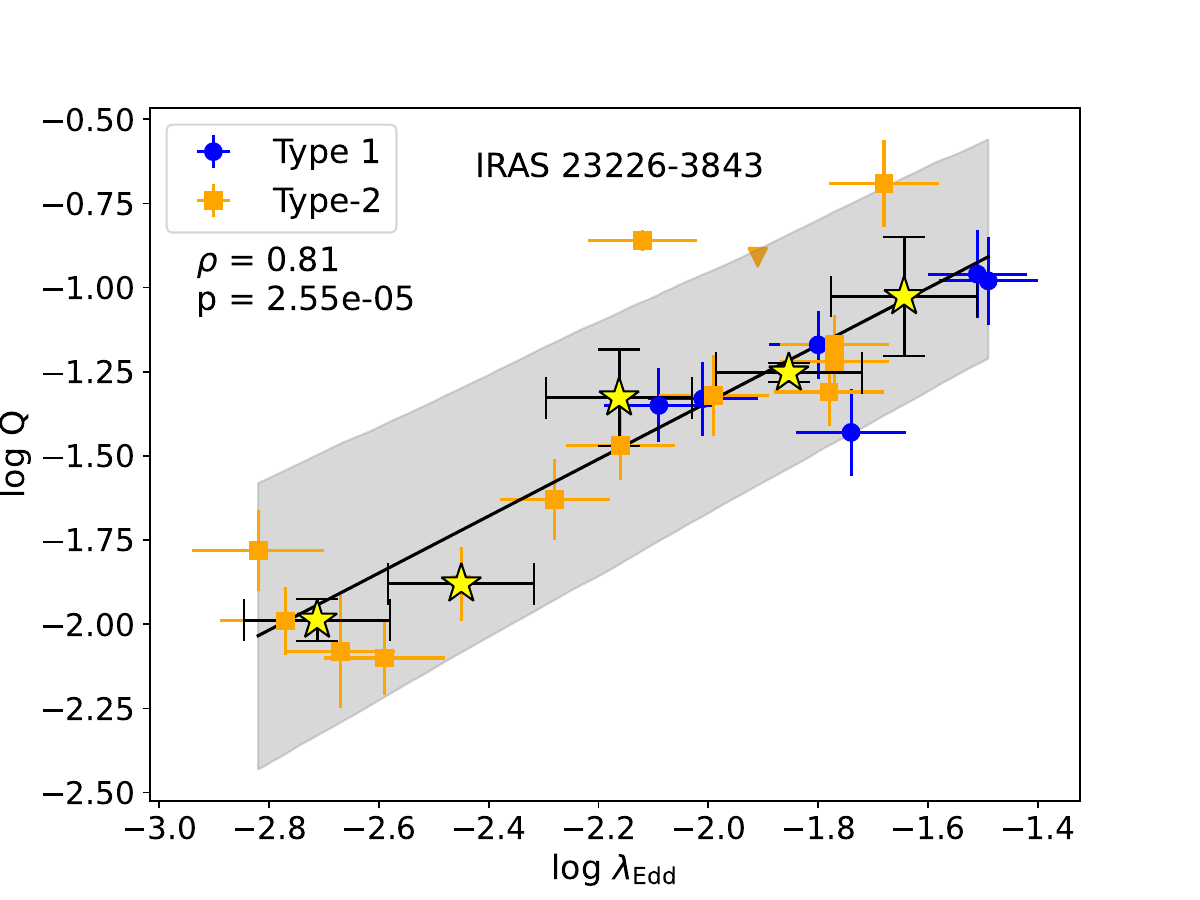}
\caption{Variation of soft-excess strength ($Q$) as a function of Eddington ratio ($\lambda_{\rm Edd}$). The black solid line represents the linear best-fit. The gray region marks the $1\sigma$ scatter.}
\label{fig:i22_qed}
\end{figure}

Figure~\ref{fig:i22_ged} displays the relation between $\Gamma$ and $\lambda_{\rm Edd}$. The $\Gamma-\lambda_{\rm Edd}$ shows a `V'-shaped relation with a break at $\log \lambda_{\rm Edd} \sim -2.1$, indicating a possible change of inner accretion flow. Figure~\ref{fig:i22_qed} shows the variation of $Q$ as a function of $\lambda_{\rm Edd}$. A positive correlation between these two parameters indicates that SE emission is strongly linked with the accretion rate. These relations are consistent with the warm Comptonization scenario as the origin of the SE.

\begin{table*}
\centering
\caption{Broadband X-ray properties of IRAS\,23226--3843}
\small
\begin{tabular}{lcccccccccc}
\hline \hline
Epoch & $\log L_{\rm PC}^{\rm 2-10}$ & $\log L_{\rm SE}^{\rm 0.5-2}$ & $\Gamma$ & $E_{\rm cut}$ & $\log R_{\rm S}$ & $kT_{\rm BB}$ & $\log Q$ & $\log \lambda_{\rm Edd}$ & $\chi^2/{\rm dof}$ \\
      &   &    &   & (keV) &   &  (eV) &  & \\
(1) & (2) & (3) & (4) & (5) & (6) & (7) & (8) & (9) & (10) \\
\hline
E1	&$	42.49	\pm	0.06	$&$	<41.14			$&$	1.89	\pm	0.03	$&$	>51					$&$	-0.44	\pm	0.09	$&$	120	^*		$&$	-1.35	\pm	0.06	$&$	-2.34	\pm	0.08	$&	158/201	\\
E2	&$	42.10	\pm	0.03	$&$	<41.10			$&$	1.94	\pm	0.05	$&$	200	^*				$&$		-		$&$	120	^*		$&$	<-1.00			$&$	-2.74	\pm	0.06	$&	86/85	\\
E3	&$	42.02	\pm	0.03	$&$	41.03	\pm	0.43	$&$	2.04	\pm	0.03	$&$	>55					$&$	-1.74	\pm	0.19	$&$	112	\pm	19	$&$	-0.99	\pm	0.43	$&$	-2.82	\pm	0.06	$&	157/168	\\
E4	&$	43.10	\pm	0.01	$&$	42.06	\pm	0.14	$&$	1.91	\pm	0.02	$&$	136	^{+99}_{-84}				$&$	-0.40	\pm	0.09	$&$	133	\pm	13	$&$	-1.04	\pm	0.14	$&$	-1.66	\pm	0.05	$&	1569/1489	\\
\hline
\end{tabular}
\label{tab:i22}
\end{table*}

\section{Soft-excess model}
\label{sec:se-mod}
We modeled SE emission using a phenomenological blackbody (BB) model. We chose this model as we wanted to test both warm corona and ionized reflection scenarios as the origin of the SE. 
We note that in warm Comptonization scenario, a Comptonization model such as \textsc{comptt} or \textsc{nthcomp} would provide a more physical description of the SE emission. To test this, we replaced the blackbody component with \textsc{nthcomp} and found that the flux in the 0.5--2\,keV range remains unchanged. However, when integrating over the broader 0.001--10\,keV range, \textsc{nthcomp} yields fluxes higher by $\sim$0.2\,dex compared to the blackbody model. This difference does not significantly affect $L_{\rm bol}$, as the bulk of the emission originates from the \textsc{diskbb} and power-law components.

\end{document}